%% file: main.tex
\documentclass[sigconf]{acmart}

\graphicspath{{img/}}  %% Le répertoire pour vos images
\def\venue{IHM~'20'21} %% Le nom abrégé de la conférence

\usepackage[french,english]{babel}
\newcommand{\en}[1]{\foreignlanguage{english}{#1}}
\newcommand{\fr}[1]{\foreignlanguage{french}{#1}}

\AtBeginDocument{%
  \providecommand\BibTeX{{%
    \normalfont B\kern-0.5em{\scshape i\kern-0.25em b}\kern-0.8em\TeX}}}

\setcopyright{acmcopyright}
\copyrightyear{2021}
\acmYear{2021}
\acmDOI{10.1145/XXXXXXXXX.XXXXXXXX}

\acmConference[\venue{}]{32\textsuperscript{e} conférence Francophone
  sur l'Interaction Homme-Machine}{April 13--16, 2021}{Metz, France}
\acmBooktitle{\venue{}~: 32\textsuperscript{e} conférence Francophone
  sur l'Interaction Homme-Machine, April 13--16, 2021, Metz, France}
\acmPrice{15.00}
\acmISBN{XXX-X-XXXX-XXXX-X/XX/XX}

\usepackage{capt-of}
\usepackage{accsupp}
\usepackage{pdfcomment}

\usepackage{balance}
\usepackage[utf8]{inputenc}

\begin{document}
\selectlanguage{english}
\sloppy

%%
%% TITRE : OBLIGATOIRE en français ET en anglais
\def\titreEN{"Can I Touch This?": Survey of Virtual Reality Interactions via Haptic Solutions}
\def\titreFR{Revue de Littérature des Interactions en Réalité Virtuelle par le biais de Solutions Haptiques}
\def\titreSHORT{"Can I Touch This?"} %% Si le titre ANGLAIS de l'article est trop long pour les entêtes ou pieds des pages

\title[\en{\titreSHORT}]{\en{\titreEN}}
\subtitle{\fr{\titreFR}}


%\author{Elodie Bouzbib$^{1,2}$ 
%              Gilles Bailly$^{1}$ 
%              Sinan Haliyo$^{1}$ 
%              Pascal Frey$^{2}$}
%\affiliation{$^{1}$Sorbonne Universit\'e, CNRS, ISIR. Paris, France $^{2}$Sorbonne Universit\'e, ISCD. Paris, France}
% \affiliation{$^{2}$ Sorbonne Universit\'e, ISCD. Paris, France}

% \numberofaffiliations{1}
% \author{
% \auteurs{4em}{Elodie Bouzbib$^{1,2}$\AND 
%               Gilles Bailly$^{1}$ \AND 
%               Sinan Haliyo$^{1}$ \AND 
%               Pascal Frey$^{2}$}
%   \affil{
%     \affaddr{$^{1}$Sorbonne Universit\'e, CNRS, ISIR. Paris, France}\\
%     \affaddr{$^2$Sorbonne Universit\'e, ISCD. Paris, France}%
% \email{}
% \affiliation{%
%   \institution{}
%   \city{}
%   \country{}
% }

%\author{Elodie Bouzbib}
%% \authornotemark[1]
%% \email{c.dupont@universite-ihm.ca}
%\affiliation{%
%	\institution{ISIR. Sorbonne Universit\'e}
%	\institution{ISCD. Sorbonne Universit\'e}
%	\city{Paris}
%	%   \state{Super Province}
%	\country{France}
%}
\author{Elodie Bouzbib}
% \email{c.dupont@universite-ihm.ca}
\affiliation{%
	\institution{ISIR. Sorbonne Universit\'e}
	\institution{ISCD. Sorbonne Universit\'e}
	\city{Paris}
	%   \state{Super Province}
	\country{France}
}
\author{Gilles Bailly}
\author{Sinan Haliyo}
\affiliation{%
	\institution{ISIR. Sorbonne Universit\'e}
	\city{Paris}
	%   \state{Super Province}
	\country{France}
}

\author{Pascal Frey}
% \email{c.dupont@universite-ihm.ca}
\affiliation{%
	\institution{ISCD. Sorbonne Universit\'e}
	\city{Paris}
	%   \state{Super Province}
	\country{France}
}

%\author{Sinan Haliyo}
%% \authornotemark[1]
%% \email{c.dupont@universite-ihm.ca}
%\affiliation{%
%	\institution{ISIR. Sorbonne Universit\'e}
%	\city{Paris}
%	%   \state{Super Province}
%	\country{France}
%}

%\author{Pascal Frey}
%% \authornotemark[1]
%% \email{c.dupont@universite-ihm.ca}
%\affiliation{%
%	\institution{ISCD. Sorbonne Universit\'e}
%	\city{Paris}
%	%   \state{Super Province}
%	\country{France}
%}

% \author{Claude Dupont}
% \email{c.dupont@universite-ihm.ca}
% \affiliation{%
%   \institution{Université de l'IHM}
%   \city{Super Ville}
%   \state{Super Province}
%   \country{Canada}
% }

% \author{Camille Anova}
% \email{camille@hci-corporation.ch}
% \affiliation{%
%   \institution{HCI Corporation}
%   \city{Green Town}
%   \country{Suisse}
% }

% \author{Charlie Goms}
% \email{charlie@hci-consortium.be}
% \affiliation{%
%   \institution{HCI Consortium}
% }


%% Si la liste des auteurs est trop longue en entête ou pied de page
\renewcommand{\shortauthors}{Bouzbib et al.}

\begin{abstract} %% OBLIGATOIRE
\en{
Haptic feedback has become crucial to enhance the user experiences in Virtual Reality (VR).
This justifies the sudden burst of novel haptic solutions proposed these past years in the HCI community. This article is a survey of Virtual Reality interactions, relying on haptic devices.
% From a \textit{technological} perspective, 
We propose two dimensions to describe and compare the current haptic solutions: their degree of physicality, as well as their degree of actuation. 
% From an \textit{interaction} perspective, we describe the users' human factors, and their involvement in each design space.
We depict a compromise between the user and the designer, highlighting how the range of required or proposed stimulation in VR is opposed to the haptic interfaces flexibility and their deployment in real-life use-cases.
% the operational cost
% of haptic solutions in their use-cases, and justifying the current lack of deployment outside research labs.
This paper (1) outlines the variety of haptic solutions and provides a novel perspective for analysing their associated interactions,
(2) highlights the limits of the current evaluation criteria regarding these interactions, and finally (3)
% generates recommendations regarding the use of "Robotic Graphics" or "encountered-type of haptic interfaces" as future research leads.
reflects the interaction, operation and conception potentials of "encountered-type of haptic devices".
}
\end{abstract}

%% french abstract %%%%%%%%%%%%%%%%%%%%%%%%%%%%%%%%%%%%%%%%%%%%%%%%%%%%%%%%%%
\begin{resume} %% OBLIGATOIRE
\fr{
Le retour haptique est devenu essentiel pour améliorer l'expérience utilisateur en Réalité Virtuelle (RV). 
C'est pourquoi nous observons une explosion du nombre de solutions haptiques proposées ces dernières années en IHM. Cet article est une revue de littérature des interactions en RV s'appuyant sur des dispositifs haptiques.
% D'un point de vue technologique, n
Nous proposons deux dimensions pour décrire et comparer les solutions haptiques : leur degré de physicalité ainsi que leur degré de robotisation.
% D'un point de vue interactionnel, %nous discutons de
%l'intégration sensorielle de l'utilisateur en fonction des tâches disponibles lors d'une expérience.
Nous formulons un compromis utilisateur/concepteur, reflétant la variété des stimulations requises/proposées en RV, en opposition à la flexibilité des interfaces et leur déploiement en situation réelle.
Ce travail (1) offre un panorama des solutions haptiques en RV ainsi qu'un cadre d'analyse pour étudier les interactions associées, 
(2) souligne les limites des critères d'évaluation actuels pour ce type d'interactions, et finalement (3) reflète les potentiels interactionnel, opérationnel et conceptuel des interfaces haptiques "à contacts intermittents".
}
\end{resume}
%%%%%%%%%%%%%%%%%%%%%%%%%%%%%%%%%%%%%%%%%%%%%%%%%%%%%%%%%%%%%%%%%%%%%%%%%%%%%

\begin{CCSXML}
<ccs2012>
<concept>
<concept_id>10003120.10003121.10003124.10010866</concept_id>
<concept_desc>Human-centered computing~Virtual reality</concept_desc>
<concept_significance>500</concept_significance>
</concept>
<concept>
<concept_id>10003120.10003121.10003125.10011752</concept_id>
<concept_desc>Human-centered computing~Haptic devices</concept_desc>
<concept_significance>500</concept_significance>
</concept>
<concept>
<concept_id>10003120.10003123.10011758</concept_id>
<concept_desc>Human-centered computing~Interaction design theory, concepts and paradigms</concept_desc>
<concept_significance>500</concept_significance>
</concept>
<concept>
<concept_id>10003120.10003123.10010860.10010859</concept_id>
<concept_desc>Human-centered computing~User centered design</concept_desc>
<concept_significance>500</concept_significance>
</concept>
</ccs2012>
\end{CCSXML}

\ccsdesc[500]{Human-centered computing~Virtual reality}
\ccsdesc[500]{Human-centered computing~Haptic devices}
\ccsdesc[500]{Human-centered computing~Interaction design theory, concepts and paradigms}
%\ccsdesc[500]{Human-centered computing~User centered design}

%
% Keywords. The author(s) should pick words that accurately describe the work being
% presented. Separate the keywords with commas.

% \keywords{haptics, Virtual Reality, presence, human factors, tangible interface, haptic devices}%, physical presence}
\keywords{haptics, Virtual Reality, human factors, haptic devices}
%% french keywords %%%%%%%%%%%%%%%%%%%%%%%%%%%%%%%%%%%%%%%%%%%%%%%%%%%%%%%%%%
% \motscles{\fr{haptique, Réalité Virtuelle, présence, facteurs humains, interface tangible, dispositif haptique}}

\motscles{\fr{haptique, Réalité Virtuelle, facteurs humains, dispositif haptique}}
%, présence physique}} %% OBLIGATOIRE
%%%%%%%%%%%%%%%%%%%%%%%%%%%%%%%%%%%%%%%%%%%%%%%%%%%%%%%%%%%%%%%%%%%%%%%%%%%%%

\maketitle

%%==============================================
%% Le corps de l'article

\input{Intro.tex}
\input{Definitions}

% \input{Solutions}
\input{Opportunities}

\input{Challenges}

\section{Conclusion}
We analysed in this paper haptic interactions in VR and their corresponding haptic solutions. We analyzed them from both the user and designer perspectives by considering interaction opportunities and visuo-haptic consistency, as well as implementation and operation costs. We proposed a novel framework to classify haptic displays, through a two-dimension design space: the interfaces' degree of physicality and degree of actuation.
%also presented a design space relying on two dimensions: degree of physicality and degree of actuation. %We especially highlighted 

We then evaluated these latter solutions from an interaction and conception perspectives.
Implementation-wise, we evaluated the interfaces robustness, their ease of use as well as their safety considerations. 
From an operation perspective, we also evaluated the costs of the proposed solutions. 
% if the proposed solutions required an additional operator cost.%, on top of the hardware/software ones. 

This survey highlights the variety of props, tasks and haptic features that a haptic solution can potentially provide in VR.
This survey can be used to analytically evaluate the existing haptic interactions. It can also help VR designers to choose the desired haptic interaction and/or haptic solution depending on their needs (tasks, workspace, use-cases etc).

We believe that combining multiple haptic solutions benefits the user experience, as it optimises the above criteria. 
Encountered-type of haptic interfaces were then highlighted as they already combine multiple interaction techniques: they displace passive props in potentially
large VR arenas and allow for numerous tasks, such as navigation, exploration, manipulation, and even allow the user to be interacted with. %With the appropriate physical:virtual mapping, they also re

\bibliographystyle{ACM-Reference-Format}
\bibliography{references}

\appendix

\end{document}

%% file: Intro.tex
\section{Introduction}
%Combining haptics and VR is a timely topic, pursuing the Ultimate Display quest \cite{sutherland_ultimate_1965}. 
% In the last few years, the terms "Virtual Reality" and "Haptics" have been amongst the most quoted keywords in HCI conferences such as ACM CHI or ACM UIST. Indeed, Head-Mounted Displays (HMDs) are now affordable and provide high quality visual and audio feedback. Enhancing VR experience through the sense of touch (haptic feedback), is thus becoming a main challenge. It results that a large variety of haptic solutions have recently been proposed but it is difficult to have a clear understanding of the design possibilities, to ascertain their similarities and differences.

In the last few years, the terms "Virtual Reality" and "Haptics" have been amongst the most quoted keywords in HCI conferences such as ACM CHI or ACM UIST. Indeed, Head-Mounted Displays (HMDs) are now affordable and provide high quality visual and audio feedback, but augmenting the experience by enhancing VR through the sense of touch (haptic feedback) has become a main challenge. A large variety of haptic solutions has currently been proposed, nonetheless they have highly different scopes, due to the wide range of haptic features. It is hence difficult to compare their similarities and differences and have a clear understanding of the design possibilities.
% \\In this paper, we present a survey of existing \textit{haptic interactions} in VR. We use the terms "haptic interactions" rather than "haptic devices" to reflect our focus on interactions, i.e. how  technologies influence users' behavior in VR (e.g. mobility) rather than on the technologies themselves. 
\\In this paper, we present a survey of existing \textit{haptic interactions} in VR. We use the terms "haptic interactions" %rather than "haptic devices" 
to emphasize the focus on the users actions, and to analyse how % technologies influence users' behavior in VR (e.g. mobility) rather than on the technologies themselves. 
the "haptic devices" influence their behaviours in VR.
\\\\We provide a synthesis %and an analysis 
of existing research on haptic interactions in VR
and depict, from the required haptic features stimulation and interaction opportunities, a design space discussing and classifying the associated haptic solutions
%and their associated solutions. To achieve this, we 
%present a design space (Figure X) to 
% discuss them 
according to two dimensions: their \textit{degree of physicality}, %of a haptic solution
i.e. their physical consistency and level of resemblance as to replicating an object, and their \textit{degree of actuation}, i.e. whether they rely on a motor-based hardware implementation enabling autonomous displacements of the interface (eg changing its shape or position) (Table \ref{table:table_dimensions}).\\
This design space is useful to characterize, classify and compare haptic interactions and the corresponding haptic solutions. 
We also propose two criteria, User experience and Conception costs, highlighting the implicit trade-offs between the quality of the \textit{user} experience and the %cost 
intricacy for the \textit{designer} %to build, deploy and maintain  %researcher/practitioner
% to conceive, 
to implement
these solutions. %in  and operational terms. 
Both of the user's and designer's perspectives are hence considered in a novel framework to evaluate haptic interactions.\\Finally, we illustrate the utility of our design space by analyzing and comparing four haptic solutions.
% In this survey, we provide a synthesis and an analysis of existing research on haptic interactions. To achieve this, we present a design space (Figure X) to discuss haptic interactions according to two dimensions: The \textit{Degree of physicality} of an haptic solution, i.e. XXX  and Degree of actuation, i.e. XX . This design space is useful to characterize, classify and compare interactions relying on haptic solutions. It also highlights the too often implicit trade-offs between the quality of the \textit{user} experience and the cost for the \textit{designer} to build, deploy and maintain these solutions. A key aspect of this work is thus to consider both the user and designer perspective offering a novel framework to evaluate haptic interactions. Finally, we illustrate the utility of our design space by analyzing and comparing four haptic solutions.
This analysis indicates that (1) the use of real props in a virtual environment benefits the user experience, but limits the interactions to the existing props available within the VR arena; (2) the use of robotised interfaces enables more various interactions; (3) combining them offers the best user experience/design cost trade-off; (4) current evaluation methods do not allow a fair representation and comparison of haptic solutions.
\\\\We hence propose  %recommendations including 
guidelines to evaluate haptic interactions from both the user and designer perspectives. %as well as recommendations such as intertwining interfaces to expand the users haptic opportunities.
We also outline how intertwining interfaces can expand haptic opportunities, by conducting a deeper investigation on Robotic Graphics interfaces \cite{mcneely_robotic_1993} . 
%or conducting a deeper investigation of the opportunities provided by Robotic Graphics \cite{mcneely_robotic_1993}. 
Indeed, in the quest of the Ultimate Display \cite{sutherland_ultimate_1965}, these show (a) the largest variety of interactions, (b) the most reliable interfaces through their automation, and (c) the most natural interactions as they encounter the users at their positions of interest without further notice.

%% file: Definitions.tex
\section{Background}

Surveys in Virtual Reality consider the technology itself and its limits \cite{zhao_survey_2009, zhou_virtual_2009}, or more specifically its use-case scenarios. VR is indeed used in industries \cite{berg_industry_2017-1, zimmermann_virtual_2008}, healthcare \cite{moline_virtual_1997}, or in gaming. In gaming, the concerns are mainly regarding the evaluation protocols \cite{merino_evaluating_2020}, ie the presence \cite{schuemie_research_2001} and its related questionnaires \cite{schwind_using_2019, usoh_using_2000}. Surveys for instance compare the results whenever the questionnaires are asked in VR or in the real world \cite{alexandrovsky_examining_2020, putze_breaking_2020}. The user behaviour in VR is also analysed, through gesture recognition \cite{sagayam_hand_2017} or system control techniques (eg menus) \cite{bowman_design_2001}.
\\The research areas are coincidentally almost similar in haptics. Indeed, surveys analyse haptics themselves \cite{varalakshmi_haptics_2012}, haptic devices \cite{seifi_haptipedia:_2019, rakkolainen_survey_2020, talvas_survey_2014, hayward_it_2007} or examine the scenarios which benefit from a stimulation of the haptic cues. Haptics are used in telemanipulation \cite{galambos_vibrotactile_2012}, for training in the industry \cite{xia_haptics_2016, bloomfield_taxonomy_2003} or for healthcare purposes \cite{coles_role_2011, rangarajan_systematic_2020}, or in gaming \cite{kim_defining_2020}.

Finally, some surveys have been proposed at the intersection of VR and haptics and focus either on specific methods (pseudo-haptic feedback) \cite{lecuyer_simulating_2009}, technology according to stimulated haptic features (temperature, shape, skin stretch, pressure)  \cite{wang_multimodal_2020, dominjon_novel_2007} or the motivations and applications of each haptic device category \cite{wang_haptic_2020}. In contrast our survey outlines the variety of haptic interactions and technologies in VR and provides a framework to analyse them.
%their associated solutions.

%which is included in our design space.

% \\ \\Haptics and VR are hence combined in multiple fields, as they benefit from each other and are naturally interlinked. For instance, Lécuyer et al. exploit virtual reality to enhance the users vision and analyse how it affects haptic feedback
% % pseudo-haptic feedback 
% \cite{lecuyer_simulating_2009}. This technique, "\textit{pseudo-haptic feedback}", tricks the users' perception into feeling virtual objects' stiffness, texture, mass. Many more haptic features can be stimulated, such as temperature, shape, skin stretch, pressure. In these regards, Wang et al. utilise haptic features to classify haptic devices and actuation methods \cite{wang_multimodal_2020, dominjon_novel_2007}, to emphasise afterwards the motivations and applications of each haptic device category \cite{wang_haptic_2020}.\\

% Our survey outlines the variety of haptic interactions in VR and provides a novel perspective to analyse their associated solutions. 

%We discriminate the interactions %through the methods proposed by their respective solutions.
%depending on human factors and currently proposed methods, then classify these respective solutions according to two dimensions: their degree of physicality and actuation.

\section{Scope and Definitions}

% Since the democratization of Virtual Reality through affordable HMDs with Oculus or HTC, it does not only consists in gaming anymore, but extends its fields to simulation rooms, training, or even therapy. %While controllers might be sufficient for gaming purposes, training for instance requires more natural interactions. 
% While presence heavily relies on the human senses, the haptics cues in controllers only consists in small vibrations and are not explored to their fullest yet. Haptics include multiple features other than vibrations, but also stimulation can be increased to a larger scale than the users' hands.

% We first explain the scope of this paper, then define the multiple aspects behind \textit{haptics}, to finally understand how research in VR can lead to research in haptics and why this relation is reciprocal.

% We will define this in a first subsection. In a second part, we will define the different ways to render a physical interaction in VR. This includes hardware/mechanical techniques (devices) but also software ones (graphics).
% We will then refine the interactions that are currently enabled in VR, through tasks definitions and scenario types. We will finally depict how common presence questionnaires and experimental protocols are outdated and unable to fairly represent the advantages and drawbacks of the novel interaction techniques.

% \subsection{Scope}

The scope of this article is to analyse how a single user interacts and is provided with believable haptic feedback in Virtual Reality \cite{magnenat-thalmann_believability_2005}. We thus define the terms "virtual reality" and "haptics" and how they are related.
%In this section, we thus define Virtual re
%This paper is at the intersection of Virtual Reality and Haptics.

%, through both actuated interfaces and the use of real objects.

\subsection{Virtual Reality}

Virtual reality corresponds to a 3D artificial numeric environment in which users are immersed in. The environment can be projected onto a large screen, in a simulation platform for instance, or multiple ones, such as with CAVE technology (where the image is projected onto at least 3 distinct walls of a room-scale arena). 
In this survey, we consider an artificial reality \cite{wexelblat_virtual_1993} where users do not perceive their physical vicinity: the outside world is not noticeable and users are fully immersed through a head-mounted display (HMD). For instance, augmented reality (AR), where the physical environment is augmented with virtual artefacts, is out of our scope.
\\Through a Head Mounted Display (HMD), Virtual reality creates immersive experiences for the users. These are only limited by the designers' imagination, and are evaluated through \textit{presence}. Presence is defined as the "subjective experience of being in one place, even when one is physically situated in another" \cite{witmer_measuring_1998, slater_measuring_1999}. %\sout{, as opposed to \textit{immersion}, "an objective description of the system, such a field of view and display resolution"} \cite{schuemie_research_2001}.
It quantifies the users' involvement and naturalness of interactions through control, sensory, distraction and realism factors. This heavily relies on the sensory input and output channels, however, as VR was mainly integrating audio and visual cues, quantifying the haptic contribution in an experience remains difficult.

\subsection{Haptics: Tactile vs Kinesthetic Perception}

Haptics is the general term for the sense of touch. They are a combination of two cues: tactile and kinesthetic. 
The tactile cues are developed through the skin, while the kinesthetic ones come from proprioception and are through the muscles and the tendons.

\subsubsection{Tactile cues:} The skin is composed of four types of mechanoreceptors \cite{lederman_haptic_2009}. The first ones, "Merkel nerve endings", transmit mechanical pressure, position and shapes or edges. They are stimulated whilst reading Braille for instance. The second ones, "Ruffini corpuscle end-organ", are sensitive to skin stretch and provide both pressure and slippage information. The third ones are the "Pacinian corpuscles", which are sensitive to vibration and pressure. The last ones, "Meissner's corpuscles", are highly sensitive and provide light touch and vibrations information.
It also contains thermoreceptors, which transmit information about temperature: the Ruffini endings respond to warmth, while the Krause ones detect cold. % (see Figure \ref{fig:skin}).
Through tactile cues, the human can hence feel shapes or edges, pressure, vibrations or temperature changes.

% \begin{figure}[h]
%   \centering
% \pdftooltip{  \includegraphics[width=0.5\linewidth]{img/Skin.png}}{A cut section of the skin is displayed: it describes the positions or all the above mechanoreceptors within the epidermis and the dermis.}
%   \caption{Schematic Cut Section of the Skin \cite{noauthor_362a_2018}: Four types mechanoreceptors (Merkel, Ruffini, Pacinian, Meissner) are located within the epidermis and the dermis, along with thermoreceptors (Ruffini, Krause).}
%   \label{fig:skin}
% \end{figure}

\subsubsection{Kinesthetic cues:} The kinesthetic cues rely on proprioception, ie the perception and the awareness of our own body parts positions and movements. Mechanoreceptors into the muscles, the "spindles", communicate to the nervous system information the forces muscle generate, as well as their length change \cite{jones_kinesthetic_2000}. The primary type of spindle is sensitive to the velocity and acceleration of a muscle contraction or limb movement, while the second type provides information about static muscle length or limb positions. 
Kinesthetic cues hence allow to feel forces, as well as perceiving weights or inertia.

\subsection{VR \& Haptics}
Whenever we touch or manipulate an object, the combination of these two previous cues allows to understand its material, but also its shape and the constraints it implies to the user. On the one side, adding physical presence \cite{lepecq_afforded_2008} through haptic feedback in VR
%integrating haptics and having a physical presence \cite{lepecq_afforded_2008} in VR experiences is timely. It is 
enhances the users' immersion, even at an emotional and physiological scale: the heart rate of a user can literally increase with the use of haptics through real objects \cite{insko_passive_2001}. Haptics are also required for interacting with the environment: the user needs to control the changes in the environment \cite{held_telepresence_1992} and to be aware of the modifications he physically has made (eg moving virtual objects, pushing a button). On the other side, haptics can benefit from VR. For instance, Lécuyer et al. leverage the users vision and analyse how it affects their haptic feedback
% pseudo-haptic feedback 
\cite{lecuyer_simulating_2009}. This approach, "\textit{pseudo-haptic feedback}", tricks the users' perception into feeling virtual objects' stiffness, texture, mass. Many more haptic features can be stimulated, such as temperature, shape, skin stretch, pressure.

\section{Analyzing haptic interactions}
The main objective of this survey is to provide analytical tools to evaluate and compare haptic interactions. %To achieve this, we present a design space and a hierarchical set of evaluation criteria. 

\subsection{Design space}
We propose a two-dimension framework to discuss and classify haptic solutions in VR (see Table \ref{table:table_dimensions}). %The design space relies on two dimensions. 

The first dimension is their \textit{degree of physicality}, ie how the haptic perception is tangible/physically consistent/resembling with the virtual objects.
This dimension is drawn as a continuum, from "no physicality" to "real objects" (see Figure \ref{fig:physicality}). We find that this continuum can be discretised as a two-category section: whether they use real objects or not.

The second orthogonal dimension is their \textit{degree of actuation}, ie whether haptic solutions rely on a motor-based hardware implementation enabling autonomous displacements (eg enabling to change its shape, position etc). % Passive haptics à enlever ou redefinir

%Haptic solutions can then be classified according to this two-dimension design space in Table \ref{table:table_dimensions}. 
%This two-dimension space is illustrated Table \ref{table:table_dimensions}.

\subsection{Analysis criteria}
We consider two main criteria to analyse haptic interactions in VR. They cover both the user and designer perspectives. 

The \textbf{User} experience is the first criterion and includes two aspects: interaction opportunities and visuo-haptic consistency/discrepancy. Interaction opportunities represent to which extent haptic solutions allow \textit{users} to interact/act (e.g navigate, explore, manipulate) in a VR scene as opposed as in the real world. Visuo-haptic consistency/discrepancy refers to the tactile and kinesthetic perceptual rendering of these interactions. These two sub-criteria are complementary focusing on both action and perception.

%This criterion thus goes beyond the quality of haptic rendering as it also includes how it constrains users' behavior.

The second criterion is the conception cost, i.e. the challenges \textbf{Designers} should address when designing haptic interactions. We distinguish implementation and operation costs. Implementation costs include several technical aspects related to the acceptability of a haptic solution such as safety, robustness and ease-of-use \cite{dominjon_novel_2007}. Operation costs include the financial and human costs required to deploy these technologies. 
%These criteria thus address both users and designers perspectives.

%We present, compare and analyse haptic solutions using this two-dimension space in Table \ref{table:table_dimensions} as well as evaluation criteria.

%enables us to evaluate and analyse more distinctly the challenges when designing a haptic solution (Section \ref{Actuation}), and consequently to understand the trade-offs between users opportunities and designers' challenges. 

% pas displacements % pas ça comme propriété pour justifier degres d'actuation
% Indeed, the real world is constituted from physical robust constrains, that affect humans' behaviours. 
% Replicating these constrains through actuation and robotics provides potentially more robust interactions, with a certain reliability and repeatability. 
%These displacements can create various constraints and stimulation to the users, and widen the interfaces' design spaces. They broaden interactions to complex shapes that would not be available using only simple props from the physical environment, or even allow for one physical object for overlay multiple ones \cite{he_physhare:_2017}.
%Haptic solutions are classified according to this two-dimension design space in Table \ref{table:table_dimensions}.
% Indeed, even though real props are an advantage in VR, their main drawback is they cannot displace themselves or modify their haptic stimulation. 

%\gb{This dimension is especially useful when designing haptic solutions to understand the trade-offs between users opportunities and designers' challenges. }

\subsection{Application}
We rely on this design space and criteria to highlight and understand the trade-offs between the user's interactions opportunities in VR, and the designers' challenges in conception. This survey offers a novel perspective for researchers to study haptic interactions in VR. It can be used to compare and analytically evaluate existing haptic interactions. For a given application, designers can evaluate the most adapted haptic interaction. For a given technique, they can evaluate a haptic solution depending on their needs (tasks, workspace, use-cases etc).\\
.%, (2) harmoniously integrated \cite{kim_defining_2020} 
% with a natural mode of control \cite{witmer_measuring_1998}, for instance with bare-hands, 
% their conceptual and implementation/operational costs, and their use-cases and deployment opportunities.
We first discuss haptic interactions from the User perspective (Section \ref{Opportunities} - Interaction opportunities, Section \ref{Physicality} - Visuo-Haptic Consistency/Discrepancy). We then adopt the designer perspective in Section \ref{Actuation}. We use our design space on Sections \ref{Physicality} and \ref{Actuation}, which emphasize haptic solutions.

%now present the interaction opportunities (first criterion) to motivate the need of novel haptic interactions. We then describe the different technologies aiming to enable these haptic interactions. We finally analyze these technologies regarding the conception cost (second criterion).
%\newpage
%challenge.
%present  cover the interaction techniques enabling users to physically interact with a virtual environment and detail the opportunities through the available tasks.

% We first discuss user interactions opportunities to secondly analyse the challenges in devices implementation.
% and to compare devices implementation to discuss perspectives and challenges of creating interactions with believable haptic feedback in VR \cite{magnenat-thalmann_believability_2005}. 
% Our research interests include both actuated interfaces and the use of real props. It covers the interaction techniques enabling to physically interact with a virtual environment and detail the user opportunities through the available tasks.

\begin{table}[t]
	\centering
	\pdftooltip{  \includegraphics[width=0.8\linewidth]{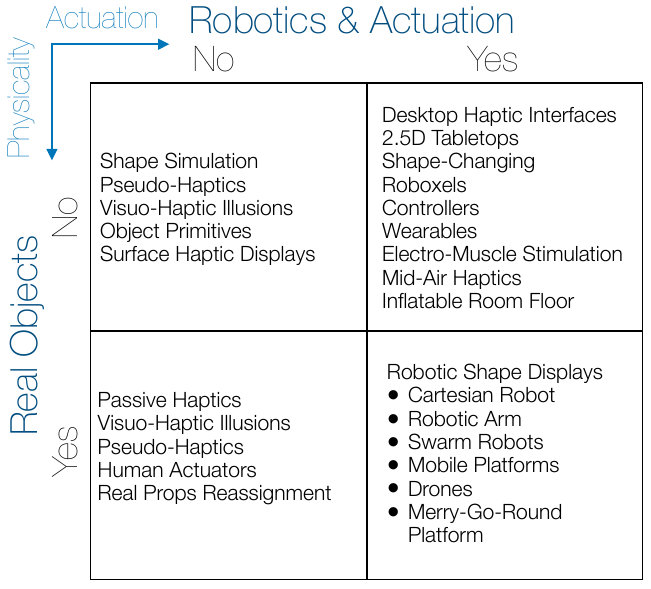}}{We propose two dimensions to classify current technologies: the degree of physicality as well as the degree of actuation. Four categories are hence drawn in this figure: Top Left: No robotics, No real objects; Top Right: Robotics, No real objects; Bottom Right: Robotics, Real objects; Bottom Left: No robotics, Real objects. We respectively displayed current technologies and interaction techniques in the category they belong to.}
	\caption{We propose two dimensions to classify current technologies: their degrees of physicality and actuation. %The previously mentioned technologies %(simulating or exploiting objects) 
	%Technologies types are hence displayed in these four distinct categories.
	}
	\label{table:table_dimensions}
\end{table}

% Haptics can be felt through the entire body, nonetheless current controllers only simulate the users' hands with simple vibrations. If a user were to touch a wall for instance, even though a collision could be noticed visually, nothing would prevent the user to go through it, as no kinesthetic constrain can be felt. Techniques that are developed in the next subsections hence tackle this lack of haptic feedback by enhancing physical interactions between the user and the tangible interfaces.

%% file: Opportunities.tex
\begin{figure*}[t]
  \centering
\pdftooltip{  \includegraphics[width=\linewidth]{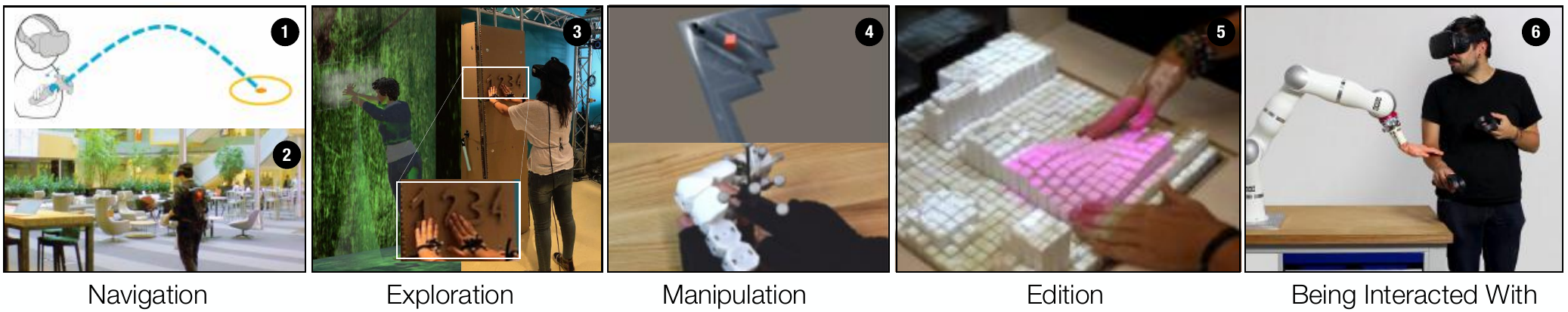}}{}
  \caption{Tasks in VR: (1) Navigation through Point \& Teleport \cite{funk_assessing_2019}; (2) Navigation through a building, using redirection \cite{cheng_vroamer:_2019}; (3) Exploration with Bare-Hands: A user finds an invisible haptic code \cite{bouzbib_covr_2020}; (4) Manipulation: Haptic proxies rearrange themselves to form a plane the user can manipulate \cite{zhao_robotic_2017}; (5) Edition: the user changes the shape of a haptic 2.5D tabletop \cite{nakagaki_materiable:_2016}; (6) The user is interacted with by a robotic arm to feel emotions \cite{teyssier_conveying_2020}.}
  \label{fig:tasks}
\end{figure*}

\section{Interaction Opportunities} \label{Opportunities}
In the real world, users \textbf{move freely} without constraints, \textbf{pick any object of their environment} and then \textbf{interact with their bare-hands}. They also can \textbf{be interacted with}, from the environment (wind, unexpected obstacles) or from other users, for instance to catch their attention or to lead them somewhere. A natural environment also naturally physically constrains users through their \textbf{entire body}. 

In this section, we discuss the interaction opportunities in VR and the methods available to provide them. In particular, we discuss them through four main tasks: navigation, exploration, manipulation and edition. %It is important as several haptic solutions do not support all these interaction opportunities. 

\subsection{Navigation} %Immersion in VR first goes through the vision. 
We qualify a navigation task as the exploration of the environment through the vision and the ability to navigate through it via the users displacements. We identify three main techniques to navigate in VR. The two firsts rely on controllers and push buttons, where the users do not necessary physically move. The last one is more natural as it allows the users to walk in the VR arena.

% {Sans marche, on penserait que ca n'evoque pas l'haptique. Pourtant, pour naviguer lorsque la marche reelle n'est dispo, la navigation est faite avec des solutions haptiques. }

\subsubsection{Panning:} With grounded desktop haptic solutions, such as the Virtuose \cite{haption_virtuose_2019}, users need to push a button to clutch the devices and hence move within the environment.

\subsubsection{Point \& Teleport:} With ungrounded solutions, such as controllers, the common technique is teleportation. Users point their controllers \cite{baloup_pointing_2018} to predetermined teleportation target areas, and are displaced in position but also in orientation \cite{funk_assessing_2019} (Figure \ref{fig:tasks} - 1).

\subsubsection{Real Walking:} Real walking in VR, \textit{"perambulation"}, has shown the best immersion and presence results \cite{usoh_walking_1999, steinicke_human_2013} because it relies on proprioception and kinesthetic feedback through the legs and gait awareness.
Nonetheless, VR arenas are not infinite and HMD have a limited tracking space, hence methods need to be developed for the user to be able to move to any location of interest. One approach is 
to mount the previously discussed grounded desktop haptic solutions over mobile \cite{satler_control_2011, lee_mobile_2007, formaglio_performance_2005, lee_system_2009, nitzsche_design_2003, pavlik_interacting_2013} or wearable \cite{barnaby_mantis:_2019} interfaces. 
%to combine walking and the previous solutions, ie 
%When walking is not available, these methods are unexpectedly relying on haptic solutions, through push buttons or controllers.
% It also does not require the users to hold or wear an interface. However, obstacles or limited tracking space imply the development of novel techniques. 
%In these regards, for the user to be able to freely move, 
%the previously discussed grounded desktop haptic solutions are for instance mounted on wearable \cite{barnaby_mantis:_2019} or robotic ungrounded mobile interfaces \cite{satler_control_2011, lee_mobile_2007, formaglio_performance_2005, lee_system_2009, nitzsche_design_2003, pavlik_interacting_2013}.
Users however still have to continuously maintain the handle in their palm. Other interfaces hence allow for free-hands Room-Scale VR \cite{bouzbib_covr_2020, wang_movevr_2020, yixian_zoomwalls_2020}. For the users to perambulate in an infinite workspace, the virtual environment can also visually be warped for the users to unconsciously modify their trajectory or avoid obstacles \cite{cheng_vroamer:_2019, razzaque_eurographics_2001, yang_dreamwalker:_2019} (Figure \ref{fig:tasks} - 2). This infinite redirection can also be provided from Electro-Muscle Stimulation (EMS) on the users' legs \cite{auda_around_2019}, with wearable electrodes. The user can also wear actuated stilts to perceive staircases \cite{schmidt_level-ups_2015} or a vibrating shoe to perceive virtual materials \cite{strohmeier_barefoot_2020}.
% Users can also wear electrodes on their legs, to be infinitely redirected through Electro-Muscle Stimulation (EMS) \cite{auda_around_2019}, actuated stilts under their feet, to perceive staircases \cite{schmidt_level-ups_2015} or even receive vibrations from a shoe to perceive virtual materials \cite{strohmeier_barefoot_2020}. 
To remain unencumbered from these wearable techniques, the VR arena can also include robotised techniques: users can for instance walk on treadmills \cite{vonach_vrrobot:_2017, frissen_enabling_2013}, or on movable platforms that encounter their feet \cite{iwata_circulafloor_2005, iwata_locomotion_2013}.
% VRSneaky? walking with sound - tu peux te faufiler discretos pcq ils ont rajoute du son.. nul

% \begin{figure*}[t]
%   \centering
% \pdftooltip{  \includegraphics[width=\linewidth]{img/Tasks.pdf}}{}
%   \caption{Tasks in VR: (1) Navigation through Point \& Teleport \cite{funk_assessing_2019}; (2) Navigation through a building, using redirection \cite{cheng_vroamer:_2019}; (3) Exploration with Bare-Hands: A user finds a haptic code, hidden through the vision \cite{bouzbib_covr_2020}; (4) Manipulation: Haptic proxies rearranging themselves to form a handle, which position changes the light intensity \cite{zhao_robotic_2017}; (5) Edition: the user changes the shape of a haptic 2.5D tabletop rendering with his bare-hands \cite{nakagaki_materiable:_2016}; (6) The user can also be interacted with by a robotic arm to feel emotions \cite{teyssier_conveying_2020}.}
%   \label{fig:tasks}
% \end{figure*}

\subsection{Hand Interactions} %Haptics in a virtual environment need to be coherent with the vision. Both of these renderings have to be complementary for users to feel what they see \cite{yokokohji_wysiwyf_1999}. 
In the real world, bare-hands interaction is important to execute everyday tasks (exploration, manipulation, edition). However, in VR, users commonly have to hold controllers, wearables or handles, which create a discrepancy between what the users feel and see \cite{yokokohji_wysiwyf_1999}. These exploit the God-object principle \cite{zilles_constraint-based_1995}, as opposed to bare-hands Real-touch interactions.

%We discuss two approaches that will be useful for the three next tasks (exploration, manipulation and edition).

% from desktop haptic interfaces
%Everyday tasks (manipulation, exploration) in the real world are usually performed with the bare-hands. However, instead of interacting with their bare-hands \eb{in VR}, users more commonly have to hold controllers, wearables or handles% from desktop haptic interfaces
%. These latter exploit the god-object principle \cite{zilles_constraint-based_1995}. 
\subsubsection{God-Object:} 
The controller is considered as a continuity of the users' hands, represented by a proxy that does not undergo physics or rigid collisions, and is attached to a complementary rigid object with a spring-damper model. This latter hence moves along with the proxy, but is constrained by the environment. Whenever it does collide with an object of interest, the users perceive the previous spring-damper stiffness through kinesthetic feedback.
\\Users hence interact though a proxy, like a desktop mouse, which position is not co-located with the users' vision. Bare-hands interactions are not necessarily needed depending on the use-cases. For instance, in healthcare and surgery training, users are more likely to interact with a tool, such as a scalpel or a clamp. Continuously holding the god-object is hence not a constrain, however the co-location of vision and haptics is recommended \cite{ortega_prop-based_2005}.
\subsubsection{Real Touch:} In other scenarios, such as gaming, industry or tool training \cite{winther_design_2020, strandholt_knock_2020}, using the appropriate tools through props and real objects is more natural. The users however need to be able to reach them whenever required. Some interfaces (e.g. Robotic Graphics; see Section \ref{RoboticGraphics}) are hence developed in these regards, to encounter the users whenever they feel like interacting.

\subsection{Exploration} As opposed to the previous definition of "navigation", based on vision cues, an "exploration" task consists in the ability to touch the environment and understand its constraints. Exploring thoroughly an environment in VR can be done through different haptic features, and can improve the users depth perception \cite{makin_tactile_2019} or distances to an object. The different methods for exploring the environment are detailed in Section \ref{Physicality}.
\\Whenever a user is exploring the environment, shapes or textures are felt through his body displacements. %In order to explore properly, the user
He needs to move for his skin to stretch (through tactile cues) or his muscles to contract (through kinesthetic cues). 

%Whenever the exploration is direct, these stimuli are focused on the hands. With the god-object principle, the stimuli are focused on the arms muscles.
%, however the user can also explore the environment through his palm, or even other body parts.
\subsubsection{Through Tactile cues:} Whenever real props or material patches are available, users can naturally interact with their fingertips to feel different materials \cite{degraen_enhancing_2019, araujo_snake_2016}, textures \cite{benko_normaltouch_2016, lo_rollingstone_2018}, temperatures \cite{ziat_enhancing_2014} or to feel shapes and patterns through their bare-hands \cite{bouzbib_covr_2020, cheng_sparse_2017} (Figure \ref{fig:tasks} - 3).
When no physicality is available, a stimulation can still be performed. As seen in Surface haptic displays \cite{bau_teslatouch:_2010}, vibrations between 80 to 400 Hz are felt through the skin, hence users perceive stickiness, smoothness, pleasure, vibration or friction, and for instance explore a 3D terrain or volumetric data \cite{sinclair_touchmover_2014}. 
%To quantify the quality of the rendering, empirical studies measuring a stimuli JND (Just Noticeable Difference) enable to understand the required frequency thresholds for a majority of users to feel them.
Vibrations can then be combined with auditory and vision cues to render collisions in VR \cite{boldt_you_2018}.

\subsubsection{Through Kinesthetic cues:} Exploring the environment can also be done through kinesthetic cues: the users can literally be physically constrained to feel a wall, using electro-muscle stimulation (EMS) for instance \cite{lopes_providing_2017}.
With the god-object principle, users can also explore the environments' constraints through force-feedback. In this configuration, the users' arms are constrained by haptic desktop interfaces, providing strong enough forces to simulate a physical collision and discriminate shapes.
%The quality of the rendering in these conditions is measured through the force-feedback the device can provide: its translational force (usually $<$ 100N) and its rotational torque ($<$ 2 Nm) \cite{seifi_haptipedia:_2019}.

% \begin{figure*}[h]
%   \centering
% \pdftooltip{  \includegraphics[width=\linewidth]{img/physicality_maigre.pdf}}{}
%   \caption{Degree of physicality continuum in VR.%The different haptic technologies provide different degrees of physicality. 
%   (1) Haptic desktop devices enable to explore the environment through a handle \cite{lee_system_2009} with the god-object principle; (2) A controller \cite{benko_normaltouch_2016} or (3) a wearable \cite{fang_wireality_2020} simulate objects for exploration tasks; (4) Mid-air technology \cite{rakkolainen_survey_2020} create vibrations through the users hand to simulate an object; (5) Passive proxies are oriented for the user to feel objects' primitives with their hands \cite{cheng_sparse_2017}; (6) Objects from the environment are assigned to virtual props with the same primitives \cite{hettiarachchi_annexing_2016}; (7) Real objects or passive props can be manipulated and interacted with each other \cite{bouzbib_covr_2020}.}
%   \label{fig:physicality}
% \end{figure*}

% \vspace{-0.8cm}
\subsection{Manipulation} A manipulation task is performed whenever modifying the position and orientation of an object. 
%We discuss direct vs. pseudo-haptic manipulation.

\subsubsection{Direct Manipulation:} In VR, we distinguish the \textit{direct manipulation} \cite{bryson_direct_2005}, \textit{"the ability for a user to control objects in a virtual environment in a direct and natural way, much as objects are manipulated in the real world"} from pointing/selecting an object with controllers. A direct manipulation relies on the ability to hold an object with kinesthetic feedback, feel its weight \cite{lopes_providing_2017, zenner_drag::_2019, heo_thors_2018, sagheb_swish:_2019, zenner_shifty:_2017, shigeyama_transcalibur:_2019}, shape \cite{follmer_inform:_2013, sun_pacapa:_2019, kovacs_haptic_2020}, and constrains from the virtual environment, for instance when making objects interact with each other \cite{bouzbib_covr_2020}.
Changing a virtual object position or orientation can be used as an input in the virtual environment: in \cite{zhao_robotic_2017} for instance, the user modifies a light intensity by moving a handle prop in the real environment. 
By transposing \cite{lopes_affordance++:_2015} in VR, an object could even communicate its dynamic use to the user.
% An object could also directly communicate its dynamic use to the user, by transposing \cite{lopes_affordance++:_2015} in VR.
\subsubsection{Pseudo-Haptic Manipulation:} 
Leveraging vision over haptics allows to move an object with different friction, weights or force perceptions \cite{rietzler_breaking_2018, samad_pseudo-haptic_2019, pusch_pseudo-haptics:_2011, rietzler_virtual_2019}. For instance, visually reducing the speed of a virtual prop displacement leads to an increase in the users' forces to move it, modifying their friction/weight perceptions.% of friction or weights.

\subsection{Edition} We qualify an Edition task as a modification of an object property, other than its orientation or position (for example through its scale \cite{xia_spacetime:_2018} or shape). %We distinguish physical vs. pseudo-haptic edition.

\subsubsection{Physical Edition:} 
Editing an interface in VR requires it to be fully equipped with sensors. With wearables for instance, the hand phalanges positions are known, and can be tightly linked with an object property \cite{villa_salazar_altering_2020}. 
Knowing their own position, modular interfaces can be rearranged to provide stretching or bending tasks \cite{feick_tangi_2020}, or be pushed on with a tool to reduce in size \cite{teng_tilepop:_2019}.
\\Shape-changing interfaces have been developed to dynamically modify material properties \cite{nakagaki_materiable:_2016} (Figure \ref{fig:tasks} - 5) or augment the interactions in Augmented Reality (AR) \cite{leithinger_sublimate_2013}, however these techniques only consider HMDs and VR as future work directions. 
\\These interfaces are relevant as 2.5D tabletops are already used in VR. Physically editing the virtual world through them could be implemented in a near future, by intertwining these interfaces with 3D modelling techniques \cite{de_araujo_mockup_2013}.

\subsubsection{Pseudo-Haptic Edition:}
%Real objects cannot benefit from an edition task. 
The difficulty behind changing a real object property is to track it in real-time. This is why pseudo-techniques are relevant: they visually change the object properties such as their shape \cite{achibet_flexifingers:_2017}, compliance \cite{lee_torc:_2019, sinclair_capstancrunch:_2019}, or their bending curvature \cite{heo_pseudobend:_2019} without physically editing the object.

\subsection{Scenario-based Interactions}
In the real world, humans are free to interact with any object without further notice. 
In this regard, common controllers enable interactions with any object through pointing, but they display a high visuo-haptic discrepancy. In more advanced haptically rendered Virtual environments, users are often constrained to scenario-based interactions: only a few interactable objects are available, accordingly with the scenario's progress.
\\The greater the virtual:physical haptic consistency, the harder it is to enhance non-deterministic scenarios, where the user is free to interact with any object with no regards to the scenario's progress. High quality haptic rendering in non-deterministic scenarios can be achieved through three methods: (a) numerous objects and primitives are available for interactions \cite{hettiarachchi_annexing_2016}; (b) the users' intentions are to be predicted prior to interaction to make it occur \cite{bouzbib_covr_2020, cheng_sparse_2017}; (c) props modify their own topology to match the users expected haptic rendering \cite{siu_shapeshift:_2018}.
%
%
%
%However in VR, it is common that users can only interact with a subset of the visually available virtual objects. Moreover, the sequence of actions might be constrained to ensure that the haptic feedback is inline with the contact points of the next interaction. So users should not be constrained to "\textit{scenario-based experiences}". They should be able to perform non-deterministic scenarios where the users can interact with any object of interest with no regards to the scenario’s progress.

%\subsubsection{Determistic Scenarios:} Subset of interactable props. Order to follow
%
%\subsubsection{Non-Determistic Scenarios:} Interactions with any object blabla. Manettes OK, OR Intentions predicted for physical interactions.
%
%, ie to execute the previous tasks, with all the objects around them and in any order they wish. However, in VR, it is common that users can only interact with a subset of the visually available virtual objects. Moreover, the sequence of actions might be constrained to ensure that the haptic feedback is inline with the contact points of the next interaction. So users should not be constrained to "\textit{scenario-based experiences}". They should be able to perform non-deterministic scenarios where the users can interact with any object of interest with no regards to the scenario’s progress.

\subsection{Environment-Initiated Interactions} In both real and virtual environments with tangible interfaces, users usually are the decision makers and get to choose their points of contact during the next interaction. However, users themselves can be considered as tangible interfaces: uncontrolled interactions, such as being touched by a colleague, or feeling a temperature change in the environment \cite{shaw_heat_2019, ziat_enhancing_2014}, are part of everyday interactions that can be transposed in Virtual Reality.
Replicating a social touch interaction in VR for instance increases presence \cite{hoppe_human_2020} or invokes emotions \cite{teyssier_conveying_2020}. 
\\This type of interactions are recurrent in sports simulations, where the user is undergoing forces from his environment and perceiving impacts (jumping into space \cite{gugenheimer_gyrovr_2016}, shooting a soccer ball \cite{wang_gaiters_2020}, goalkeeping in a soccer game \cite{tsai_elastimpact_2019}, paragliding \cite{ye_pull-ups:_2019}, intercepting a volleyball \cite{gunther_pneumovolley_2020}, flying \cite{cheng_haptic_2014}).
\\These interactions are involving multiple force types: tension, traction, reaction, resistance, impact that help enhancing the user experience in VR \cite{wang_movevr_2020}. These can be strong enough to even lead the user through forces \cite{bouzbib_covr_2020}.

\subsection{Whole-Body Involvement} All the previous subsections evoke interactions that mainly involve the hands or the fingers. This paradigm is revoked in \cite{zielasko_either_2020}: a user should be able to choose his posture. This is currently only enabled in room-scale VR applications, where users experience sitting, standing, climbing or crouching \cite{teng_tilepop:_2019, suzuki_roomshift_2020, bouzbib_covr_2020, danieau_hapseat:_2012} and interact with their whole-body.%\gb{j'aurai bien fusionné hand interaction and whole body}\\

%In conclusion, haptic solutions should allow users to experience these six interaction opportunities (summarized in the table \ref{table:Examples}). However, very few of them allow this due to technical limitations. We now detail these different technologies. 

\begin{figure*}[t]
	\centering
	\pdftooltip{  \includegraphics[width=\linewidth]{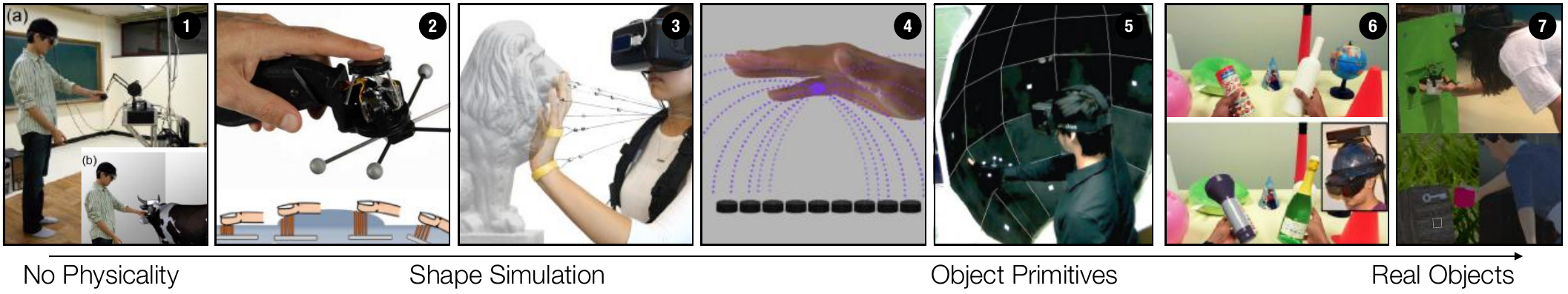}}{}
	\caption{Degree of physicality continuum in VR.%The different haptic technologies provide different degrees of physicality. 
		(1) Haptic desktop devices enable to explore the environment through a handle \cite{lee_system_2009} with the god-object principle; (2) A controller \cite{benko_normaltouch_2016} or (3) a wearable \cite{fang_wireality_2020} simulate objects for exploration tasks; (4) Mid-air technology \cite{rakkolainen_survey_2020} create vibrations through the user's hand to simulate an object; (5) Passive proxies are oriented for the user to feel objects' primitives with their hands \cite{cheng_sparse_2017}; (6) Objects from the environment are assigned to virtual props with the same primitives \cite{hettiarachchi_annexing_2016}; (7) Real objects or passive props can be manipulated and interacted with each other \cite{bouzbib_covr_2020}.}
	\label{fig:physicality}
\end{figure*}

\begin{figure}[b]
  \centering
\pdftooltip{  \includegraphics[width=0.9\linewidth]{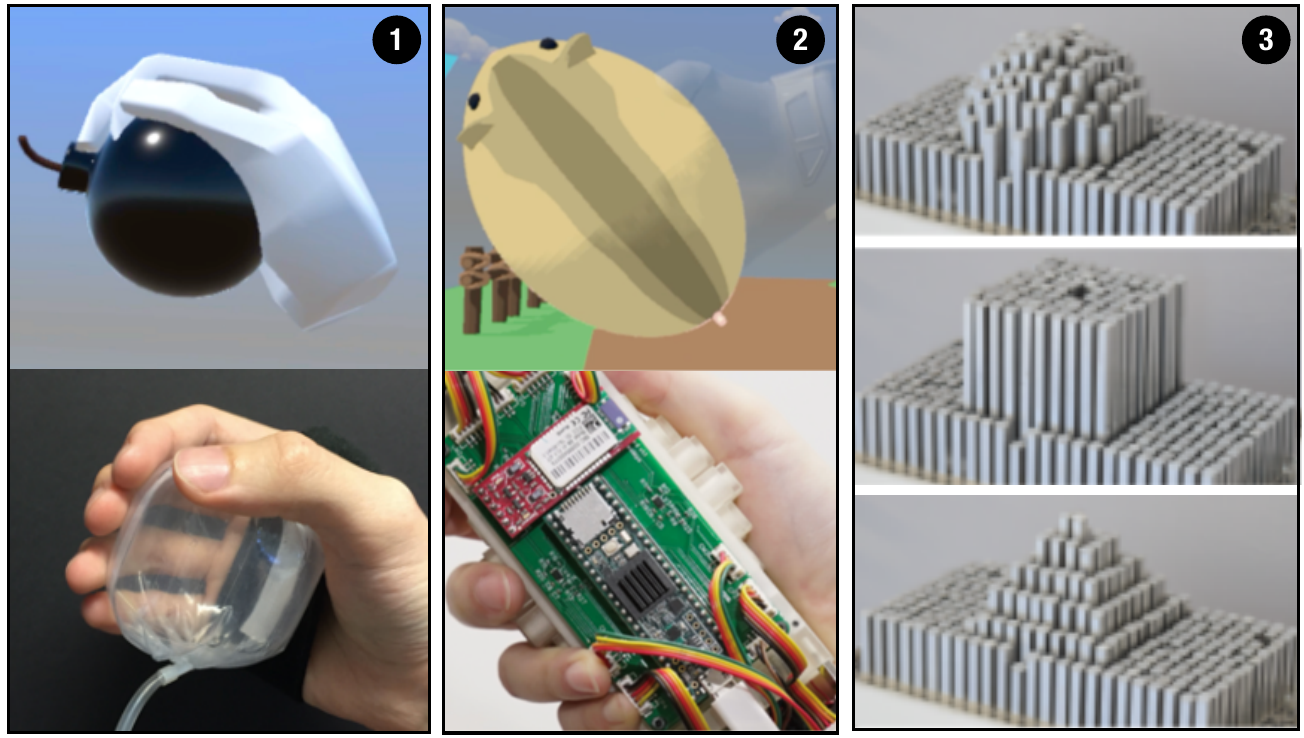}}{Three physical devices simulating objects are shown with their virtual counterparts. (1) An inflatable prop in the user's palm simulates holding a bomb. (2) A pin-based interface shaped as a ball interacts in the user palm to replicate a hamster. (3) Different primitives (ball, cube, pyramid) are displayed on a 2.5D tabletop.}
  \caption{Simulating Objects. (1) A controller with an inflatable prop in the user's palm simulates holding a bomb \cite{teng_pupop_2018}. (2) A pin-based interface shaped as a ball interacts in the user palm to replicate a hamster \cite{yoshida_pocopo_2020}. (3) Different primitives (ball, cube, pyramid) are displayed on a 2.5D tabletop \cite{siu_shapeshift:_2018}.}
  \label{fig:simulatingObjects}
\end{figure}

\section{Visuo-Haptic Consistency/Discrepancy} \label{Physicality}

Visuo-Haptic Consistency is the second aspect of the user experience. We exploit the dimension \textit{degree of physicality} of our design space (Table \ref{table:table_dimensions}) to discuss the different haptic solutions. In particular, we distinguish whether these solutions use real objects (exploiting real objects) or not (simulating objects). %We also highlight the implications of using these solutions on bare hands and whole body interaction (section \ref{Opportunities}).

\subsection{Simulating Objects}

Object properties that need to be simulated are their shape, texture, temperature, weight.
\subsubsection{No Physicality, (Figure \ref{fig:physicality} - 1)}
Currently, grounded haptic devices such as the Virtuose \cite{haption_virtuose_2019} or the PHaNToM \cite{massie_phantom_1994} simulate objects through their shapes (Figure \ref{fig:physicality} - 1). The rendering is only done through kinesthetic feedback via a proxy. Conceptually, the ideal link between the users and this proxy is a massless, infinitely rigid stick, which would be an equivalent to moving the proxy directly \cite{hayward_it_2007, sato_spidar_2002}.
These solutions only provide stimulation at the hand-scale, with no regards to the rest of the body. 
\subsubsection{Shape Simulation, (Figure \ref{fig:physicality} - 2-3-4)}
In the same regard, gloves or controllers provide some physicality (Figure \ref{fig:physicality} - 2-3). Gloves or exoskeletons literally constrain the users hands for simulating shapes \cite{fang_wireality_2020, gu_dexmo:_2016, noauthor_cybergrasp_2019, amirpour_design_2019, choi_grabity_2017, tsai_elasticvr:_2018, achibet_elastic-arm:_2015, choi_wolverine_2016, nakagaki_chainform:_2016, achibet_virtual_2014, provancher_contact_2005}, or stimulate other haptic features such as stiffness, friction \cite{villa_salazar_altering_2020} or slippage \cite{tsagarakis_slip_2005}. These can be extended to overall body suits for users to feel impacts or even temperature changes \cite{noauthor_teslasuit_2019, danieau_hfx_2018}, or even intertwined with grounded devices to extend their use-cases \cite{steed_docking_2020}.
\\Customised controllers are currently designed to be either stimulating the palm \cite{sun_pacapa:_2019, yoshida_pocopo_2020, de_tinguy_weatavix_2020} (Figure \ref{fig:simulatingObjects} - 1, 2), or held in the palm while providing haptic feedback on the fingertips. For instance, \cite{whitmire_haptic_2018} proposes interchangeable haptic wheels with different textures or shapes, while \cite{benko_normaltouch_2016} enables textures and shapes and \cite{lee_torc:_2019} displays compliance changes.
%In these configurations, users hold a single controller, however bi-manual interactions can be created by combining \eb{two controllers}. Their link can hence transmit kinesthetic feedback, as their respective positions are constrained from each other \cite{strasnick_haptic_2018, wei_elastilinks_2020}.
In these configurations, users hold a single controller, however bi-manual interactions can be created by combining two controllers. Their link transmits kinesthetic feedback, and constrain their respective positions to each other \cite{strasnick_haptic_2018, wei_elastilinks_2020}.
\\Contactless technology has also been developed for simulating shapes. While studies demonstrated that interacting with bare-hands increased the user's cognitive load \cite{galais_natural_2019}, combining bare-hands interactions with haptic feedback actually enhances the users involvement. Since haptic feedback does require contact, "contactless" technology defines an interaction where the users are unencumbered, as per Krueger's postulate \cite{wexelblat_virtual_1993}, and ultrasounds are sent to their hands, for them to perceive shapes on their skin, without a physical prop contact \cite{rakkolainen_survey_2020} (Figure \ref{fig:physicality} - 4).
\\ These unencumbered methods are also achieved through shape-changing interfaces, for instance with balloons arrays \cite{takizawa_encountered-type_2017} or 2.5D tabletops (Figure \ref{fig:simulatingObjects} - 3, Figure \ref{fig:tasks} - 5) \cite{follmer_inform:_2013, siu_shapeshift:_2018, iwata_project_2001}. These latter are constituted from pins, that raise and lower themselves to replicate different shapes. In the same regard, swarm interfaces rearrange themselves to display different shapes. These have mainly been developed in the real world \cite{le_goc_zooids:_2016, suzuki_dynablock:_2018, kim_user-defined_2020, suzuki_shapebots:_2019, ducatelle_self-organized_2011, marquardt_haptic_2009} but slowly take off as VR user interfaces \cite{zhao_robotic_2017} (Figure \ref{fig:tasks} - 4).
Indeed, while these latter devices are used as desktop interfaces, the swarm robot idea has extended to the air, with drones for instance \cite{gomes_bitdrones_2016, rubens_bitdrones:_2015, knierim_tactile_2017, hoppe_vrhapticdrones:_2018, tsykunov_wiredswarm:_2019}.
\\All of these previous interfaces embrace the \textbf{Roboxel} principle enunciated in \textit{Robotic Graphics} \cite{mcneely_robotic_1993}: "\textit{cellular robots that dynamically configure themselves into the desired shape and size}".

\subsubsection{Object Primitives, (Figure \ref{fig:physicality} - 5)}
Finally, a user can interact with object primitives. These represent the simplest geometries available: circle, cube, pyramid, cylinder, torus. Simply feeling an orientation through the fingertips provides the required information to understand an object shape, in an exploration task for instance. Panels with diverse orientations can hence be displayed for a user to explore various objects in a virtual environment \cite{cheng_sparse_2017} (Figure \ref{fig:physicality} - 5) or directly encounter the user at their position of interest \cite{yokokohji_haptic_2005, yokokohji_path_2001}. 
% \vspace{-0.5cm}
\\On the opposite, a bare-hands manipulation task requires multiple primitives to be available at the same time within the hand vicinity. This is why the exploitation of real objects is necessary.

\subsection{Exploiting Real Objects}
Passive haptics \cite{insko_passive_2001}, ie the use of passive props, consist in placing real objects corresponding to their exact virtual match at their virtual position. Insko demonstrated that passive haptics enhanced the virtual environment \cite{insko_passive_2001}. Nonetheless, this does suffer from a main limitation: substituting the physical environment for a virtual one \cite{simeone_substitutional_2015} requires a thorough mapping of objects shapes, sizes, textures, and requires numerous props \cite{pair_flatworld:_2003}. This can be done with real objects in simulation rooms for instance (e.g plane cockpit, motorcycle), but cheaper methods need to be implemented to facilitate their use in other fields.
%physically mapping a various types of virtual objects shapes, sizes or textures requires the use of numerous props.

\subsubsection{Object Primitives, (Figure \ref{fig:physicality} - 6)}
One solution is to extract the primitives of the objects that are already available in the physical environment, to map virtual objects of the approximate same primitive over them \cite{hettiarachchi_annexing_2016} (Figure \ref{fig:physicality} - 6).

\subsubsection{Visuo-Proprioceptive Illusions \& Pseudo Haptics}
The number of props within the environment can also be reduced, while letting the users interact at different positions of the physical world. It is possible to leverage the vision over haptics and modify the users' proprioception to redirect their trajectory \cite{kohli_redirected_2010, kohli_redirected_2012, kohli_redirected_2013, azmandian_haptic_2016, gonzalez_investigating_2019, han_evaluating_2018}. A user might perceive multiple distinct cubes for instance, while interacting with a single one. On the same principle, the user hand displacement can be redirected at an angle, up-/down-scaled \cite{abtahi_visuo-haptic_2018, bergstrom_resized_2019}, or slowed down for friction or weight perception \cite{samad_pseudo-haptic_2019, praveena_supporting_2020}. These techniques also allow for the exploration and manipulation of various shapes: models can for instance be added to enable complex virtual shapes to be mapped over real physical objects boundaries \cite{zhao_functional_2018}. The user can also be redirected to pinch a multi-primitive object (cubic, pyramidal and cylindrical) from different locations, which theoretically widens the variety of available props with a single one \cite{de_tinguy_toward_2019}. On the same principle, pseudo-haptics allow to modify the users' shape \cite{ban_modifying_2012-1, ban_modifying_2012} or texture \cite{degraen_enhancing_2019} perceptions when interacting with a physical prop.

\subsubsection{Displacing Objects, (Figure \ref{fig:physicality} - 7)}
Whenever objects are indeed available within the environment, various directions are available to displace them. This displacement allows for mapping one physical object over multiple ones, but also to display a multitude of props.
These directions embrace the \textit{Robotic Shape Display} principle from \textit{Robotic Graphics} 
\cite{mcneely_robotic_1993}: "a robot that can reach any location of a virtual desktop with an end-effector" and matches the user's object of interest.
\\Their usability have been validated through a Wizard-of-Oz implementation, where human operators move real objects or even people around a Room-scale VR arena to encounter the users \cite{cheng_turkdeck:_2015} (Figure \ref{fig:robotisation} - 2). The users themselves can also reconfigure and actuate real props \cite{cheng_iturk:_2018}.
\\Robotic Shape Displays, RSDs, are also called \textbf{encountered-type of haptic devices}, as they literally encounter the users at their object of interest to provide haptic feedback. They allow to display real pieces of material \cite{araujo_snake_2016, abtahi_beyond_2019}, physical props to simulate walls \cite{bouzbib_covr_2020, kim_encountered-type_2018, yamaguchi_non-grounded_2016}, or even display furniture \cite{suzuki_roomshift_2020} or untethered objects \cite{he_physhare:_2017, he_robotic_2017, huang_haptic-go-round_2020, bouzbib_covr_2020}, that can be interacted with each other.

%% file: Challenges.tex
\begin{figure*}[t]
	\centering
	\pdftooltip{  \includegraphics[width=\linewidth]{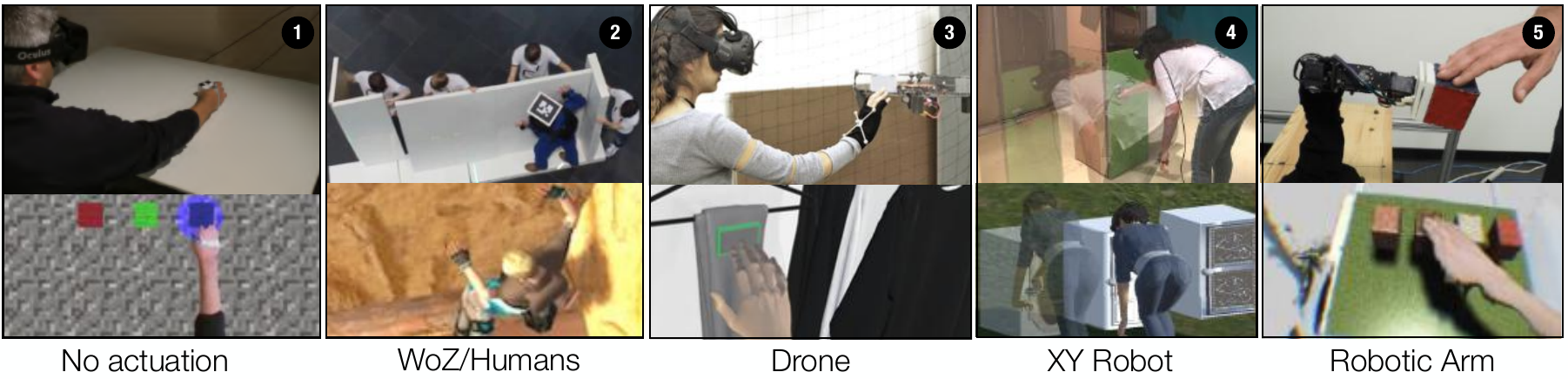}}{}
	\caption{Degree of Actuation. (1) No actuation is available. The user's hand is redirected to touch a passive prop that cannot move \cite{azmandian_haptic_2016}. The implementation of this technique relies exclusively on a software development leveraging the vision cues; (2) Human actuators are used to illustrate the Robotic Graphics \cite{mcneely_robotic_1993} principle with a Wizard of Oz technique \cite{cheng_turkdeck:_2015}. They carry props for the user to feel a real continuous wall; Encountered-type of haptic devices (3-5): (3) A drone %covers a 2 $m^3$ arena and 
		encounters the users' hand for exploring passive props; (4) A cartesian robot %covers a 30 $m^3$ workspace and 
		displaces itself autonomously for users to interact with physical props \cite{bouzbib_covr_2020}; (5) A robotic arm with multiple degrees of freedom displaces itself to encounter the users' hand, and rotates its shape-approximation device to provide the right material \cite{araujo_snake_2016}.}
	\label{fig:robotisation}
\end{figure*}

%\begin{table}[b]
%  \centering
%\pdftooltip{  \includegraphics[width=0.36\linewidth]{img/Table-Dimensions-Ref.pdf}}{We propose two dimensions to classify current technologies: the degree of physicality as well as the degree of actuation. Four categories are hence drawn in this figure: Top Left: No robotics, No real objects; Top Right: Robotics, No real objects; Bottom Right: Robotics, Real objects; Bottom Left: No robotics, Real objects. We respectively displayed current technologies and interaction techniques in the category they belong to.}
%  \caption{We propose two dimensions to classify current technologies: their degree of physicality as well as their degree of actuation. The previously mentioned technologies %(simulating or exploiting objects) 
%  are hence displayed in these four distinct categories.}
%  \label{table:table_dimensions}
%\end{table}

\section{Conception cost} \label{Actuation}

%In this section, we describe the designers' challenges when developing a user interface.%, both conceptually and operationally. 
%When designing haptic interaction solutions, "\textit{the criteria for success include theoretical efficiency and user-oriented factors}" \cite{dominjon_novel_2007}. In practice, designers also have to trade-off their interaction design space with the implementation and operational costs.
In practice, designers have to trade-off their interaction design space with implementation and operational costs in the conception phase. Implementation costs include technical aspects related to the acceptability of an haptic solution such as safety, robustness and ease-of-use \cite{dominjon_novel_2007}.
For instance, actuated haptic solutions require a special attention regarding this criterion. Operation costs include the financial and human cost for using a haptic solution. The financial cost is measured through the cost of the haptic device and additional elements such as motion capture systems to precisely track the users' hand or the prior preparation of required props. Human cost refers to both labour time and number of human operators required during the user's interactions. For instance, actuated haptic solutions generally do not require human operators (low human cost) but might be mechanically expensive.

In this section, we use our two-dimension design space (Table \ref{table:table_dimensions}) to discuss haptic solutions according to their conception cost. As non-actuated solutions globally share the same approaches and have a low implementation cost, we discuss them together in the "No Robotics" subsection.

\subsection{No Robotics}
Regarding implementation costs, all non-actuated haptic solutions are safe, robust and easy-to-use. We depict here an important design choice when opting for these solutions: either the designer relies on \textbf{graphics solutions}, leveraging vision cues over haptic ones, or needs \textbf{operators} to displace or change the interactable props (see Table \ref{table:table_dimensions}).
%\gb{2 sous-catégories: Graphics solutions and operator-based solutions?}

\subsubsection{Passive Props.} Passive props \cite{insko_passive_2001} only consist in placing real objects corresponding to their exact virtual match at their virtual position. They provide a natural way of interacting through the objects' natural affordances \cite{norman_design_2013}. They however are limited to the available objects within the scene as they are not actuated. They only can be used in a scenario-based experience, where the target is known in advance. The environment hence requires a prop for each available virtual object. %: \gb{the designer can then reassign them with a virtual:physical \cite{he_physhare:_2017} mapping.}\gb{C'est quoi la difference avec 7.15}

\subsubsection{Shape Simulation, Pseudo-Haptics, Visuo-Haptic Illusions, Object Primitives.}
For graphics solutions, users are redirected towards their object of interest \cite{azmandian_haptic_2016} using visuo-haptic illusions. However, physically overlaying a prop or primitive over a virtual object has a tracking cost,
% mapping a prop or primitive physical position to a virtual one has a tracking cost, 
which usually relies on trackers which can be operationally costly (eg Optitrack \cite{optitrack_motion_2019} or HTC Trackers).
\\Otherwise, the users intentions have to be predicted for the interaction to occur. The users hands are then redirected to the appropriate motionless prop, for them to explore their object of interest \cite{cheng_sparse_2017}. Operationally, the cost only relies on the proxy fabrication (Figure \ref{fig:physicality} - 5). These implementations offer various scenarios in terms of interaction (even non-deterministic), at an affordable cost.

\subsubsection{Surface Haptic Displays.} These techniques exclusively allow for exploration through multiple 
haptic features such as friction or textures. They also can integrate a tablet or a smartphone \cite{savino_virtual_2020}, on which the user can interact at any location.

\subsubsection{Human Actuators.}
This technique consists in using human operators to displace props in the VR arena. %With operators, 
The designers however come across reliability and speed issues with these operators. Even though they only are used in scenario-based experiences, %eventually,
\textit{delay mechanisms} based on graphics need to be implemented \cite{cheng_turkdeck:_2015} (Figure \ref{fig:robotisation} - 2) to overcome these issues. Conceptually, they broaden the interaction scope, however this solution is operationally very costly.

\subsubsection{Real Props Reassignment.}
Instead of using a tracking system for passive props, a depth camera for instance allows to reassign props to different virtual objects of the same primitive \cite{hettiarachchi_annexing_2016} (Figure \ref{fig:physicality} - 6). The objects are hence all available to be interacted with. This drastically reduces the operational costs as they only rely on computer vision. This enables non-deterministic scenarios as the real world is literally substituted for a virtual one \cite{simeone_substitutional_2015} and objects can be reassigned with virtual:physical \cite{he_physhare:_2017} mappings.

\subsection{Robotics \& No Real Objects}
This section gathers technologies simulating the virtual environment through actuation: they replicate it to constrain the users.
%Robotics solutions for haptic interactions in VR can be evaluated through their levels of \textbf{safety, robustness and ease of use} \cite{dominjon_novel_2007}. 
%We also analyse them through our third proposed dimension, eg whether these solutions are adapted to scenario-based experiences or non-deterministic ones.

\subsubsection{Desktop Haptic Interfaces.} The SPIDAR \cite{sato_spidar_2002}, the Virtuose \cite{haption_virtuose_2019} and other classic desktop haptic interfaces are already compared in multiple surveys \cite{dominjon_novel_2007, seifi_haptipedia:_2019, wang_haptic_2020} (see Figure \ref{fig:physicality} - 1). They are safe as they are controlled by the user and only constrain their arm movements with kinesthetic feedback and adapt to any available object from the virtual scene (non-deterministic scenarios). They show a high perceived stiffness and robustness, but remain really expensive (>10k\$).

\subsubsection{Shape-Changing Interfaces, Roboxels, 2.5D Tabletops.} 
These technologies present a high perceived stiffness and change their shapes accordingly with the virtual environment \cite{fitzgerald_mediate_2018, leithinger_sublimate_2013}. They hence do not require any operator and allow for non-deterministic scenarios whenever their displacements are enabled \cite{siu_shapeshift:_2018} (see Figure \ref{fig:simulatingObjects} - 3). They are however complex to build and require multiple motors as they are composed of arrays of numerous pins, which define their haptic fidelity resolution. Even though they present high voltages, they remain safe around the users. As they require bare-hands interactions, they hence show a high ease of use. %This last reference (\cite{siu_shapeshift:_2018}) also belongs to the Robotic Shape Display category, as it tracks the users' hands to encounters them. %, but is currently only used at a desktop scale.

\subsubsection{Wearables, Controllers, EMS.} These rely on small torques, which are sufficient to constrain the users body parts. They are safe and easy to use, but in return are not robust enough to resist to users' actions. As they are continuously changing the users' haptic perception, they do allow non-deterministic scenarios and change their rendered stiffness and rigidity as a function of the distance to a virtual prop \cite{de_tinguy_weatavix_2020, kovacs_haptic_2020}. 
%In terms of conception costs, a customised controller usually relies on 3D printed parts and small servomotors and can be easily replicated \cite{sun_pacapa:_2019} (Figure \ref{fig:physicality} - 2,3; Figure \ref{fig:simulatingObjects} - 1,2).
A customised controller usually relies on 3D printed parts and small servomotors and can be easily replicated \cite{sun_pacapa:_2019} (Figure \ref{fig:physicality} - 2,3; Figure \ref{fig:simulatingObjects} - 1,2).

\subsubsection{Mid-Air Haptics.} Providing contactless interactions, mid-air haptics also provide a high level of safety around the user. They however do not allow to navigate the VR environment, and hence cannot consider non-deterministic scenarios. Their robustness is very low, as they send ultrasounds to the users and do not physically constrain them \cite{rakkolainen_survey_2020}. 

\subsubsection{Inflatable Floor.} The floor topology can be modified and inflated to create interactions at the body-scale \cite{teng_tilepop:_2019}. The users cannot inflate them, however they can push some tiles down and hence, \textit{edit} them. These are safe, though they do not provide a wide range of interactions, but offer multiple static body postures.

\subsection{Robotics \& Real Objects} \label{RoboticGraphics}
In this subsection, we detail the different types of Robotic Shapes Displays - otherwise known as "\textbf{encountered-type of haptic devices}", mentioned in the Table \ref{table:table_dimensions}. 
First, these interfaces move to encounter the users: this feature optimises their ease of use. Second, as these interfaces move within the user vicinity, safety concerns are raised in this section, depending on the interfaces robustness. 
%We once again describe their use in non-deterministic scenarios, and detail whether their use require a supplementary operator cost.
\\Encountered-type of haptic devices combine different types of interaction techniques: they can provide the users with passive props, textures or primitives, and allow navigation, exploration, manipulation tasks. Their mechanical implementations offer a good repeatability and reliability.
%Thanks to their displacements, they encounter the users at their object of interest to facilitate their ease of use, and potentially show a high robustness and perceived stiffness if required. They even can initiate the interactions. Even though they require an operator cost, \eb{their} mechanical implementations offer a good repeatability and reliability.
%They all belong to the encountered-type of haptics devices category, which optimises their ease of use. 
% These interfaces literally "display" shapes. In order to change  
%RSD! Moving baby All kinda require operator to change objects to display (makes sense) (or are limited to one use: wall for instance)
\subsubsection{Cartesian Robot:} In \cite{bouzbib_covr_2020}, CoVR, a physical column mounted over a Cartesian XY ceiling robot enables interactions at any height and any position of a room-scale VR arena (see Figure \ref{fig:tasks} - 2; Figure \ref{fig:robotisation} - 4). This implementation presents a high perceived stiffness, and because it carries passive props around the arena, enables a high fidelity haptic rendering. It displays high accuracy and speed, and presents an algorithm which optimises %optimising
the column's displacements as a function of the users intentions. It hence enables non-deterministic scenarios. Safety measures have been validated in the field. In practice, the column's celerity is decreasing around the user, as it is repulsed by this latter. Its software implementation ensures a safe environment for the user to perambulate in the arena without unexpected collision. %Even though a collision were to happen, the reduced speed hence ensures a safe environment for the user to perambulate in the arena.
However, in order to display many props in different scenarios, an operator is required to create panels and modify them. The materials however remain cheap, and even though its structure and motors are more expensive than 3D printed cases and servomotors, as per customised controllers for instance, this solution provides a wide range of interactions.

\subsubsection{Robotic Arm:} A robotic arm provides more degrees of freedom than the previous Cartesian robot. This primarily means a higher cost and a higher safety risk. For instance, H-Wall, using a Kuka LBR Iiwa robot, presents high motor torques and can hence increase the safety risks around the users. This implementation hence does not allow non-deterministic scenarios, and presents either a wall or a revolving door to the user, with a high robustness. Implementations with smaller torques, such as \cite{vonach_vrrobot:_2017, araujo_snake_2016} are safer but display a reduced perceived stiffness. The use-cases for all these interactions are hence drastically different: H-Wall simulates a rigid wall while VRRobot \cite{vonach_vrrobot:_2017} and Snake Charmer \cite{araujo_snake_2016} (Figure \ref{fig:robotisation} - 5) present more interaction opportunities. This latter is also the single Robotic Shape Display that autonomously changes its end-effector, without an operator.

\subsubsection{Drones, Swarm Robots, Mobile Platforms:}%, Merry-Go-Round:} 
With drones, the interactions are limited to the available props, for instance with a single wall at a given position \cite{yamaguchi_non-grounded_2016}. Going from an active mode (flying) to a passive one (graspable by the user) has a long delay (10s) \cite{abtahi_beyond_2019}, which on top of the safety concerns, does not allow non-deterministic scenarios. \cite{tsykunov_slingdrone:_2019} however allows the user to change the drone trajectory to fetch and magnetically recover an object of interest.
Their accuracy and speed are limited \cite{gomes_bitdrones_2016, rubens_bitdrones:_2015} compared to the previous grounded interfaces, and can require dynamic redirection techniques to improve their performances \cite{abtahi_beyond_2019}. As they are ungrounded, they do not have a high robustness nor perceived stiffness. This is also valid for mobile robots, such as \cite{he_physhare:_2017, gonzalez_reach_2020}, which only display passive props.
To decrease the conception cost, existing vacuuming robots are used as mobile platforms in \cite{wang_movevr_2020, yixian_zoomwalls_2020}. Designers can choose to duplicate them, as swarm robots, to enable non-deterministic scenarios \cite{suzuki_roomshift_2020}. These are safe to use around the users, as their speed and robustness are limited.
Instead of swarm mobile interfaces, a merry-go-round platform can also be designed to display various props at an equidistant position from the user \cite{huang_haptic-go-round_2020}.
All of the previous interfaces require an operator cost on top of their mechanical and software ones, to modify the interactable props available, depending on the use-cases.
\\On the opposite, \cite{zhao_robotic_2017} proposes autonomous reconfigurable interfaces intertwining both Robotic Shape Displays and Roboxels \cite{mcneely_robotic_1993} principles to get rid of the operator cost (see Figure \ref{fig:tasks} - 4). These small robotic volume elements reconfigure themselves into the users objects of interest. They have a sufficient perceived stiffness to represent objects, but are not robust enough to resist to body-scaled forces, for instance to simulate a rigid wall.
%\\Encountered-type of haptic devices combine different types of other interaction techniques: they can provide the users with passive props, textures or primitives, and allow navigation, exploration, manipulation tasks. Thanks to their displacements, they encounter the users at their object of interest to facilitate their ease of use, and potentially show a high robustness and perceived stiffness if required. They even can initiate the interactions. Even though they require an operator cost, \eb{their} mechanical implementations offer a good repeatability and reliability.

\section{Evaluation Protocols}
On top of choosing from the different trade-offs between conception and interaction opportunities, the designer also needs to pick-up an evaluation protocol. %We depicted in the previous sections how different the interactions opportunities and stimulated haptic features are, depending on the technologies. %These technologies also are widely different, from the degree of physicality and the degree of actuation they provide. 
These protocols depend on the VR use-cases.
For instance, the haptic benefits for medical or industrial assembly training can be evaluated against a real experience condition \cite{poyade_validation_2012}, with criteria such as completion time, number of errors, user cognitive load \cite{gutierrez_ima-vr_2010}. On the opposite, the haptic benefits for a gaming experience are more likely to be evaluated through immersion and presence, comparing "with/without haptics" conditions \cite{cheng_turkdeck:_2015}.
Although some papers do compare multiple haptic displays \cite{escobar-castillejos_review_2016, ullrich_haptic_2012}, we point out the lack of referenced evaluation protocols for evaluating haptic solutions in VR.

\subsection{Current Reference Evaluation Methods}
% \subsection{Current Referenced Evaluation Methods}
% However, the
% Categories: comparaison reel/virtuel; comparaison haptic/no haptic; comparaison multiple haptic devices
% use cases: 
%Current formal and most used evaluation methods \cite{schwind_using_2019} available for VR experiences do not provide a sufficient level of detail to be able to compare a system's haptic contribution. 

The most common evaluation methods in VR are the SUS or WS presence questionnaires \cite{witmer_measuring_1998, slater_depth_1994}.
%\\
These questionnaires mainly focus on graphics rendering and only two Likert-scale questions actually focus on haptic feedback: "How well could you actively survey the VE using touch?" and "How well could you manipulate objects in the VE?". Besides, most of the above technologies are evaluated against "no haptic feedback", hence the results can seem biased and most of all, \textit{expected}.
This justifies why some implementations provide results on single parts of the questionnaire, or arbitrarily combine their results \cite{choi_claw:_2018} with new subsections (eg "ability to examine/act") or tasks specific questions (eg "How realistic was it to feel different textures?).% (eg "How realistic was it to feel different textures?"). % subsections, or add some task specific questions (eg "How realistic was it to feel different textures?").

\begin{table*}[th]
	\centering
	\includegraphics[width=\linewidth]{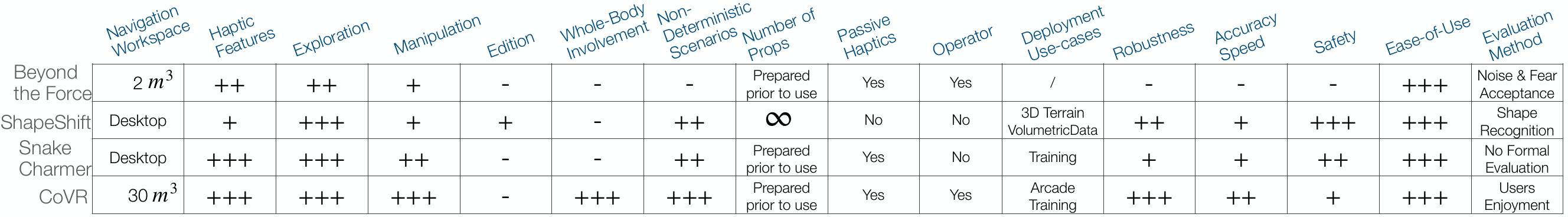}{}
	\caption{Comparison \& Evaluation of 4 Encountered-type of Haptic Devices, according to the "Evaluation section" parameters.}
	\label{table:Examples}
\end{table*}

\subsection{Evaluation Recommendations}
% \subsection{Evaluation Recommendations}
Haptics should be more incorporated into the different factors enunciated in \cite{witmer_measuring_1998} ("Control, Sensory, Distraction, Realism"). 
%\\
In this direction, Kim et al. defined the \textit{Haptic Experience model} \cite{kim_defining_2020}%(Figure \ref{fig:HX})
, where they take into account both of the designer and user experiences. % and the user one.
It depicts how Design parameters ("timeliness, intensity, density and timbre") impact Usability requirements ("utility, causality, consistency, saliency") and target Experiential dimensions ("harmony, expressivity, autotelics, immersion, realism") on the user's side. 
% This latter's experience is also rated in terms of "Control, Sensory, Distraction and Realism" factors \cite{witmer_measuring_1998}.
\\In the same regards, we propose additional guidelines to evaluate haptic solutions in VR experiments  (see Table \ref{table:Examples}). We believe that the different elements of interaction opportunities should be added to the \textit{users control parameters}.

%We believe the \textbf{number of props} manipulated and the \textbf{ability to perform non-deterministic scenarios} should be added to the users \textit{control parameters}, as well as the \textbf{variety of tasks}. 
In the \textit{sensory factors}, the \textbf{number of haptic features} available should be added (eg shape, texture, friction, temperature), in line with their quality, in terms of "timeliness, intensity, density and timbre".
The \textbf{usability} requirements should \textbf{identify the use-cases and number of scenarios} with the proposed solutions. Hence, a good evaluation of the interface timeliness and usability should anticipate future deployments and avoid unnecessary developments.

\section{Examples: Encountered-Type of Haptic Devices}

% As we previously highlighted the encountered-type of haptic devices, we propose in this section to compare four of these solutions:
We propose in this section to compare four encountered-type of haptic devices:
Beyond the Force (BTF) drone \cite{abtahi_beyond_2019} (Figure \ref{fig:robotisation} - 3), ShapeShift \cite{siu_shapeshift:_2018} (Figure \ref{fig:simulatingObjects} - 3), Snake Charmer \cite{araujo_snake_2016} (Figure \ref{fig:robotisation} - 5), and CoVR \cite{bouzbib_covr_2020} (Figure \ref{fig:robotisation} - 4).

In terms of interactions and number of props, the drone is the most limited one. Indeed, because of both safety and implementation limitations, it only enables free navigation in a reduced workspace. It also allows exploration (through textures) and manipulation tasks. However, the manipulation task is at the moment limited to a single light object as BTF cannot handle large embedded masses yet. Whenever grabbed, it does not provide a haptic transparency \cite{hayward_it_2007} during the interactions because of its thrust and inertia. For the users to perform different tasks, an operator needs to manually change the drone configuration.
Its mechanical implementation does not provide a sufficient speed for overlaying virtual props in non-deterministic scenarios, but its accuracy is also unsatisfactory and requires dynamic redirection techniques for the interactions to occur. It also provides %seems to provide a lot of
unwanted noise and wind, which %can reduce the immersion and the interaction realism.
reduces the interaction realism.
\\ShapeShift \cite{siu_shapeshift:_2018} is drastically different: it is a 2.5D desktop interface that displaces itself. Even though a drone is theoretically available in an infinite workspace, in practice they do share approximately the same one. As \cite{siu_shapeshift:_2018} relies on a shape-changing interface, no operator is required and it shape changes itself to overlay the users' virtual objects of interest, in non-deterministic scenarios. It allows a free navigation at a desktop scale, as well as bimanual manipulation and exploration. 
Both of these devices haptic transparency are limited as they are ungrounded solutions. We believe that ShapeShift could be updated to allow Edition tasks, by synchronising the users force actions with the actuated pins stiffness. In terms of haptic features, it simulates shapes and stimulates both tactile and kinesthetic cues. As per all 2.5D tabletops, it can be used in various applications: 3D terrain exploration, volumetric data etc. Its resolution seems promising as its studies shows successful object recognition and haptic search.
%\\Combining such a shape-changing interface to one side of the CoVR Cartesian robot \cite{bouzbib_covr_2020} would be promising: indeed, it would get rid of some operator work and represent a multitude of props. With the right robotic control, the haptic transparency could be alright baby
\\The same interactions are available at a desktop scale with Snake Charmer \cite{araujo_snake_2016}, which provides a wide range of props and stimulation, as each of its end-effector include 6 faces with various interaction opportunities (textures to explore, buttons to push, heater and fan to perceive temperature, handle and lightbulb to grasp and manipulate...). It also can change its shape approximation device, SAD (ie its end-effector), autonomously, using magnets. It follows the user hand and orient the expected interaction face of its \textit{SAD} prior to the interactions: it hence enables non-deterministic scenarios. Besides, Snake Charmer has a promising future regarding its deployment: \textit{LobbyBot} \cite{noauthor_renault_nodate}, is already in the Renault industry research lab, to enable VR haptic feedback in the automotive industry.
\\Finally, CoVR \cite{bouzbib_covr_2020} enables the largest workspace as well as the highest range of interactions. The user is free to navigate in a 30 $m^3$ VR arena, and CoVR predicts and physically overlays his object of interest prior to interaction. These interactions include tactile exploration, manipulation of untethered objects (full haptic transparency), %interaction between objects, and 
body postures. Indeed, CoVR is robust enough to resist body-scaled users, and shows over a 100N perceived stiffness and can carry over 80kg of embedded mass.
CoVR can also initiate the interactions with the users, and is strong enough to lead the users through forces or even to transport them.
Moreover, with the appropriate physical:virtual mapping \cite{he_physhare:_2017}, one physical prop can overlay multiple virtual ones of the same approximate primitive without redirection techniques. It however requires an operator to create, assemble and display panels on its sides.
%It remains costly for a single use, but for arcade rooms or training purposes for instance, it represents a long-term investment, compared to the Shake Charmer. 
\\ \\Room-scale VR becomes more and more relevant, and Snake Charmer could benefit from being attached to an interface such as CoVR. Similarly, intertwining CoVR with a robotic arm autonomously changing its SAD like Snake Charmer or with a shape-changing interface could reduce its operational costs. This would %represent all the capabilities of Robotic Graphics concept.
display all of the Robotics Graphics concept capabilities.

%% file: main.bbl

\begin{thebibliography}{195}

%%% the Web address.

\ifx \showCODEN    \undefined \def \showCODEN     #1{\unskip}     \fi
\ifx \showDOI      \undefined \def \showDOI       #1{#1}\fi
\ifx \showISBNx    \undefined \def \showISBNx     #1{\unskip}     \fi
\ifx \showISBNxiii \undefined \def \showISBNxiii  #1{\unskip}     \fi
\ifx \showISSN     \undefined \def \showISSN      #1{\unskip}     \fi
\ifx \showLCCN     \undefined \def \showLCCN      #1{\unskip}     \fi
\ifx \shownote     \undefined \def \shownote      #1{#1}          \fi
\ifx \showarticletitle \undefined \def \showarticletitle #1{#1}   \fi
\ifx \showURL      \undefined \def \showURL       {\relax}        \fi
% invisible to TeX
\providecommand\bibfield[2]{#2}
\providecommand\bibinfo[2]{#2}
\providecommand\natexlab[1]{#1}
\providecommand\showeprint[2][]{arXiv:#2}

\bibitem[\protect\citeauthoryear{??}{noa}{[n.d.]}]%
        {noauthor_renault_nodate}
 \bibinfo{year}{[n.d.]}\natexlab{}.
\newblock \bibinfo{title}{renault}.
\newblock
\newblock
\urldef\tempurl%
\url{https://www.clarte-lab.fr/component/tags/tag/renault}
\showURL{%
\tempurl}


\bibitem[\protect\citeauthoryear{??}{noa}{2019a}]%
        {noauthor_cybergrasp_2019}
 \bibinfo{year}{2019}\natexlab{a}.
\newblock \bibinfo{title}{{CyberGrasp}}.
\newblock
\newblock
\urldef\tempurl%
\url{http://www.cyberglovesystems.com/cybergrasp}
\showURL{%
\tempurl}


\bibitem[\protect\citeauthoryear{??}{noa}{2019b}]%
        {noauthor_teslasuit_2019}
 \bibinfo{year}{2019}\natexlab{b}.
\newblock \bibinfo{title}{Teslasuit {\textbar} {Full} body haptic {VR} suit for
  motion capture and training}.
\newblock
\newblock
\urldef\tempurl%
\url{https://teslasuit.io/}
\showURL{%
\tempurl}


\bibitem[\protect\citeauthoryear{Abtahi and Follmer}{Abtahi and
  Follmer}{2018}]%
        {abtahi_visuo-haptic_2018}
\bibfield{author}{\bibinfo{person}{Parastoo Abtahi} {and} \bibinfo{person}{Sean
  Follmer}.} \bibinfo{year}{2018}\natexlab{}.
\newblock \showarticletitle{Visuo-{Haptic} {Illusions} for {Improving} the
  {Perceived} {Performance} of {Shape} {Displays}}. In
  \bibinfo{booktitle}{\emph{Proceedings of the 2018 {CHI} {Conference} on
  {Human} {Factors} in {Computing} {Systems} - {CHI} '18}}.
  \bibinfo{publisher}{ACM Press}, \bibinfo{address}{Montreal QC, Canada},
  \bibinfo{pages}{1--13}.
\newblock
\showISBNx{978-1-4503-5620-6}
\urldef\tempurl%
\url{https://doi.org/10.1145/3173574.3173724}
\showDOI{\tempurl}


\bibitem[\protect\citeauthoryear{Abtahi, Landry, Yang, Pavone, Follmer, and
  Landay}{Abtahi et~al\mbox{.}}{2019}]%
        {abtahi_beyond_2019}
\bibfield{author}{\bibinfo{person}{Parastoo Abtahi}, \bibinfo{person}{Benoit
  Landry}, \bibinfo{person}{Jackie~(Junrui) Yang}, \bibinfo{person}{Marco
  Pavone}, \bibinfo{person}{Sean Follmer}, {and} \bibinfo{person}{James~A.
  Landay}.} \bibinfo{year}{2019}\natexlab{}.
\newblock \showarticletitle{Beyond {The} {Force}: {Using} {Quadcopters} to
  {Appropriate} {Objects} and the {Environment} for {Haptics} in {Virtual}
  {Reality}}. In \bibinfo{booktitle}{\emph{Proceedings of the 2019 {CHI}
  {Conference} on {Human} {Factors} in {Computing} {Systems} - {CHI} '19}}.
  \bibinfo{publisher}{ACM Press}, \bibinfo{address}{Glasgow, Scotland Uk},
  \bibinfo{pages}{1--13}.
\newblock
\showISBNx{978-1-4503-5970-2}
\urldef\tempurl%
\url{https://doi.org/10.1145/3290605.3300589}
\showDOI{\tempurl}


\bibitem[\protect\citeauthoryear{Achibet, Girard, Talvas, Marchal, and
  Lecuyer}{Achibet et~al\mbox{.}}{2015}]%
        {achibet_elastic-arm:_2015}
\bibfield{author}{\bibinfo{person}{Merwan Achibet}, \bibinfo{person}{Adrien
  Girard}, \bibinfo{person}{Anthony Talvas}, \bibinfo{person}{Maud Marchal},
  {and} \bibinfo{person}{Anatole Lecuyer}.} \bibinfo{year}{2015}\natexlab{}.
\newblock \showarticletitle{Elastic-{Arm}: {Human}-scale passive haptic
  feedback for augmenting interaction and perception in virtual environments}.
  In \bibinfo{booktitle}{\emph{2015 {IEEE} {Virtual} {Reality} ({VR})}}.
  \bibinfo{publisher}{IEEE}, \bibinfo{address}{Arles, Camargue, Provence,
  France}, \bibinfo{pages}{63--68}.
\newblock
\showISBNx{978-1-4799-1727-3}
\urldef\tempurl%
\url{https://doi.org/10.1109/VR.2015.7223325}
\showDOI{\tempurl}


\bibitem[\protect\citeauthoryear{Achibet, Le~Gouis, Marchal, Leziart,
  Argelaguet, Girard, Lecuyer, and Kajimoto}{Achibet et~al\mbox{.}}{2017}]%
        {achibet_flexifingers:_2017}
\bibfield{author}{\bibinfo{person}{Merwan Achibet}, \bibinfo{person}{Benoit
  Le~Gouis}, \bibinfo{person}{Maud Marchal}, \bibinfo{person}{Pierre-Alexandre
  Leziart}, \bibinfo{person}{Ferran Argelaguet}, \bibinfo{person}{Adrien
  Girard}, \bibinfo{person}{Anatole Lecuyer}, {and} \bibinfo{person}{Hiroyuki
  Kajimoto}.} \bibinfo{year}{2017}\natexlab{}.
\newblock \showarticletitle{{FlexiFingers}: {Multi}-finger interaction in {VR}
  combining passive haptics and pseudo-haptics}. In
  \bibinfo{booktitle}{\emph{2017 {IEEE} {Symposium} on {3D} {User} {Interfaces}
  ({3DUI})}}. \bibinfo{publisher}{IEEE}, \bibinfo{address}{Los Angeles, CA,
  USA}, \bibinfo{pages}{103--106}.
\newblock
\showISBNx{978-1-5090-6716-9}
\urldef\tempurl%
\url{https://doi.org/10.1109/3DUI.2017.7893325}
\showDOI{\tempurl}


\bibitem[\protect\citeauthoryear{Achibet, Marchal, Argelaguet, and
  Lecuyer}{Achibet et~al\mbox{.}}{2014}]%
        {achibet_virtual_2014}
\bibfield{author}{\bibinfo{person}{Merwan Achibet}, \bibinfo{person}{Maud
  Marchal}, \bibinfo{person}{Ferran Argelaguet}, {and} \bibinfo{person}{Anatole
  Lecuyer}.} \bibinfo{year}{2014}\natexlab{}.
\newblock \showarticletitle{The {Virtual} {Mitten}: {A} novel interaction
  paradigm for visuo-haptic manipulation of objects using grip force}. In
  \bibinfo{booktitle}{\emph{2014 {IEEE} {Symposium} on {3D} {User} {Interfaces}
  ({3DUI})}}. \bibinfo{publisher}{IEEE}, \bibinfo{address}{MN, USA},
  \bibinfo{pages}{59--66}.
\newblock
\showISBNx{978-1-4799-3624-3}
\urldef\tempurl%
\url{https://doi.org/10.1109/3DUI.2014.6798843}
\showDOI{\tempurl}


\bibitem[\protect\citeauthoryear{Alexandrovsky, Putze, Bonfert, Höffner,
  Michelmann, Wenig, Malaka, and Smeddinck}{Alexandrovsky
  et~al\mbox{.}}{2020}]%
        {alexandrovsky_examining_2020}
\bibfield{author}{\bibinfo{person}{Dmitry Alexandrovsky},
  \bibinfo{person}{Susanne Putze}, \bibinfo{person}{Michael Bonfert},
  \bibinfo{person}{Sebastian Höffner}, \bibinfo{person}{Pitt Michelmann},
  \bibinfo{person}{Dirk Wenig}, \bibinfo{person}{Rainer Malaka}, {and}
  \bibinfo{person}{Jan~David Smeddinck}.} \bibinfo{year}{2020}\natexlab{}.
\newblock \showarticletitle{Examining {Design} {Choices} of {Questionnaires} in
  {VR} {User} {Studies}}. In \bibinfo{booktitle}{\emph{Proceedings of the 2020
  {CHI} {Conference} on {Human} {Factors} in {Computing} {Systems}}}.
  \bibinfo{publisher}{ACM}, \bibinfo{address}{Honolulu HI USA},
  \bibinfo{pages}{1--21}.
\newblock
\showISBNx{978-1-4503-6708-0}
\urldef\tempurl%
\url{https://doi.org/10.1145/3313831.3376260}
\showDOI{\tempurl}


\bibitem[\protect\citeauthoryear{Amirpour, Savabi, Saboukhi, Gorii, Ghafarirad,
  Fesharakifard, and Rezaei}{Amirpour et~al\mbox{.}}{2019}]%
        {amirpour_design_2019}
\bibfield{author}{\bibinfo{person}{E. Amirpour}, \bibinfo{person}{M. Savabi},
  \bibinfo{person}{A. Saboukhi}, \bibinfo{person}{M.~Rahimi Gorii},
  \bibinfo{person}{H. Ghafarirad}, \bibinfo{person}{R. Fesharakifard}, {and}
  \bibinfo{person}{S.~Mehdi Rezaei}.} \bibinfo{year}{2019}\natexlab{}.
\newblock \showarticletitle{Design and {Optimization} of a {Multi}-{DOF} {Hand}
  {Exoskeleton} for {Haptic} {Applications}}. In \bibinfo{booktitle}{\emph{2019
  7th {International} {Conference} on {Robotics} and {Mechatronics}
  ({ICRoM})}}. \bibinfo{pages}{270--275}.
\newblock
\urldef\tempurl%
\url{https://doi.org/10.1109/ICRoM48714.2019.9071884}
\showDOI{\tempurl}
\newblock
\shownote{ISSN: 2572-6889.}


\bibitem[\protect\citeauthoryear{Araujo, Jota, Perumal, Yao, Singh, and
  Wigdor}{Araujo et~al\mbox{.}}{2016}]%
        {araujo_snake_2016}
\bibfield{author}{\bibinfo{person}{Bruno Araujo}, \bibinfo{person}{Ricardo
  Jota}, \bibinfo{person}{Varun Perumal}, \bibinfo{person}{Jia~Xian Yao},
  \bibinfo{person}{Karan Singh}, {and} \bibinfo{person}{Daniel Wigdor}.}
  \bibinfo{year}{2016}\natexlab{}.
\newblock \showarticletitle{Snake {Charmer}: {Physically} {Enabling} {Virtual}
  {Objects}}. In \bibinfo{booktitle}{\emph{Proceedings of the {TEI} '16:
  {Tenth} {International} {Conference} on {Tangible}, {Embedded}, and
  {Embodied} {Interaction} - {TEI} '16}}. \bibinfo{publisher}{ACM Press},
  \bibinfo{address}{Eindhoven, Netherlands}, \bibinfo{pages}{218--226}.
\newblock
\showISBNx{978-1-4503-3582-9}
\urldef\tempurl%
\url{https://doi.org/10.1145/2839462.2839484}
\showDOI{\tempurl}


\bibitem[\protect\citeauthoryear{Auda, Pascher, and Schneegass}{Auda
  et~al\mbox{.}}{2019}]%
        {auda_around_2019}
\bibfield{author}{\bibinfo{person}{Jonas Auda}, \bibinfo{person}{Max Pascher},
  {and} \bibinfo{person}{Stefan Schneegass}.} \bibinfo{year}{2019}\natexlab{}.
\newblock \showarticletitle{Around the ({Virtual}) {World}: {Infinite}
  {Walking} in {Virtual} {Reality} {Using} {Electrical} {Muscle}
  {Stimulation}}. In \bibinfo{booktitle}{\emph{Proceedings of the 2019 {CHI}
  {Conference} on {Human} {Factors} in {Computing} {Systems} - {CHI} '19}}.
  \bibinfo{publisher}{ACM Press}, \bibinfo{address}{Glasgow, Scotland Uk},
  \bibinfo{pages}{1--8}.
\newblock
\showISBNx{978-1-4503-5970-2}
\urldef\tempurl%
\url{https://doi.org/10.1145/3290605.3300661}
\showDOI{\tempurl}


\bibitem[\protect\citeauthoryear{Azmandian, Hancock, Benko, Ofek, and
  Wilson}{Azmandian et~al\mbox{.}}{2016}]%
        {azmandian_haptic_2016}
\bibfield{author}{\bibinfo{person}{Mahdi Azmandian}, \bibinfo{person}{Mark
  Hancock}, \bibinfo{person}{Hrvoje Benko}, \bibinfo{person}{Eyal Ofek}, {and}
  \bibinfo{person}{Andrew~D. Wilson}.} \bibinfo{year}{2016}\natexlab{}.
\newblock \showarticletitle{Haptic {Retargeting}: {Dynamic} {Repurposing} of
  {Passive} {Haptics} for {Enhanced} {Virtual} {Reality} {Experiences}}. In
  \bibinfo{booktitle}{\emph{Proceedings of the 2016 {CHI} {Conference} on
  {Human} {Factors} in {Computing} {Systems} - {CHI} '16}}.
  \bibinfo{publisher}{ACM Press}, \bibinfo{address}{Santa Clara, California,
  USA}, \bibinfo{pages}{1968--1979}.
\newblock
\showISBNx{978-1-4503-3362-7}
\urldef\tempurl%
\url{https://doi.org/10.1145/2858036.2858226}
\showDOI{\tempurl}


\bibitem[\protect\citeauthoryear{Baloup, Oudjail, Pietrzak, and Casiez}{Baloup
  et~al\mbox{.}}{2018}]%
        {baloup_pointing_2018}
\bibfield{author}{\bibinfo{person}{Marc Baloup}, \bibinfo{person}{Veïs
  Oudjail}, \bibinfo{person}{Thomas Pietrzak}, {and} \bibinfo{person}{Géry
  Casiez}.} \bibinfo{year}{2018}\natexlab{}.
\newblock \showarticletitle{Pointing techniques for distant targets in virtual
  reality}. In \bibinfo{booktitle}{\emph{Proceedings of the 30th {Conference}
  on l'{Interaction} {Homme}-{Machine} - {IHM} '18}}. \bibinfo{publisher}{ACM
  Press}, \bibinfo{address}{Brest, France}, \bibinfo{pages}{100--107}.
\newblock
\showISBNx{978-1-4503-6078-4}
\urldef\tempurl%
\url{https://doi.org/10.1145/3286689.3286696}
\showDOI{\tempurl}


\bibitem[\protect\citeauthoryear{Ban, Kajinami, Narumi, Tanikawa, and
  Hirose}{Ban et~al\mbox{.}}{2012a}]%
        {ban_modifying_2012-1}
\bibfield{author}{\bibinfo{person}{Y. Ban}, \bibinfo{person}{T. Kajinami},
  \bibinfo{person}{T. Narumi}, \bibinfo{person}{T. Tanikawa}, {and}
  \bibinfo{person}{M. Hirose}.} \bibinfo{year}{2012}\natexlab{a}.
\newblock \showarticletitle{Modifying an identified curved surface shape using
  pseudo-haptic effect}. In \bibinfo{booktitle}{\emph{2012 {IEEE} {Haptics}
  {Symposium} ({HAPTICS})}}. \bibinfo{pages}{211--216}.
\newblock
\urldef\tempurl%
\url{https://doi.org/10.1109/HAPTIC.2012.6183793}
\showDOI{\tempurl}


\bibitem[\protect\citeauthoryear{Ban, Narumi, Tanikawa, and Hirose}{Ban
  et~al\mbox{.}}{2012b}]%
        {ban_modifying_2012}
\bibfield{author}{\bibinfo{person}{Yuki Ban}, \bibinfo{person}{Takuji Narumi},
  \bibinfo{person}{Tomohiro Tanikawa}, {and} \bibinfo{person}{Michitaka
  Hirose}.} \bibinfo{year}{2012}\natexlab{b}.
\newblock \showarticletitle{Modifying an identified position of edged shapes
  using pseudo-haptic effects}. In \bibinfo{booktitle}{\emph{Proceedings of the
  18th {ACM} symposium on {Virtual} reality software and technology - {VRST}
  '12}}. \bibinfo{publisher}{ACM Press}, \bibinfo{address}{Toronto, Ontario,
  Canada}, \bibinfo{pages}{93}.
\newblock
\showISBNx{978-1-4503-1469-5}
\urldef\tempurl%
\url{https://doi.org/10.1145/2407336.2407353}
\showDOI{\tempurl}


\bibitem[\protect\citeauthoryear{Barnaby and Roudaut}{Barnaby and
  Roudaut}{2019}]%
        {barnaby_mantis:_2019}
\bibfield{author}{\bibinfo{person}{Gareth Barnaby} {and} \bibinfo{person}{Anne
  Roudaut}.} \bibinfo{year}{2019}\natexlab{}.
\newblock \showarticletitle{Mantis: {A} {Scalable}, {Lightweight} and
  {Accessible} {Architecture} to {Build} {Multiform} {Force} {Feedback}
  {Systems}}. In \bibinfo{booktitle}{\emph{Proceedings of the 32nd {Annual}
  {ACM} {Symposium} on {User} {Interface} {Software} and {Technology} - {UIST}
  '19}}. \bibinfo{publisher}{ACM Press}, \bibinfo{address}{New Orleans, LA,
  USA}, \bibinfo{pages}{937--948}.
\newblock
\showISBNx{978-1-4503-6816-2}
\urldef\tempurl%
\url{https://doi.org/10.1145/3332165.3347909}
\showDOI{\tempurl}


\bibitem[\protect\citeauthoryear{Bau, Poupyrev, Israr, and Harrison}{Bau
  et~al\mbox{.}}{2010}]%
        {bau_teslatouch:_2010}
\bibfield{author}{\bibinfo{person}{Olivier Bau}, \bibinfo{person}{Ivan
  Poupyrev}, \bibinfo{person}{Ali Israr}, {and} \bibinfo{person}{Chris
  Harrison}.} \bibinfo{year}{2010}\natexlab{}.
\newblock \showarticletitle{{TeslaTouch}: electrovibration for touch surfaces}.
  In \bibinfo{booktitle}{\emph{Proceedings of the 23nd annual {ACM} symposium
  on {User} interface software and technology - {UIST} '10}}.
  \bibinfo{publisher}{ACM Press}, \bibinfo{address}{New York, New York, USA},
  \bibinfo{pages}{283}.
\newblock
\showISBNx{978-1-4503-0271-5}
\urldef\tempurl%
\url{https://doi.org/10.1145/1866029.1866074}
\showDOI{\tempurl}


\bibitem[\protect\citeauthoryear{Benko, Holz, Sinclair, and Ofek}{Benko
  et~al\mbox{.}}{2016}]%
        {benko_normaltouch_2016}
\bibfield{author}{\bibinfo{person}{Hrvoje Benko}, \bibinfo{person}{Christian
  Holz}, \bibinfo{person}{Mike Sinclair}, {and} \bibinfo{person}{Eyal Ofek}.}
  \bibinfo{year}{2016}\natexlab{}.
\newblock \showarticletitle{{NormalTouch} and {TextureTouch}: {High}-fidelity
  {3D} {Haptic} {Shape} {Rendering} on {Handheld} {Virtual} {Reality}
  {Controllers}}. In \bibinfo{booktitle}{\emph{Proceedings of the 29th {Annual}
  {Symposium} on {User} {Interface} {Software} and {Technology} - {UIST} '16}}.
  \bibinfo{publisher}{ACM Press}, \bibinfo{address}{Tokyo, Japan},
  \bibinfo{pages}{717--728}.
\newblock
\showISBNx{978-1-4503-4189-9}
\urldef\tempurl%
\url{https://doi.org/10.1145/2984511.2984526}
\showDOI{\tempurl}


\bibitem[\protect\citeauthoryear{Berg and Vance}{Berg and Vance}{2017}]%
        {berg_industry_2017-1}
\bibfield{author}{\bibinfo{person}{Leif~P. Berg} {and} \bibinfo{person}{Judy~M.
  Vance}.} \bibinfo{year}{2017}\natexlab{}.
\newblock \showarticletitle{Industry use of virtual reality in product design
  and manufacturing: a survey}.
\newblock \bibinfo{journal}{\emph{Virtual Reality}} \bibinfo{volume}{21},
  \bibinfo{number}{1} (\bibinfo{date}{March} \bibinfo{year}{2017}),
  \bibinfo{pages}{1--17}.
\newblock
\showISSN{1359-4338, 1434-9957}
\urldef\tempurl%
\url{https://doi.org/10.1007/s10055-016-0293-9}
\showDOI{\tempurl}


\bibitem[\protect\citeauthoryear{Bergström, Mottelson, and Knibbe}{Bergström
  et~al\mbox{.}}{2019}]%
        {bergstrom_resized_2019}
\bibfield{author}{\bibinfo{person}{Joanna Bergström}, \bibinfo{person}{Aske
  Mottelson}, {and} \bibinfo{person}{Jarrod Knibbe}.}
  \bibinfo{year}{2019}\natexlab{}.
\newblock \showarticletitle{Resized {Grasping} in {VR}: {Estimating}
  {Thresholds} for {Object} {Discrimination}}. In
  \bibinfo{booktitle}{\emph{Proceedings of the 32nd {Annual} {ACM} {Symposium}
  on {User} {Interface} {Software} and {Technology}}}.
  \bibinfo{publisher}{ACM}, \bibinfo{address}{New Orleans LA USA},
  \bibinfo{pages}{1175--1183}.
\newblock
\showISBNx{978-1-4503-6816-2}
\urldef\tempurl%
\url{https://doi.org/10.1145/3332165.3347939}
\showDOI{\tempurl}


\bibitem[\protect\citeauthoryear{Bloomfield, {Yu Deng}, Wampler, Rondot, Harth,
  McManus, and Badler}{Bloomfield et~al\mbox{.}}{2003}]%
        {bloomfield_taxonomy_2003}
\bibfield{author}{\bibinfo{person}{A. Bloomfield}, \bibinfo{person}{{Yu Deng}},
  \bibinfo{person}{J. Wampler}, \bibinfo{person}{P. Rondot},
  \bibinfo{person}{D. Harth}, \bibinfo{person}{M. McManus}, {and}
  \bibinfo{person}{N. Badler}.} \bibinfo{year}{2003}\natexlab{}.
\newblock \showarticletitle{A taxonomy and comparison of haptic actions for
  disassembly tasks}. In \bibinfo{booktitle}{\emph{{IEEE} {Virtual} {Reality},
  2003. {Proceedings}.}} \bibinfo{publisher}{IEEE Comput. Soc},
  \bibinfo{address}{Los Angeles, CA, USA}, \bibinfo{pages}{225--231}.
\newblock
\showISBNx{978-0-7695-1882-4}
\urldef\tempurl%
\url{https://doi.org/10.1109/VR.2003.1191143}
\showDOI{\tempurl}


\bibitem[\protect\citeauthoryear{Boldt, Liu, Nguyen, Panova, Singh,
  Steenbergen, Malaka, Smeddinck, Bonfert, Lehne, Cahnbley, Korschinq, Bikas,
  Finke, Hanci, and Kraft}{Boldt et~al\mbox{.}}{2018}]%
        {boldt_you_2018}
\bibfield{author}{\bibinfo{person}{Mette Boldt}, \bibinfo{person}{Boxuan Liu},
  \bibinfo{person}{Tram Nguyen}, \bibinfo{person}{Alina Panova},
  \bibinfo{person}{Ramneek Singh}, \bibinfo{person}{Alexander Steenbergen},
  \bibinfo{person}{Rainer Malaka}, \bibinfo{person}{Jan Smeddinck},
  \bibinfo{person}{Michael Bonfert}, \bibinfo{person}{Inga Lehne},
  \bibinfo{person}{Melina Cahnbley}, \bibinfo{person}{Kim Korschinq},
  \bibinfo{person}{Loannis Bikas}, \bibinfo{person}{Stefan Finke},
  \bibinfo{person}{Martin Hanci}, {and} \bibinfo{person}{Valentin Kraft}.}
  \bibinfo{year}{2018}\natexlab{}.
\newblock \showarticletitle{You {Shall} {Not} {Pass}: {Non}-{Intrusive}
  {Feedback} for {Virtual} {Walls} in {VR} {Environments} with {Room}-{Scale}
  {Mapping}}. In \bibinfo{booktitle}{\emph{2018 {IEEE} {Conference} on
  {Virtual} {Reality} and {3D} {User} {Interfaces} ({VR})}}.
  \bibinfo{publisher}{IEEE}, \bibinfo{address}{Reutlingen},
  \bibinfo{pages}{143--150}.
\newblock
\showISBNx{978-1-5386-3365-6}
\urldef\tempurl%
\url{https://doi.org/10.1109/VR.2018.8446177}
\showDOI{\tempurl}


\bibitem[\protect\citeauthoryear{Bouzbib, Bailly, Haliyo, and Frey}{Bouzbib
  et~al\mbox{.}}{2020}]%
        {bouzbib_covr_2020}
\bibfield{author}{\bibinfo{person}{Elodie Bouzbib}, \bibinfo{person}{Gilles
  Bailly}, \bibinfo{person}{Sinan Haliyo}, {and} \bibinfo{person}{Pascal
  Frey}.} \bibinfo{year}{2020}\natexlab{}.
\newblock \showarticletitle{{CoVR}: {A} {Large}-{Scale} {Force}-{Feedback}
  {Robotic} {Interface} for {Non}-{Deterministic} {Scenarios} in {VR}}. In
  \bibinfo{booktitle}{\emph{Proceedings of the 33rd {Annual} {ACM} {Symposium}
  on {User} {Interface} {Software} and {Technology}}}.
  \bibinfo{publisher}{ACM}, \bibinfo{address}{Virtual Event USA},
  \bibinfo{pages}{209--222}.
\newblock
\showISBNx{978-1-4503-7514-6}
\urldef\tempurl%
\url{https://doi.org/10.1145/3379337.3415891}
\showDOI{\tempurl}


\bibitem[\protect\citeauthoryear{Bowman and Wingrave}{Bowman and
  Wingrave}{2001}]%
        {bowman_design_2001}
\bibfield{author}{\bibinfo{person}{D.A. Bowman} {and} \bibinfo{person}{C.A.
  Wingrave}.} \bibinfo{year}{2001}\natexlab{}.
\newblock \showarticletitle{Design and evaluation of menu systems for immersive
  virtual environments}. In \bibinfo{booktitle}{\emph{Proceedings {IEEE}
  {Virtual} {Reality} 2001}}. \bibinfo{publisher}{IEEE Comput. Soc},
  \bibinfo{address}{Yokohama, Japan}, \bibinfo{pages}{149--156}.
\newblock
\showISBNx{978-0-7695-0948-8}
\urldef\tempurl%
\url{https://doi.org/10.1109/VR.2001.913781}
\showDOI{\tempurl}


\bibitem[\protect\citeauthoryear{Bryson}{Bryson}{2005}]%
        {bryson_direct_2005}
\bibfield{author}{\bibinfo{person}{Steve Bryson}.}
  \bibinfo{year}{2005}\natexlab{}.
\newblock \showarticletitle{Direct {Manipulation} in {Virtual} {Reality}}.
\newblock In \bibinfo{booktitle}{\emph{Visualization {Handbook}}}.
  \bibinfo{publisher}{Elsevier}, \bibinfo{pages}{413--430}.
\newblock
\showISBNx{978-0-12-387582-2}
\urldef\tempurl%
\url{https://doi.org/10.1016/B978-012387582-2/50023-X}
\showDOI{\tempurl}


\bibitem[\protect\citeauthoryear{Cheng}{Cheng}{2019}]%
        {cheng_vroamer:_2019}
\bibfield{author}{\bibinfo{person}{Lung-Pan Cheng}.}
  \bibinfo{year}{2019}\natexlab{}.
\newblock \showarticletitle{{VRoamer}: {Generating} {On}-{The}-{Fly} {VR}
  {Experiences} {While} {Walking} inside {Large}, {Unknown} {Real}-{World}
  {Building} {Environments}}.
\newblock  (\bibinfo{year}{2019}), \bibinfo{pages}{8}.
\newblock


\bibitem[\protect\citeauthoryear{Cheng, Chang, Marwecki, and Baudisch}{Cheng
  et~al\mbox{.}}{2018}]%
        {cheng_iturk:_2018}
\bibfield{author}{\bibinfo{person}{Lung-Pan Cheng}, \bibinfo{person}{Li Chang},
  \bibinfo{person}{Sebastian Marwecki}, {and} \bibinfo{person}{Patrick
  Baudisch}.} \bibinfo{year}{2018}\natexlab{}.
\newblock \showarticletitle{{iTurk}: {Turning} {Passive} {Haptics} into
  {Active} {Haptics} by {Making} {Users} {Reconfigure} {Props} in {Virtual}
  {Reality}}. In \bibinfo{booktitle}{\emph{Proceedings of the 2018 {CHI}
  {Conference} on {Human} {Factors} in {Computing} {Systems} - {CHI} '18}}.
  \bibinfo{publisher}{ACM Press}, \bibinfo{address}{Montreal QC, Canada},
  \bibinfo{pages}{1--10}.
\newblock
\showISBNx{978-1-4503-5620-6}
\urldef\tempurl%
\url{https://doi.org/10.1145/3173574.3173663}
\showDOI{\tempurl}


\bibitem[\protect\citeauthoryear{Cheng, Lühne, Lopes, Sterz, and
  Baudisch}{Cheng et~al\mbox{.}}{2014}]%
        {cheng_haptic_2014}
\bibfield{author}{\bibinfo{person}{Lung-Pan Cheng}, \bibinfo{person}{Patrick
  Lühne}, \bibinfo{person}{Pedro Lopes}, \bibinfo{person}{Christoph Sterz},
  {and} \bibinfo{person}{Patrick Baudisch}.} \bibinfo{year}{2014}\natexlab{}.
\newblock \showarticletitle{Haptic {Turk}: a {Motion} {Platform} {Based} on
  {People}}.
\newblock  (\bibinfo{year}{2014}), \bibinfo{pages}{11}.
\newblock


\bibitem[\protect\citeauthoryear{Cheng, Ofek, Holz, Benko, and Wilson}{Cheng
  et~al\mbox{.}}{2017}]%
        {cheng_sparse_2017}
\bibfield{author}{\bibinfo{person}{Lung-Pan Cheng}, \bibinfo{person}{Eyal
  Ofek}, \bibinfo{person}{Christian Holz}, \bibinfo{person}{Hrvoje Benko},
  {and} \bibinfo{person}{Andrew~D. Wilson}.} \bibinfo{year}{2017}\natexlab{}.
\newblock \showarticletitle{Sparse {Haptic} {Proxy}: {Touch} {Feedback} in
  {Virtual} {Environments} {Using} a {General} {Passive} {Prop}}. In
  \bibinfo{booktitle}{\emph{Proceedings of the 2017 {CHI} {Conference} on
  {Human} {Factors} in {Computing} {Systems} - {CHI} '17}}.
  \bibinfo{publisher}{ACM Press}, \bibinfo{address}{Denver, Colorado, USA},
  \bibinfo{pages}{3718--3728}.
\newblock
\showISBNx{978-1-4503-4655-9}
\urldef\tempurl%
\url{https://doi.org/10.1145/3025453.3025753}
\showDOI{\tempurl}


\bibitem[\protect\citeauthoryear{Cheng, Roumen, Rantzsch, Köhler, Schmidt,
  Kovacs, Jasper, Kemper, and Baudisch}{Cheng et~al\mbox{.}}{2015}]%
        {cheng_turkdeck:_2015}
\bibfield{author}{\bibinfo{person}{Lung-Pan Cheng}, \bibinfo{person}{Thijs
  Roumen}, \bibinfo{person}{Hannes Rantzsch}, \bibinfo{person}{Sven Köhler},
  \bibinfo{person}{Patrick Schmidt}, \bibinfo{person}{Robert Kovacs},
  \bibinfo{person}{Johannes Jasper}, \bibinfo{person}{Jonas Kemper}, {and}
  \bibinfo{person}{Patrick Baudisch}.} \bibinfo{year}{2015}\natexlab{}.
\newblock \showarticletitle{{TurkDeck}: {Physical} {Virtual} {Reality} {Based}
  on {People}}. In \bibinfo{booktitle}{\emph{Proceedings of the 28th {Annual}
  {ACM} {Symposium} on {User} {Interface} {Software} \& {Technology} - {UIST}
  '15}}. \bibinfo{publisher}{ACM Press}, \bibinfo{address}{Daegu, Kyungpook,
  Republic of Korea}, \bibinfo{pages}{417--426}.
\newblock
\showISBNx{978-1-4503-3779-3}
\urldef\tempurl%
\url{https://doi.org/10.1145/2807442.2807463}
\showDOI{\tempurl}


\bibitem[\protect\citeauthoryear{Choi, Culbertson, Miller, Olwal, and
  Follmer}{Choi et~al\mbox{.}}{2017}]%
        {choi_grabity_2017}
\bibfield{author}{\bibinfo{person}{Inrak Choi}, \bibinfo{person}{Heather
  Culbertson}, \bibinfo{person}{Mark~R. Miller}, \bibinfo{person}{Alex Olwal},
  {and} \bibinfo{person}{Sean Follmer}.} \bibinfo{year}{2017}\natexlab{}.
\newblock \showarticletitle{Grabity: {A} {Wearable} {Haptic} {Interface} for
  {Simulating} {Weight} and {Grasping} in {Virtual} {Reality}}. In
  \bibinfo{booktitle}{\emph{Proceedings of the 30th {Annual} {ACM} {Symposium}
  on {User} {Interface} {Software} and {Technology} - {UIST} '17}}.
  \bibinfo{publisher}{ACM Press}, \bibinfo{address}{Qu\&\#233;bec City, QC,
  Canada}, \bibinfo{pages}{119--130}.
\newblock
\showISBNx{978-1-4503-4981-9}
\urldef\tempurl%
\url{https://doi.org/10.1145/3126594.3126599}
\showDOI{\tempurl}


\bibitem[\protect\citeauthoryear{Choi, Hawkes, Christensen, Ploch, and
  Follmer}{Choi et~al\mbox{.}}{2016}]%
        {choi_wolverine_2016}
\bibfield{author}{\bibinfo{person}{Inrak Choi}, \bibinfo{person}{Elliot~W.
  Hawkes}, \bibinfo{person}{David~L. Christensen},
  \bibinfo{person}{Christopher~J. Ploch}, {and} \bibinfo{person}{Sean
  Follmer}.} \bibinfo{year}{2016}\natexlab{}.
\newblock \showarticletitle{Wolverine: {A} wearable haptic interface for
  grasping in virtual reality}. In \bibinfo{booktitle}{\emph{2016 {IEEE}/{RSJ}
  {International} {Conference} on {Intelligent} {Robots} and {Systems}
  ({IROS})}}. \bibinfo{publisher}{IEEE}, \bibinfo{address}{Daejeon, South
  Korea}, \bibinfo{pages}{986--993}.
\newblock
\showISBNx{978-1-5090-3762-9}
\urldef\tempurl%
\url{https://doi.org/10.1109/IROS.2016.7759169}
\showDOI{\tempurl}


\bibitem[\protect\citeauthoryear{Choi, Ofek, Benko, Sinclair, and Holz}{Choi
  et~al\mbox{.}}{2018}]%
        {choi_claw:_2018}
\bibfield{author}{\bibinfo{person}{Inrak Choi}, \bibinfo{person}{Eyal Ofek},
  \bibinfo{person}{Hrvoje Benko}, \bibinfo{person}{Mike Sinclair}, {and}
  \bibinfo{person}{Christian Holz}.} \bibinfo{year}{2018}\natexlab{}.
\newblock \showarticletitle{{CLAW}: {A} {Multifunctional} {Handheld} {Haptic}
  {Controller} for {Grasping}, {Touching}, and {Triggering} in {Virtual}
  {Reality}}. In \bibinfo{booktitle}{\emph{Proceedings of the 2018 {CHI}
  {Conference} on {Human} {Factors} in {Computing} {Systems} - {CHI} '18}}.
  \bibinfo{publisher}{ACM Press}, \bibinfo{address}{Montreal QC, Canada},
  \bibinfo{pages}{1--13}.
\newblock
\showISBNx{978-1-4503-5620-6}
\urldef\tempurl%
\url{https://doi.org/10.1145/3173574.3174228}
\showDOI{\tempurl}


\bibitem[\protect\citeauthoryear{Coles, Meglan, and John}{Coles
  et~al\mbox{.}}{2011}]%
        {coles_role_2011}
\bibfield{author}{\bibinfo{person}{Timothy~R. Coles}, \bibinfo{person}{Dwight
  Meglan}, {and} \bibinfo{person}{Nigel~W. John}.}
  \bibinfo{year}{2011}\natexlab{}.
\newblock \showarticletitle{The {Role} of {Haptics} in {Medical} {Training}
  {Simulators}: {A} {Survey} of the {State} of the {Art}}.
\newblock \bibinfo{journal}{\emph{IEEE Transactions on Haptics}}
  \bibinfo{volume}{4}, \bibinfo{number}{1} (\bibinfo{date}{Jan.}
  \bibinfo{year}{2011}), \bibinfo{pages}{51--66}.
\newblock
\showISSN{1939-1412}
\urldef\tempurl%
\url{https://doi.org/10.1109/TOH.2010.19}
\showDOI{\tempurl}


\bibitem[\protect\citeauthoryear{Danieau, Fleureau, Guillotel, Mollet,
  Lécuyer, and Christie}{Danieau et~al\mbox{.}}{2012}]%
        {danieau_hapseat:_2012}
\bibfield{author}{\bibinfo{person}{Fabien Danieau}, \bibinfo{person}{Julien
  Fleureau}, \bibinfo{person}{Philippe Guillotel}, \bibinfo{person}{Nicolas
  Mollet}, \bibinfo{person}{Anatole Lécuyer}, {and} \bibinfo{person}{Marc
  Christie}.} \bibinfo{year}{2012}\natexlab{}.
\newblock \showarticletitle{{HapSeat}: producing motion sensation with multiple
  force-feedback devices embedded in a seat}. In
  \bibinfo{booktitle}{\emph{Proceedings of the 18th {ACM} symposium on
  {Virtual} reality software and technology - {VRST} '12}}.
  \bibinfo{publisher}{ACM Press}, \bibinfo{address}{Toronto, Ontario, Canada},
  \bibinfo{pages}{69}.
\newblock
\showISBNx{978-1-4503-1469-5}
\urldef\tempurl%
\url{https://doi.org/10.1145/2407336.2407350}
\showDOI{\tempurl}


\bibitem[\protect\citeauthoryear{Danieau, Guillotel, Dumas, Lopez, Leroy, and
  Mollet}{Danieau et~al\mbox{.}}{2018}]%
        {danieau_hfx_2018}
\bibfield{author}{\bibinfo{person}{Fabien Danieau}, \bibinfo{person}{Philippe
  Guillotel}, \bibinfo{person}{Olivier Dumas}, \bibinfo{person}{Thomas Lopez},
  \bibinfo{person}{Bertrand Leroy}, {and} \bibinfo{person}{Nicolas Mollet}.}
  \bibinfo{year}{2018}\natexlab{}.
\newblock \showarticletitle{{HFX} studio: haptic editor for full-body immersive
  experiences}. In \bibinfo{booktitle}{\emph{Proceedings of the 24th {ACM}
  {Symposium} on {Virtual} {Reality} {Software} and {Technology} - {VRST}
  '18}}. \bibinfo{publisher}{ACM Press}, \bibinfo{address}{Tokyo, Japan},
  \bibinfo{pages}{1--9}.
\newblock
\showISBNx{978-1-4503-6086-9}
\urldef\tempurl%
\url{https://doi.org/10.1145/3281505.3281518}
\showDOI{\tempurl}


\bibitem[\protect\citeauthoryear{De~Araújo, Casiez, Jorge, and
  Hachet}{De~Araújo et~al\mbox{.}}{2013}]%
        {de_araujo_mockup_2013}
\bibfield{author}{\bibinfo{person}{Bruno~R. De~Araújo}, \bibinfo{person}{Géry
  Casiez}, \bibinfo{person}{Joaquim~A. Jorge}, {and} \bibinfo{person}{Martin
  Hachet}.} \bibinfo{year}{2013}\natexlab{}.
\newblock \showarticletitle{Mockup {Builder}: {3D} modeling on and above the
  surface}.
\newblock \bibinfo{journal}{\emph{Computers \& Graphics}} \bibinfo{volume}{37},
  \bibinfo{number}{3} (\bibinfo{date}{May} \bibinfo{year}{2013}),
  \bibinfo{pages}{165--178}.
\newblock
\showISSN{00978493}
\urldef\tempurl%
\url{https://doi.org/10.1016/j.cag.2012.12.005}
\showDOI{\tempurl}


\bibitem[\protect\citeauthoryear{de~Tinguy, Howard, Pacchierotti, Marchal, and
  Lécuyer}{de~Tinguy et~al\mbox{.}}{2020}]%
        {de_tinguy_weatavix_2020}
\bibfield{author}{\bibinfo{person}{Xavier de Tinguy}, \bibinfo{person}{Thomas
  Howard}, \bibinfo{person}{Claudio Pacchierotti}, \bibinfo{person}{Maud
  Marchal}, {and} \bibinfo{person}{Anatole Lécuyer}.}
  \bibinfo{year}{2020}\natexlab{}.
\newblock \showarticletitle{{WeATaViX}: {WEarable} {Actuated} {TAngibles} for
  {VIrtual} reality {eXperiences}}.
\newblock  (\bibinfo{year}{2020}), \bibinfo{pages}{9}.
\newblock


\bibitem[\protect\citeauthoryear{de~Tinguy, Pacchierotti, Marchal, and
  Lecuyer}{de~Tinguy et~al\mbox{.}}{2019}]%
        {de_tinguy_toward_2019}
\bibfield{author}{\bibinfo{person}{Xavier de Tinguy}, \bibinfo{person}{Claudio
  Pacchierotti}, \bibinfo{person}{Maud Marchal}, {and} \bibinfo{person}{Anatole
  Lecuyer}.} \bibinfo{year}{2019}\natexlab{}.
\newblock \showarticletitle{Toward {Universal} {Tangible} {Objects}:
  {Optimizing} {Haptic} {Pinching} {Sensations} in {3D} {Interaction}}. In
  \bibinfo{booktitle}{\emph{2019 {IEEE} {Conference} on {Virtual} {Reality} and
  {3D} {User} {Interfaces} ({VR})}}. \bibinfo{publisher}{IEEE},
  \bibinfo{address}{Osaka, Japan}, \bibinfo{pages}{321--330}.
\newblock
\showISBNx{978-1-72811-377-7}
\urldef\tempurl%
\url{https://doi.org/10.1109/VR.2019.8798205}
\showDOI{\tempurl}


\bibitem[\protect\citeauthoryear{Degraen, Zenner, and Krüger}{Degraen
  et~al\mbox{.}}{2019}]%
        {degraen_enhancing_2019}
\bibfield{author}{\bibinfo{person}{Donald Degraen}, \bibinfo{person}{André
  Zenner}, {and} \bibinfo{person}{Antonio Krüger}.}
  \bibinfo{year}{2019}\natexlab{}.
\newblock \showarticletitle{Enhancing {Texture} {Perception} in {Virtual}
  {Reality} {Using} {3D}-{Printed} {Hair} {Structures}}. In
  \bibinfo{booktitle}{\emph{Proceedings of the 2019 {CHI} {Conference} on
  {Human} {Factors} in {Computing} {Systems} - {CHI} '19}}.
  \bibinfo{publisher}{ACM Press}, \bibinfo{address}{Glasgow, Scotland Uk},
  \bibinfo{pages}{1--12}.
\newblock
\showISBNx{978-1-4503-5970-2}
\urldef\tempurl%
\url{https://doi.org/10.1145/3290605.3300479}
\showDOI{\tempurl}


\bibitem[\protect\citeauthoryear{Dominjon, Perret, and Lécuyer}{Dominjon
  et~al\mbox{.}}{2007}]%
        {dominjon_novel_2007}
\bibfield{author}{\bibinfo{person}{Lionel Dominjon}, \bibinfo{person}{Jérôme
  Perret}, {and} \bibinfo{person}{Anatole Lécuyer}.}
  \bibinfo{year}{2007}\natexlab{}.
\newblock \showarticletitle{Novel devices and interaction techniques for
  human-scale haptics}.
\newblock \bibinfo{journal}{\emph{The Visual Computer}} \bibinfo{volume}{23},
  \bibinfo{number}{4} (\bibinfo{date}{March} \bibinfo{year}{2007}),
  \bibinfo{pages}{257--266}.
\newblock
\showISSN{0178-2789, 1432-2315}
\urldef\tempurl%
\url{https://doi.org/10.1007/s00371-007-0100-4}
\showDOI{\tempurl}


\bibitem[\protect\citeauthoryear{Ducatelle, Di Caro, Pinciroli, and
  Gambardella}{Ducatelle et~al\mbox{.}}{2011}]%
        {ducatelle_self-organized_2011}
\bibfield{author}{\bibinfo{person}{Frederick Ducatelle},
  \bibinfo{person}{Gianni~A. Di Caro}, \bibinfo{person}{Carlo Pinciroli},
  {and} \bibinfo{person}{Luca~M. Gambardella}.}
  \bibinfo{year}{2011}\natexlab{}.
\newblock \showarticletitle{Self-organized cooperation between robotic swarms}.
\newblock \bibinfo{journal}{\emph{Swarm Intelligence}} \bibinfo{volume}{5},
  \bibinfo{number}{2} (\bibinfo{date}{June} \bibinfo{year}{2011}),
  \bibinfo{pages}{73--96}.
\newblock
\showISSN{1935-3812, 1935-3820}
\urldef\tempurl%
\url{https://doi.org/10.1007/s11721-011-0053-0}
\showDOI{\tempurl}


\bibitem[\protect\citeauthoryear{Escobar-Castillejos, Noguez, Neri, Magana, and
  Benes}{Escobar-Castillejos et~al\mbox{.}}{2016}]%
        {escobar-castillejos_review_2016}
\bibfield{author}{\bibinfo{person}{David Escobar-Castillejos},
  \bibinfo{person}{Julieta Noguez}, \bibinfo{person}{Luis Neri},
  \bibinfo{person}{Alejandra Magana}, {and} \bibinfo{person}{Bedrich Benes}.}
  \bibinfo{year}{2016}\natexlab{}.
\newblock \showarticletitle{A {Review} of {Simulators} with {Haptic} {Devices}
  for {Medical} {Training}}.
\newblock \bibinfo{journal}{\emph{Journal of Medical Systems}}
  \bibinfo{volume}{40}, \bibinfo{number}{4} (\bibinfo{date}{April}
  \bibinfo{year}{2016}), \bibinfo{pages}{1--22}.
\newblock
\showISSN{0148-5598}
\urldef\tempurl%
\url{https://doi.org/10.1007/s10916-016-0459-8}
\showDOI{\tempurl}


\bibitem[\protect\citeauthoryear{Fang, Zhang, Dworman, and Harrison}{Fang
  et~al\mbox{.}}{2020}]%
        {fang_wireality_2020}
\bibfield{author}{\bibinfo{person}{Cathy Fang}, \bibinfo{person}{Yang Zhang},
  \bibinfo{person}{Matthew Dworman}, {and} \bibinfo{person}{Chris Harrison}.}
  \bibinfo{year}{2020}\natexlab{}.
\newblock \showarticletitle{Wireality: {Enabling} {Complex} {Tangible}
  {Geometries} in {Virtual} {Reality} with {Worn} {Multi}-{String} {Haptics}}.
\newblock  (\bibinfo{year}{2020}), \bibinfo{pages}{10}.
\newblock


\bibitem[\protect\citeauthoryear{Feick, Bateman, Tang, Miede, and
  Marquardt}{Feick et~al\mbox{.}}{2020}]%
        {feick_tangi_2020}
\bibfield{author}{\bibinfo{person}{Martin Feick}, \bibinfo{person}{Scott
  Bateman}, \bibinfo{person}{Anthony Tang}, \bibinfo{person}{André Miede},
  {and} \bibinfo{person}{Nicolai Marquardt}.} \bibinfo{year}{2020}\natexlab{}.
\newblock \showarticletitle{{TanGi}: {Tangible} {Proxies} for {Embodied}
  {Object} {Exploration} and {Manipulation} in {Virtual} {Reality}}.
\newblock \bibinfo{journal}{\emph{arXiv:2001.03021 [cs]}} (\bibinfo{date}{Jan.}
  \bibinfo{year}{2020}).
\newblock
\urldef\tempurl%
\url{http://arxiv.org/abs/2001.03021}
\showURL{%
\tempurl}
\newblock
\shownote{arXiv: 2001.03021.}


\bibitem[\protect\citeauthoryear{Fitzgerald and Ishii}{Fitzgerald and
  Ishii}{2018}]%
        {fitzgerald_mediate_2018}
\bibfield{author}{\bibinfo{person}{Daniel Fitzgerald} {and}
  \bibinfo{person}{Hiroshi Ishii}.} \bibinfo{year}{2018}\natexlab{}.
\newblock \showarticletitle{Mediate: {A} {Spatial} {Tangible} {Interface} for
  {Mixed} {Reality}}. In \bibinfo{booktitle}{\emph{Extended {Abstracts} of the
  2018 {CHI} {Conference} on {Human} {Factors} in {Computing} {Systems}}}.
  \bibinfo{publisher}{ACM}, \bibinfo{address}{Montreal QC Canada},
  \bibinfo{pages}{1--6}.
\newblock
\showISBNx{978-1-4503-5621-3}
\urldef\tempurl%
\url{https://doi.org/10.1145/3170427.3188472}
\showDOI{\tempurl}


\bibitem[\protect\citeauthoryear{Follmer, Leithinger, Olwal, Hogge, and
  Ishii}{Follmer et~al\mbox{.}}{2013}]%
        {follmer_inform:_2013}
\bibfield{author}{\bibinfo{person}{Sean Follmer}, \bibinfo{person}{Daniel
  Leithinger}, \bibinfo{person}{Alex Olwal}, \bibinfo{person}{Akimitsu Hogge},
  {and} \bibinfo{person}{Hiroshi Ishii}.} \bibinfo{year}{2013}\natexlab{}.
\newblock \showarticletitle{{inFORM}: dynamic physical affordances and
  constraints through shape and object actuation}. In
  \bibinfo{booktitle}{\emph{Proceedings of the 26th annual {ACM} symposium on
  {User} interface software and technology - {UIST} '13}}.
  \bibinfo{publisher}{ACM Press}, \bibinfo{address}{St. Andrews, Scotland,
  United Kingdom}, \bibinfo{pages}{417--426}.
\newblock
\showISBNx{978-1-4503-2268-3}
\urldef\tempurl%
\url{https://doi.org/10.1145/2501988.2502032}
\showDOI{\tempurl}


\bibitem[\protect\citeauthoryear{Formaglio, Giannitrapani, Franzini,
  Prattichizzo, and Barbagli}{Formaglio et~al\mbox{.}}{2005}]%
        {formaglio_performance_2005}
\bibfield{author}{\bibinfo{person}{A. Formaglio}, \bibinfo{person}{A.
  Giannitrapani}, \bibinfo{person}{M. Franzini}, \bibinfo{person}{D.
  Prattichizzo}, {and} \bibinfo{person}{F. Barbagli}.}
  \bibinfo{year}{2005}\natexlab{}.
\newblock \showarticletitle{Performance of {Mobile} {Haptic} {Interfaces}}. In
  \bibinfo{booktitle}{\emph{Proceedings of the 44th {IEEE} {Conference} on
  {Decision} and {Control}}}. \bibinfo{pages}{8343--8348}.
\newblock
\urldef\tempurl%
\url{https://doi.org/10.1109/CDC.2005.1583513}
\showDOI{\tempurl}


\bibitem[\protect\citeauthoryear{Frissen, Campos, Sreenivasa, and
  Ernst}{Frissen et~al\mbox{.}}{2013}]%
        {frissen_enabling_2013}
\bibfield{author}{\bibinfo{person}{Ilja Frissen}, \bibinfo{person}{Jennifer~L.
  Campos}, \bibinfo{person}{Manish Sreenivasa}, {and} \bibinfo{person}{Marc~O.
  Ernst}.} \bibinfo{year}{2013}\natexlab{}.
\newblock \showarticletitle{Enabling {Unconstrained} {Omnidirectional}
  {Walking} {Through} {Virtual} {Environments}: {An} {Overview} of the
  {CyberWalk} {Project}}.
\newblock In \bibinfo{booktitle}{\emph{Human {Walking} in {Virtual}
  {Environments}: {Perception}, {Technology}, and {Applications}}},
  \bibfield{editor}{\bibinfo{person}{Frank Steinicke}, \bibinfo{person}{Yon
  Visell}, \bibinfo{person}{Jennifer Campos}, {and} \bibinfo{person}{Anatole
  Lécuyer}} (Eds.). \bibinfo{publisher}{Springer}, \bibinfo{address}{New York,
  NY}, \bibinfo{pages}{113--144}.
\newblock
\showISBNx{978-1-4419-8432-6}
\urldef\tempurl%
\url{https://doi.org/10.1007/978-1-4419-8432-6_6}
\showDOI{\tempurl}


\bibitem[\protect\citeauthoryear{Funk, Müller, Fendrich, Shene, Kolvenbach,
  Dobbertin, Günther, and Mühlhäuser}{Funk et~al\mbox{.}}{2019}]%
        {funk_assessing_2019}
\bibfield{author}{\bibinfo{person}{Markus Funk}, \bibinfo{person}{Florian
  Müller}, \bibinfo{person}{Marco Fendrich}, \bibinfo{person}{Megan Shene},
  \bibinfo{person}{Moritz Kolvenbach}, \bibinfo{person}{Niclas Dobbertin},
  \bibinfo{person}{Sebastian Günther}, {and} \bibinfo{person}{Max
  Mühlhäuser}.} \bibinfo{year}{2019}\natexlab{}.
\newblock \showarticletitle{Assessing the {Accuracy} of {Point} \& {Teleport}
  {Locomotion} with {Orientation} {Indication} for {Virtual} {Reality} using
  {Curved} {Trajectories}}. In \bibinfo{booktitle}{\emph{Proceedings of the
  2019 {CHI} {Conference} on {Human} {Factors} in {Computing} {Systems} - {CHI}
  '19}}. \bibinfo{publisher}{ACM Press}, \bibinfo{address}{Glasgow, Scotland
  Uk}, \bibinfo{pages}{1--12}.
\newblock
\showISBNx{978-1-4503-5970-2}
\urldef\tempurl%
\url{https://doi.org/10.1145/3290605.3300377}
\showDOI{\tempurl}


\bibitem[\protect\citeauthoryear{Galais, Delmas, and Alonso}{Galais
  et~al\mbox{.}}{2019}]%
        {galais_natural_2019}
\bibfield{author}{\bibinfo{person}{Thomas Galais}, \bibinfo{person}{Alexandra
  Delmas}, {and} \bibinfo{person}{Rémy Alonso}.}
  \bibinfo{year}{2019}\natexlab{}.
\newblock \showarticletitle{Natural interaction in virtual reality: impact on
  the cognitive load}. In \bibinfo{booktitle}{\emph{Proceedings of the 31st
  {Conference} on l'{Interaction} {Homme}-{Machine} {Adjunct} - {IHM} '19}}.
  \bibinfo{publisher}{ACM Press}, \bibinfo{address}{Grenoble, France},
  \bibinfo{pages}{1--9}.
\newblock
\showISBNx{978-1-4503-7027-1}
\urldef\tempurl%
\url{https://doi.org/10.1145/3366551.3370342}
\showDOI{\tempurl}


\bibitem[\protect\citeauthoryear{Galambos}{Galambos}{2012}]%
        {galambos_vibrotactile_2012}
\bibfield{author}{\bibinfo{person}{Péter Galambos}.}
  \bibinfo{year}{2012}\natexlab{}.
\newblock \showarticletitle{Vibrotactile {Feedback} for {Haptics} and
  {Telemanipulation}: {Survey}, {Concept} and {Experiment}}.
\newblock \bibinfo{journal}{\emph{Acta Polytechnica Hungarica}}
  \bibinfo{volume}{9}, \bibinfo{number}{1} (\bibinfo{year}{2012}),
  \bibinfo{pages}{25}.
\newblock


\bibitem[\protect\citeauthoryear{Gomes, Rubens, Braley, and Vertegaal}{Gomes
  et~al\mbox{.}}{2016}]%
        {gomes_bitdrones_2016}
\bibfield{author}{\bibinfo{person}{Antonio Gomes}, \bibinfo{person}{Calvin
  Rubens}, \bibinfo{person}{Sean Braley}, {and} \bibinfo{person}{Roel
  Vertegaal}.} \bibinfo{year}{2016}\natexlab{}.
\newblock \showarticletitle{{BitDrones}: {Towards} {Using} {3D} {Nanocopter}
  {Displays} as {Interactive} {Self}-{Levitating} {Programmable} {Matter}}. In
  \bibinfo{booktitle}{\emph{Proceedings of the 2016 {CHI} {Conference} on
  {Human} {Factors} in {Computing} {Systems} - {CHI} '16}}.
  \bibinfo{publisher}{ACM Press}, \bibinfo{address}{Santa Clara, California,
  USA}, \bibinfo{pages}{770--780}.
\newblock
\showISBNx{978-1-4503-3362-7}
\urldef\tempurl%
\url{https://doi.org/10.1145/2858036.2858519}
\showDOI{\tempurl}


\bibitem[\protect\citeauthoryear{Gonzalez, Abtahi, and Follmer}{Gonzalez
  et~al\mbox{.}}{2020}]%
        {gonzalez_reach_2020}
\bibfield{author}{\bibinfo{person}{Eric~J. Gonzalez}, \bibinfo{person}{Parastoo
  Abtahi}, {and} \bibinfo{person}{Sean Follmer}.}
  \bibinfo{year}{2020}\natexlab{}.
\newblock \showarticletitle{{REACH}+: {Extending} the {Reachability} of
  {Encountered}-type {Haptics} {Devices} through {Dynamic} {Redirection} in
  {VR}}. In \bibinfo{booktitle}{\emph{Proceedings of the 33rd {Annual} {ACM}
  {Symposium} on {User} {Interface} {Software} and {Technology}}}.
  \bibinfo{publisher}{ACM}, \bibinfo{address}{Virtual Event USA},
  \bibinfo{pages}{236--248}.
\newblock
\showISBNx{978-1-4503-7514-6}
\urldef\tempurl%
\url{https://doi.org/10.1145/3379337.3415870}
\showDOI{\tempurl}


\bibitem[\protect\citeauthoryear{Gonzalez and Follmer}{Gonzalez and
  Follmer}{2019}]%
        {gonzalez_investigating_2019}
\bibfield{author}{\bibinfo{person}{Eric~J. Gonzalez} {and}
  \bibinfo{person}{Sean Follmer}.} \bibinfo{year}{2019}\natexlab{}.
\newblock \showarticletitle{Investigating the {Detection} of {Bimanual}
  {Haptic} {Retargeting} in {Virtual} {Reality}}. In
  \bibinfo{booktitle}{\emph{25th {ACM} {Symposium} on {Virtual} {Reality}
  {Software} and {Technology} on - {VRST} '19}}. \bibinfo{publisher}{ACM
  Press}, \bibinfo{address}{Parramatta, NSW, Australia}, \bibinfo{pages}{1--5}.
\newblock
\showISBNx{978-1-4503-7001-1}
\urldef\tempurl%
\url{https://doi.org/10.1145/3359996.3364248}
\showDOI{\tempurl}


\bibitem[\protect\citeauthoryear{Gu, Zhang, Sun, Bian, Zhou, and
  Kristensson}{Gu et~al\mbox{.}}{2016}]%
        {gu_dexmo:_2016}
\bibfield{author}{\bibinfo{person}{Xiaochi Gu}, \bibinfo{person}{Yifei Zhang},
  \bibinfo{person}{Weize Sun}, \bibinfo{person}{Yuanzhe Bian},
  \bibinfo{person}{Dao Zhou}, {and} \bibinfo{person}{Per~Ola Kristensson}.}
  \bibinfo{year}{2016}\natexlab{}.
\newblock \showarticletitle{Dexmo: {An} {Inexpensive} and {Lightweight}
  {Mechanical} {Exoskeleton} for {Motion} {Capture} and {Force} {Feedback} in
  {VR}}. In \bibinfo{booktitle}{\emph{Proceedings of the 2016 {CHI}
  {Conference} on {Human} {Factors} in {Computing} {Systems} - {CHI} '16}}.
  \bibinfo{publisher}{ACM Press}, \bibinfo{address}{Santa Clara, California,
  USA}, \bibinfo{pages}{1991--1995}.
\newblock
\showISBNx{978-1-4503-3362-7}
\urldef\tempurl%
\url{https://doi.org/10.1145/2858036.2858487}
\showDOI{\tempurl}


\bibitem[\protect\citeauthoryear{Gugenheimer, Wolf, Eiriksson, Maes, and
  Rukzio}{Gugenheimer et~al\mbox{.}}{2016}]%
        {gugenheimer_gyrovr_2016}
\bibfield{author}{\bibinfo{person}{Jan Gugenheimer}, \bibinfo{person}{Dennis
  Wolf}, \bibinfo{person}{Eythor~R. Eiriksson}, \bibinfo{person}{Pattie Maes},
  {and} \bibinfo{person}{Enrico Rukzio}.} \bibinfo{year}{2016}\natexlab{}.
\newblock \showarticletitle{{GyroVR}: {Simulating} {Inertia} in {Virtual}
  {Reality} using {Head} {Worn} {Flywheels}}. In
  \bibinfo{booktitle}{\emph{Proceedings of the 29th {Annual} {Symposium} on
  {User} {Interface} {Software} and {Technology}}}. \bibinfo{publisher}{ACM},
  \bibinfo{address}{Tokyo Japan}, \bibinfo{pages}{227--232}.
\newblock
\showISBNx{978-1-4503-4189-9}
\urldef\tempurl%
\url{https://doi.org/10.1145/2984511.2984535}
\showDOI{\tempurl}


\bibitem[\protect\citeauthoryear{Gutierrez, Rodriguez, Velaz, Casado, Suescun,
  and Sanchez}{Gutierrez et~al\mbox{.}}{2010}]%
        {gutierrez_ima-vr_2010}
\bibfield{author}{\bibinfo{person}{T. Gutierrez}, \bibinfo{person}{J.
  Rodriguez}, \bibinfo{person}{Y. Velaz}, \bibinfo{person}{S. Casado},
  \bibinfo{person}{A. Suescun}, {and} \bibinfo{person}{E.~J. Sanchez}.}
  \bibinfo{year}{2010}\natexlab{}.
\newblock \showarticletitle{{IMA}-{VR}: {A} multimodal virtual training system
  for skills transfer in {Industrial} {Maintenance} and {Assembly} tasks}.
\newblock \bibinfo{journal}{\emph{19th International Symposium in Robot and
  Human Interactive Communication}} (\bibinfo{year}{2010}).
\newblock
\urldef\tempurl%
\url{https://www.academia.edu/15623406/IMA_VR_A_multimodal_virtual_training_system_for_skills_transfer_in_Industrial_Maintenance_and_Assembly_tasks}
\showURL{%
\tempurl}


\bibitem[\protect\citeauthoryear{Günther, Schön, Müller, Mühlhäuser, and
  Schmitz}{Günther et~al\mbox{.}}{2020}]%
        {gunther_pneumovolley_2020}
\bibfield{author}{\bibinfo{person}{Sebastian Günther},
  \bibinfo{person}{Dominik Schön}, \bibinfo{person}{Florian Müller},
  \bibinfo{person}{Max Mühlhäuser}, {and} \bibinfo{person}{Martin Schmitz}.}
  \bibinfo{year}{2020}\natexlab{}.
\newblock \showarticletitle{{PneumoVolley}: {Pressure}-based {Haptic}
  {Feedback} on the {Head} through {Pneumatic} {Actuation}}.
\newblock  (\bibinfo{year}{2020}), \bibinfo{pages}{10}.
\newblock


\bibitem[\protect\citeauthoryear{Han, Suhail, and Ragan}{Han
  et~al\mbox{.}}{2018}]%
        {han_evaluating_2018}
\bibfield{author}{\bibinfo{person}{Dustin~T. Han}, \bibinfo{person}{Mohamed
  Suhail}, {and} \bibinfo{person}{Eric~D. Ragan}.}
  \bibinfo{year}{2018}\natexlab{}.
\newblock \showarticletitle{Evaluating {Remapped} {Physical} {Reach} for {Hand}
  {Interactions} with {Passive} {Haptics} in {Virtual} {Reality}}.
\newblock \bibinfo{journal}{\emph{IEEE Transactions on Visualization and
  Computer Graphics}} \bibinfo{volume}{24}, \bibinfo{number}{4}
  (\bibinfo{date}{April} \bibinfo{year}{2018}), \bibinfo{pages}{1467--1476}.
\newblock
\showISSN{1077-2626}
\urldef\tempurl%
\url{https://doi.org/10.1109/TVCG.2018.2794659}
\showDOI{\tempurl}


\bibitem[\protect\citeauthoryear{Haption}{Haption}{2019}]%
        {haption_virtuose_2019}
\bibfield{author}{\bibinfo{person}{Haption}.} \bibinfo{year}{2019}\natexlab{}.
\newblock \bibinfo{title}{Virtuose™ {6D} - {HAPTION} {SA}}.
\newblock
\newblock
\urldef\tempurl%
\url{https://www.haption.com/en/products-en/virtuose-6d-en.html}
\showURL{%
\tempurl}


\bibitem[\protect\citeauthoryear{Hayward and Maclean}{Hayward and
  Maclean}{2007}]%
        {hayward_it_2007}
\bibfield{author}{\bibinfo{person}{Vincent Hayward} {and}
  \bibinfo{person}{Karon Maclean}.} \bibinfo{year}{2007}\natexlab{}.
\newblock \showarticletitle{Do it yourself haptics: part {I}}.
\newblock \bibinfo{journal}{\emph{IEEE Robotics \& Automation Magazine}}
  \bibinfo{volume}{14}, \bibinfo{number}{4} (\bibinfo{date}{Dec.}
  \bibinfo{year}{2007}), \bibinfo{pages}{88--104}.
\newblock
\showISSN{1070-9932}
\urldef\tempurl%
\url{https://doi.org/10.1109/M-RA.2007.907921}
\showDOI{\tempurl}


\bibitem[\protect\citeauthoryear{He, Zhu, Gaudette, and Perlin}{He
  et~al\mbox{.}}{2017b}]%
        {he_robotic_2017}
\bibfield{author}{\bibinfo{person}{Zhenyi He}, \bibinfo{person}{Fengyuan Zhu},
  \bibinfo{person}{Aaron Gaudette}, {and} \bibinfo{person}{Ken Perlin}.}
  \bibinfo{year}{2017}\natexlab{b}.
\newblock \showarticletitle{Robotic {Haptic} {Proxies} for {Collaborative}
  {Virtual} {Reality}}.
\newblock \bibinfo{journal}{\emph{arXiv:1701.08879 [cs]}} (\bibinfo{date}{Jan.}
  \bibinfo{year}{2017}).
\newblock
\urldef\tempurl%
\url{http://arxiv.org/abs/1701.08879}
\showURL{%
\tempurl}
\newblock
\shownote{arXiv: 1701.08879.}


\bibitem[\protect\citeauthoryear{He, Zhu, and Perlin}{He
  et~al\mbox{.}}{2017a}]%
        {he_physhare:_2017}
\bibfield{author}{\bibinfo{person}{Zhenyi He}, \bibinfo{person}{Fengyuan Zhu},
  {and} \bibinfo{person}{Ken Perlin}.} \bibinfo{year}{2017}\natexlab{a}.
\newblock \showarticletitle{{PhyShare}: {Sharing} {Physical} {Interaction} in
  {Virtual} {Reality}}.
\newblock \bibinfo{journal}{\emph{arXiv:1708.04139 [cs]}} (\bibinfo{date}{Aug.}
  \bibinfo{year}{2017}).
\newblock
\urldef\tempurl%
\url{http://arxiv.org/abs/1708.04139}
\showURL{%
\tempurl}
\newblock
\shownote{arXiv: 1708.04139.}


\bibitem[\protect\citeauthoryear{Held and Durlach}{Held and Durlach}{1992}]%
        {held_telepresence_1992}
\bibfield{author}{\bibinfo{person}{Richard~M. Held} {and}
  \bibinfo{person}{Nathaniel~I. Durlach}.} \bibinfo{year}{1992}\natexlab{}.
\newblock \showarticletitle{Telepresence}.
\newblock \bibinfo{journal}{\emph{Presence: Teleoperators and Virtual
  Environments}} \bibinfo{volume}{1}, \bibinfo{number}{1} (\bibinfo{date}{Jan.}
  \bibinfo{year}{1992}), \bibinfo{pages}{109--112}.
\newblock
\showISSN{1054-7460}
\urldef\tempurl%
\url{https://doi.org/10.1162/pres.1992.1.1.109}
\showDOI{\tempurl}


\bibitem[\protect\citeauthoryear{Heo, Chung, Lee, and Wigdor}{Heo
  et~al\mbox{.}}{2018}]%
        {heo_thors_2018}
\bibfield{author}{\bibinfo{person}{Seongkook Heo}, \bibinfo{person}{Christina
  Chung}, \bibinfo{person}{Geehyuk Lee}, {and} \bibinfo{person}{Daniel
  Wigdor}.} \bibinfo{year}{2018}\natexlab{}.
\newblock \showarticletitle{Thor's {Hammer}: {An} {Ungrounded} {Force}
  {Feedback} {Device} {Utilizing} {Propeller}-{Induced} {Propulsive} {Force}}.
  In \bibinfo{booktitle}{\emph{Proceedings of the 2018 {CHI} {Conference} on
  {Human} {Factors} in {Computing} {Systems} - {CHI} '18}}.
  \bibinfo{publisher}{ACM Press}, \bibinfo{address}{Montreal QC, Canada},
  \bibinfo{pages}{1--11}.
\newblock
\showISBNx{978-1-4503-5620-6}
\urldef\tempurl%
\url{https://doi.org/10.1145/3173574.3174099}
\showDOI{\tempurl}


\bibitem[\protect\citeauthoryear{Heo, Lee, and Wigdor}{Heo
  et~al\mbox{.}}{2019}]%
        {heo_pseudobend:_2019}
\bibfield{author}{\bibinfo{person}{Seongkook Heo}, \bibinfo{person}{Jaeyeon
  Lee}, {and} \bibinfo{person}{Daniel Wigdor}.}
  \bibinfo{year}{2019}\natexlab{}.
\newblock \showarticletitle{{PseudoBend}: {Producing} {Haptic} {Illusions} of
  {Stretching}, {Bending}, and {Twisting} {Using} {Grain} {Vibrations}}. In
  \bibinfo{booktitle}{\emph{Proceedings of the 32nd {Annual} {ACM} {Symposium}
  on {User} {Interface} {Software} and {Technology} - {UIST} '19}}.
  \bibinfo{publisher}{ACM Press}, \bibinfo{address}{New Orleans, LA, USA},
  \bibinfo{pages}{803--813}.
\newblock
\showISBNx{978-1-4503-6816-2}
\urldef\tempurl%
\url{https://doi.org/10.1145/3332165.3347941}
\showDOI{\tempurl}


\bibitem[\protect\citeauthoryear{Hettiarachchi and Wigdor}{Hettiarachchi and
  Wigdor}{2016}]%
        {hettiarachchi_annexing_2016}
\bibfield{author}{\bibinfo{person}{Anuruddha Hettiarachchi} {and}
  \bibinfo{person}{Daniel Wigdor}.} \bibinfo{year}{2016}\natexlab{}.
\newblock \showarticletitle{Annexing {Reality}: {Enabling} {Opportunistic}
  {Use} of {Everyday} {Objects} as {Tangible} {Proxies} in {Augmented}
  {Reality}}. In \bibinfo{booktitle}{\emph{Proceedings of the 2016 {CHI}
  {Conference} on {Human} {Factors} in {Computing} {Systems} - {CHI} '16}}.
  \bibinfo{publisher}{ACM Press}, \bibinfo{address}{Santa Clara, California,
  USA}, \bibinfo{pages}{1957--1967}.
\newblock
\showISBNx{978-1-4503-3362-7}
\urldef\tempurl%
\url{https://doi.org/10.1145/2858036.2858134}
\showDOI{\tempurl}


\bibitem[\protect\citeauthoryear{Hoppe, Knierim, Kosch, Funk, Futami,
  Schneegass, Henze, Schmidt, and Machulla}{Hoppe et~al\mbox{.}}{2018}]%
        {hoppe_vrhapticdrones:_2018}
\bibfield{author}{\bibinfo{person}{Matthias Hoppe}, \bibinfo{person}{Pascal
  Knierim}, \bibinfo{person}{Thomas Kosch}, \bibinfo{person}{Markus Funk},
  \bibinfo{person}{Lauren Futami}, \bibinfo{person}{Stefan Schneegass},
  \bibinfo{person}{Niels Henze}, \bibinfo{person}{Albrecht Schmidt}, {and}
  \bibinfo{person}{Tonja Machulla}.} \bibinfo{year}{2018}\natexlab{}.
\newblock \showarticletitle{{VRHapticDrones}: {Providing} {Haptics} in
  {Virtual} {Reality} through {Quadcopters}}. In
  \bibinfo{booktitle}{\emph{Proceedings of the 17th {International}
  {Conference} on {Mobile} and {Ubiquitous} {Multimedia} - {MUM} 2018}}.
  \bibinfo{publisher}{ACM Press}, \bibinfo{address}{Cairo, Egypt},
  \bibinfo{pages}{7--18}.
\newblock
\showISBNx{978-1-4503-6594-9}
\urldef\tempurl%
\url{https://doi.org/10.1145/3282894.3282898}
\showDOI{\tempurl}


\bibitem[\protect\citeauthoryear{Hoppe, Neumann, Streuber, Schmidt, and
  Machulla}{Hoppe et~al\mbox{.}}{2020}]%
        {hoppe_human_2020}
\bibfield{author}{\bibinfo{person}{Matthias Hoppe}, \bibinfo{person}{Daniel
  Neumann}, \bibinfo{person}{Stephan Streuber}, \bibinfo{person}{Albrecht
  Schmidt}, {and} \bibinfo{person}{Tonja-Katrin Machulla}.}
  \bibinfo{year}{2020}\natexlab{}.
\newblock \bibinfo{booktitle}{\emph{A {Human} {Touch}: {Social} {Touch}
  {Increases} the {Perceived} {Human}-likeness of {Agents} in {Virtual}
  {Reality}}}.
\newblock
\urldef\tempurl%
\url{https://doi.org/10.1145/3313831.3376719}
\showDOI{\tempurl}


\bibitem[\protect\citeauthoryear{Huang, Ning, Wang, Cheng, and Cheng}{Huang
  et~al\mbox{.}}{2020}]%
        {huang_haptic-go-round_2020}
\bibfield{author}{\bibinfo{person}{Hsin-Yu Huang}, \bibinfo{person}{Chih-Wei
  Ning}, \bibinfo{person}{Po-Yao Wang}, \bibinfo{person}{Jen-Hao Cheng}, {and}
  \bibinfo{person}{Lung-Pan Cheng}.} \bibinfo{year}{2020}\natexlab{}.
\newblock \showarticletitle{Haptic-go-round: {A} {Surrounding} {Platform} for
  {Encounter}-type {Haptics} in {Virtual} {Reality} {Experiences}}. In
  \bibinfo{booktitle}{\emph{Proceedings of the 2020 {CHI} {Conference} on
  {Human} {Factors} in {Computing} {Systems}}}. \bibinfo{publisher}{ACM},
  \bibinfo{address}{Honolulu HI USA}, \bibinfo{pages}{1--10}.
\newblock
\showISBNx{978-1-4503-6708-0}
\urldef\tempurl%
\url{https://doi.org/10.1145/3313831.3376476}
\showDOI{\tempurl}


\bibitem[\protect\citeauthoryear{Insko}{Insko}{2001}]%
        {insko_passive_2001}
\bibfield{author}{\bibinfo{person}{Brent~Edward Insko}.}
  \bibinfo{year}{2001}\natexlab{}.
\newblock \showarticletitle{Passive {Haptics} {Significantly} {Enhances}
  {Virtual} {Environments}}.
\newblock  (\bibinfo{year}{2001}), \bibinfo{pages}{111}.
\newblock


\bibitem[\protect\citeauthoryear{Iwata}{Iwata}{2005}]%
        {iwata_circulafloor_2005}
\bibfield{author}{\bibinfo{person}{Hiroo Iwata}.}
  \bibinfo{year}{2005}\natexlab{}.
\newblock \bibinfo{title}{{CirculaFloor}}.
\newblock
\newblock
\urldef\tempurl%
\url{https://ieeexplore.ieee.org/abstract/document/1381227}
\showURL{%
\tempurl}


\bibitem[\protect\citeauthoryear{Iwata}{Iwata}{2013}]%
        {iwata_locomotion_2013}
\bibfield{author}{\bibinfo{person}{Hiroo Iwata}.}
  \bibinfo{year}{2013}\natexlab{}.
\newblock \showarticletitle{Locomotion {Interfaces}}.
\newblock In \bibinfo{booktitle}{\emph{Human {Walking} in {Virtual}
  {Environments}: {Perception}, {Technology}, and {Applications}}},
  \bibfield{editor}{\bibinfo{person}{Frank Steinicke}, \bibinfo{person}{Yon
  Visell}, \bibinfo{person}{Jennifer Campos}, {and} \bibinfo{person}{Anatole
  Lécuyer}} (Eds.). \bibinfo{publisher}{Springer}, \bibinfo{address}{New York,
  NY}, \bibinfo{pages}{199--219}.
\newblock
\showISBNx{978-1-4419-8432-6}
\urldef\tempurl%
\url{https://doi.org/10.1007/978-1-4419-8432-6_9}
\showDOI{\tempurl}


\bibitem[\protect\citeauthoryear{Iwata, Yano, Nakaizumi, and Kawamura}{Iwata
  et~al\mbox{.}}{2001}]%
        {iwata_project_2001}
\bibfield{author}{\bibinfo{person}{Hiroo Iwata}, \bibinfo{person}{Hiroaki
  Yano}, \bibinfo{person}{Fumitaka Nakaizumi}, {and} \bibinfo{person}{Ryo
  Kawamura}.} \bibinfo{year}{2001}\natexlab{}.
\newblock \showarticletitle{Project {FEELEX}: adding haptic surface to
  graphics}. In \bibinfo{booktitle}{\emph{Proceedings of the 28th annual
  conference on {Computer} graphics and interactive techniques - {SIGGRAPH}
  '01}}. \bibinfo{publisher}{ACM Press}, \bibinfo{address}{Not Known},
  \bibinfo{pages}{469--476}.
\newblock
\showISBNx{978-1-58113-374-5}
\urldef\tempurl%
\url{https://doi.org/10.1145/383259.383314}
\showDOI{\tempurl}


\bibitem[\protect\citeauthoryear{Jones}{Jones}{2000}]%
        {jones_kinesthetic_2000}
\bibfield{author}{\bibinfo{person}{Lynette Jones}.}
  \bibinfo{year}{2000}\natexlab{}.
\newblock \showarticletitle{Kinesthetic {Sensing}}.
\newblock \bibinfo{journal}{\emph{Human and Machine Haptics}}
  (\bibinfo{year}{2000}).
\newblock
\urldef\tempurl%
\url{http://bdml.stanford.edu/twiki/pub/Haptics/PapersInProgress/jones00.pdf}
\showURL{%
\tempurl}


\bibitem[\protect\citeauthoryear{Kim and Schneider}{Kim and Schneider}{2020}]%
        {kim_defining_2020}
\bibfield{author}{\bibinfo{person}{Erin Kim} {and} \bibinfo{person}{Oliver
  Schneider}.} \bibinfo{year}{2020}\natexlab{}.
\newblock \showarticletitle{Defining {Haptic} {Experience}: {Foundations} for
  {Understanding}, {Communicating}, and {Evaluating} {HX}}. In
  \bibinfo{booktitle}{\emph{Proceedings of the 2020 {CHI} {Conference} on
  {Human} {Factors} in {Computing} {Systems}}}. \bibinfo{publisher}{ACM},
  \bibinfo{address}{Honolulu HI USA}, \bibinfo{pages}{1--13}.
\newblock
\showISBNx{978-1-4503-6708-0}
\urldef\tempurl%
\url{https://doi.org/10.1145/3313831.3376280}
\showDOI{\tempurl}


\bibitem[\protect\citeauthoryear{Kim, Drew, Domova, and Follmer}{Kim
  et~al\mbox{.}}{2020}]%
        {kim_user-defined_2020}
\bibfield{author}{\bibinfo{person}{Lawrence~H. Kim}, \bibinfo{person}{Daniel~S.
  Drew}, \bibinfo{person}{Veronika Domova}, {and} \bibinfo{person}{Sean
  Follmer}.} \bibinfo{year}{2020}\natexlab{}.
\newblock \showarticletitle{User-defined {Swarm} {Robot} {Control}}. In
  \bibinfo{booktitle}{\emph{Proceedings of the 2020 {CHI} {Conference} on
  {Human} {Factors} in {Computing} {Systems}}}. \bibinfo{publisher}{ACM},
  \bibinfo{address}{Honolulu HI USA}, \bibinfo{pages}{1--13}.
\newblock
\showISBNx{978-1-4503-6708-0}
\urldef\tempurl%
\url{https://doi.org/10.1145/3313831.3376814}
\showDOI{\tempurl}


\bibitem[\protect\citeauthoryear{Kim, Kim, and Kim}{Kim et~al\mbox{.}}{2018}]%
        {kim_encountered-type_2018}
\bibfield{author}{\bibinfo{person}{Yaesol Kim}, \bibinfo{person}{Hyun~Jung
  Kim}, {and} \bibinfo{person}{Young~J. Kim}.} \bibinfo{year}{2018}\natexlab{}.
\newblock \showarticletitle{Encountered-type haptic display for large {VR}
  environment using per-plane reachability maps: {Encountered}-type {Haptic}
  {Display} for {Large} {VR} {Environment}}.
\newblock \bibinfo{journal}{\emph{Computer Animation and Virtual Worlds}}
  \bibinfo{volume}{29}, \bibinfo{number}{3-4} (\bibinfo{date}{May}
  \bibinfo{year}{2018}), \bibinfo{pages}{e1814}.
\newblock
\showISSN{15464261}
\urldef\tempurl%
\url{https://doi.org/10.1002/cav.1814}
\showDOI{\tempurl}


\bibitem[\protect\citeauthoryear{Knierim, Kosch, Schwind, Funk, Kiss,
  Schneegass, and Henze}{Knierim et~al\mbox{.}}{2017}]%
        {knierim_tactile_2017}
\bibfield{author}{\bibinfo{person}{Pascal Knierim}, \bibinfo{person}{Thomas
  Kosch}, \bibinfo{person}{Valentin Schwind}, \bibinfo{person}{Markus Funk},
  \bibinfo{person}{Francisco Kiss}, \bibinfo{person}{Stefan Schneegass}, {and}
  \bibinfo{person}{Niels Henze}.} \bibinfo{year}{2017}\natexlab{}.
\newblock \showarticletitle{Tactile {Drones} - {Providing} {Immersive}
  {Tactile} {Feedback} in {Virtual} {Reality} through {Quadcopters}}. In
  \bibinfo{booktitle}{\emph{Proceedings of the 2017 {CHI} {Conference}
  {Extended} {Abstracts} on {Human} {Factors} in {Computing} {Systems} - {CHI}
  {EA} '17}}. \bibinfo{publisher}{ACM Press}, \bibinfo{address}{Denver,
  Colorado, USA}, \bibinfo{pages}{433--436}.
\newblock
\showISBNx{978-1-4503-4656-6}
\urldef\tempurl%
\url{https://doi.org/10.1145/3027063.3050426}
\showDOI{\tempurl}


\bibitem[\protect\citeauthoryear{Kohli}{Kohli}{2010}]%
        {kohli_redirected_2010}
\bibfield{author}{\bibinfo{person}{Luv Kohli}.}
  \bibinfo{year}{2010}\natexlab{}.
\newblock \showarticletitle{Redirected touching: {Warping} space to remap
  passive haptics}. In \bibinfo{booktitle}{\emph{2010 {IEEE} {Symposium} on
  {3D} {User} {Interfaces} ({3DUI})}}. \bibinfo{publisher}{IEEE},
  \bibinfo{address}{Waltham, MA, USA}, \bibinfo{pages}{129--130}.
\newblock
\showISBNx{978-1-4244-6846-1}
\urldef\tempurl%
\url{https://doi.org/10.1109/3DUI.2010.5444703}
\showDOI{\tempurl}


\bibitem[\protect\citeauthoryear{Kohli, Whitton, and Brooks}{Kohli
  et~al\mbox{.}}{2012}]%
        {kohli_redirected_2012}
\bibfield{author}{\bibinfo{person}{L. Kohli}, \bibinfo{person}{M.~C. Whitton},
  {and} \bibinfo{person}{F.~P. Brooks}.} \bibinfo{year}{2012}\natexlab{}.
\newblock \showarticletitle{Redirected touching: {The} effect of warping space
  on task performance}. In \bibinfo{booktitle}{\emph{2012 {IEEE} {Symposium} on
  {3D} {User} {Interfaces} ({3DUI})}}. \bibinfo{publisher}{IEEE},
  \bibinfo{address}{Costa Mesa, CA}, \bibinfo{pages}{105--112}.
\newblock
\showISBNx{978-1-4673-1205-9 978-1-4673-1204-2}
\urldef\tempurl%
\url{https://doi.org/10.1109/3DUI.2012.6184193}
\showDOI{\tempurl}


\bibitem[\protect\citeauthoryear{Kohli, Whitton, and Brooks}{Kohli
  et~al\mbox{.}}{2013}]%
        {kohli_redirected_2013}
\bibfield{author}{\bibinfo{person}{Luv Kohli}, \bibinfo{person}{Mary~C.
  Whitton}, {and} \bibinfo{person}{Frederick~P. Brooks}.}
  \bibinfo{year}{2013}\natexlab{}.
\newblock \showarticletitle{Redirected {Touching}: {Training} and adaptation in
  warped virtual spaces}. In \bibinfo{booktitle}{\emph{2013 {IEEE} {Symposium}
  on {3D} {User} {Interfaces} ({3DUI})}}. \bibinfo{publisher}{IEEE},
  \bibinfo{address}{Orlando, FL}, \bibinfo{pages}{79--86}.
\newblock
\showISBNx{978-1-4673-6098-2 978-1-4673-6097-5}
\urldef\tempurl%
\url{https://doi.org/10.1109/3DUI.2013.6550201}
\showDOI{\tempurl}


\bibitem[\protect\citeauthoryear{Kovacs, Ofek, Gonzalez~Franco, Siu, Marwecki,
  Holz, and Sinclair}{Kovacs et~al\mbox{.}}{2020}]%
        {kovacs_haptic_2020}
\bibfield{author}{\bibinfo{person}{Robert Kovacs}, \bibinfo{person}{Eyal Ofek},
  \bibinfo{person}{Mar Gonzalez~Franco}, \bibinfo{person}{Alexa~Fay Siu},
  \bibinfo{person}{Sebastian Marwecki}, \bibinfo{person}{Christian Holz}, {and}
  \bibinfo{person}{Mike Sinclair}.} \bibinfo{year}{2020}\natexlab{}.
\newblock \showarticletitle{Haptic {PIVOT}: {On}-{Demand} {Handhelds} in {VR}}.
  In \bibinfo{booktitle}{\emph{Proceedings of the 33rd {Annual} {ACM}
  {Symposium} on {User} {Interface} {Software} and {Technology}}}
  \emph{(\bibinfo{series}{{UIST} '20})}. \bibinfo{publisher}{Association for
  Computing Machinery}, \bibinfo{address}{New York, NY, USA},
  \bibinfo{pages}{1046--1059}.
\newblock
\showISBNx{978-1-4503-7514-6}
\urldef\tempurl%
\url{https://doi.org/10.1145/3379337.3415854}
\showDOI{\tempurl}


\bibitem[\protect\citeauthoryear{Le~Goc, Kim, Parsaei, Fekete, Dragicevic, and
  Follmer}{Le~Goc et~al\mbox{.}}{2016}]%
        {le_goc_zooids:_2016}
\bibfield{author}{\bibinfo{person}{Mathieu Le~Goc},
  \bibinfo{person}{Lawrence~H. Kim}, \bibinfo{person}{Ali Parsaei},
  \bibinfo{person}{Jean-Daniel Fekete}, \bibinfo{person}{Pierre Dragicevic},
  {and} \bibinfo{person}{Sean Follmer}.} \bibinfo{year}{2016}\natexlab{}.
\newblock \showarticletitle{Zooids: {Building} {Blocks} for {Swarm} {User}
  {Interfaces}}. In \bibinfo{booktitle}{\emph{Proceedings of the 29th {Annual}
  {Symposium} on {User} {Interface} {Software} and {Technology} - {UIST} '16}}.
  \bibinfo{publisher}{ACM Press}, \bibinfo{address}{Tokyo, Japan},
  \bibinfo{pages}{97--109}.
\newblock
\showISBNx{978-1-4503-4189-9}
\urldef\tempurl%
\url{https://doi.org/10.1145/2984511.2984547}
\showDOI{\tempurl}


\bibitem[\protect\citeauthoryear{Lederman and Klatzky}{Lederman and
  Klatzky}{2009}]%
        {lederman_haptic_2009}
\bibfield{author}{\bibinfo{person}{S.~J. Lederman} {and} \bibinfo{person}{R.~L.
  Klatzky}.} \bibinfo{year}{2009}\natexlab{}.
\newblock \showarticletitle{Haptic perception: {A} tutorial}.
\newblock \bibinfo{journal}{\emph{Attention, Perception \& Psychophysics}}
  \bibinfo{volume}{71}, \bibinfo{number}{7} (\bibinfo{date}{Oct.}
  \bibinfo{year}{2009}), \bibinfo{pages}{1439--1459}.
\newblock
\showISSN{1943-3921, 1943-393X}
\urldef\tempurl%
\url{https://doi.org/10.3758/APP.71.7.1439}
\showDOI{\tempurl}


\bibitem[\protect\citeauthoryear{Lee, Hong, Lee, Choi, Han, Kim, Choi, and
  Lee}{Lee et~al\mbox{.}}{2007}]%
        {lee_mobile_2007}
\bibfield{author}{\bibinfo{person}{Chaehyun Lee}, \bibinfo{person}{Min~Sik
  Hong}, \bibinfo{person}{In Lee}, \bibinfo{person}{Oh~Kyu Choi},
  \bibinfo{person}{Kyung-Lyong Han}, \bibinfo{person}{Yoo~Yeon Kim},
  \bibinfo{person}{Seungmoon Choi}, {and} \bibinfo{person}{Jin~S Lee}.}
  \bibinfo{year}{2007}\natexlab{}.
\newblock \showarticletitle{Mobile {Haptic} {Interface} for {Large} {Immersive}
  {Virtual} {Environments}: {PoMHI} v0.5}.
\newblock  (\bibinfo{year}{2007}), \bibinfo{pages}{2}.
\newblock


\bibitem[\protect\citeauthoryear{Lee, Hwang, Han, Choi, Choi, and Lee}{Lee
  et~al\mbox{.}}{2009}]%
        {lee_system_2009}
\bibfield{author}{\bibinfo{person}{In Lee}, \bibinfo{person}{Inwook Hwang},
  \bibinfo{person}{Kyung-Lyoung Han}, \bibinfo{person}{Oh~Kyu Choi},
  \bibinfo{person}{Seungmoon Choi}, {and} \bibinfo{person}{Jin~S. Lee}.}
  \bibinfo{year}{2009}\natexlab{}.
\newblock \showarticletitle{System improvements in {Mobile} {Haptic}
  {Interface}}. In \bibinfo{booktitle}{\emph{World {Haptics} 2009 - {Third}
  {Joint} {EuroHaptics} conference and {Symposium} on {Haptic} {Interfaces} for
  {Virtual} {Environment} and {Teleoperator} {Systems}}}.
  \bibinfo{publisher}{IEEE}, \bibinfo{address}{Salt Lake City, UT, USA},
  \bibinfo{pages}{109--114}.
\newblock
\showISBNx{978-1-4244-3858-7}
\urldef\tempurl%
\url{https://doi.org/10.1109/WHC.2009.4810834}
\showDOI{\tempurl}


\bibitem[\protect\citeauthoryear{Lee, Sinclair, Gonzalez-Franco, Ofek, and
  Holz}{Lee et~al\mbox{.}}{2019}]%
        {lee_torc:_2019}
\bibfield{author}{\bibinfo{person}{Jaeyeon Lee}, \bibinfo{person}{Mike
  Sinclair}, \bibinfo{person}{Mar Gonzalez-Franco}, \bibinfo{person}{Eyal
  Ofek}, {and} \bibinfo{person}{Christian Holz}.}
  \bibinfo{year}{2019}\natexlab{}.
\newblock \showarticletitle{{TORC}: {A} {Virtual} {Reality} {Controller} for
  {In}-{Hand} {High}-{Dexterity} {Finger} {Interaction}}. In
  \bibinfo{booktitle}{\emph{Proceedings of the 2019 {CHI} {Conference} on
  {Human} {Factors} in {Computing} {Systems} - {CHI} '19}}.
  \bibinfo{publisher}{ACM Press}, \bibinfo{address}{Glasgow, Scotland Uk},
  \bibinfo{pages}{1--13}.
\newblock
\showISBNx{978-1-4503-5970-2}
\urldef\tempurl%
\url{https://doi.org/10.1145/3290605.3300301}
\showDOI{\tempurl}


\bibitem[\protect\citeauthoryear{Leithinger, Follmer, Olwal, Luescher, Hogge,
  Lee, and Ishii}{Leithinger et~al\mbox{.}}{2013}]%
        {leithinger_sublimate_2013}
\bibfield{author}{\bibinfo{person}{Daniel Leithinger}, \bibinfo{person}{Sean
  Follmer}, \bibinfo{person}{Alex Olwal}, \bibinfo{person}{Samuel Luescher},
  \bibinfo{person}{Akimitsu Hogge}, \bibinfo{person}{Jinha Lee}, {and}
  \bibinfo{person}{Hiroshi Ishii}.} \bibinfo{year}{2013}\natexlab{}.
\newblock \showarticletitle{Sublimate: state-changing virtual and physical
  rendering to augment interaction with shape displays}. In
  \bibinfo{booktitle}{\emph{Proceedings of the {SIGCHI} {Conference} on {Human}
  {Factors} in {Computing} {Systems} - {CHI} '13}}. \bibinfo{publisher}{ACM
  Press}, \bibinfo{address}{Paris, France}, \bibinfo{pages}{1441}.
\newblock
\showISBNx{978-1-4503-1899-0}
\urldef\tempurl%
\url{https://doi.org/10.1145/2470654.2466191}
\showDOI{\tempurl}


\bibitem[\protect\citeauthoryear{Lepecq, Bringoux, Pergandi, Coyle, and
  Mestre}{Lepecq et~al\mbox{.}}{2008}]%
        {lepecq_afforded_2008}
\bibfield{author}{\bibinfo{person}{Jean-Claude Lepecq}, \bibinfo{person}{Lionel
  Bringoux}, \bibinfo{person}{Jean-Marie Pergandi}, \bibinfo{person}{Thelma
  Coyle}, {and} \bibinfo{person}{Daniel Mestre}.}
  \bibinfo{year}{2008}\natexlab{}.
\newblock \showarticletitle{Afforded {Actions} as a {Behavioral} {Assessment}
  of {Physical} {Presence}}.
\newblock  (\bibinfo{year}{2008}), \bibinfo{pages}{8}.
\newblock


\bibitem[\protect\citeauthoryear{Lo, Huang, Sun, Hou, and Chen}{Lo
  et~al\mbox{.}}{2018}]%
        {lo_rollingstone_2018}
\bibfield{author}{\bibinfo{person}{Jo-Yu Lo}, \bibinfo{person}{Da-Yuan Huang},
  \bibinfo{person}{Chen-Kuo Sun}, \bibinfo{person}{Chu-En Hou}, {and}
  \bibinfo{person}{Bing-Yu Chen}.} \bibinfo{year}{2018}\natexlab{}.
\newblock \showarticletitle{{RollingStone}: {Using} {Single} {Slip} {Taxel} for
  {Enhancing} {Active} {Finger} {Exploration} with a {Virtual} {Reality}
  {Controller}}. In \bibinfo{booktitle}{\emph{The 31st {Annual} {ACM}
  {Symposium} on {User} {Interface} {Software} and {Technology} - {UIST} '18}}.
  \bibinfo{publisher}{ACM Press}, \bibinfo{address}{Berlin, Germany},
  \bibinfo{pages}{839--851}.
\newblock
\showISBNx{978-1-4503-5948-1}
\urldef\tempurl%
\url{https://doi.org/10.1145/3242587.3242627}
\showDOI{\tempurl}


\bibitem[\protect\citeauthoryear{Lopes, Jonell, and Baudisch}{Lopes
  et~al\mbox{.}}{2015}]%
        {lopes_affordance++:_2015}
\bibfield{author}{\bibinfo{person}{Pedro Lopes}, \bibinfo{person}{Patrik
  Jonell}, {and} \bibinfo{person}{Patrick Baudisch}.}
  \bibinfo{year}{2015}\natexlab{}.
\newblock \showarticletitle{Affordance++: {Allowing} {Objects} to {Communicate}
  {Dynamic} {Use}}. In \bibinfo{booktitle}{\emph{Proceedings of the 33rd
  {Annual} {ACM} {Conference} on {Human} {Factors} in {Computing} {Systems} -
  {CHI} '15}}. \bibinfo{publisher}{ACM Press}, \bibinfo{address}{Seoul,
  Republic of Korea}, \bibinfo{pages}{2515--2524}.
\newblock
\showISBNx{978-1-4503-3145-6}
\urldef\tempurl%
\url{https://doi.org/10.1145/2702123.2702128}
\showDOI{\tempurl}


\bibitem[\protect\citeauthoryear{Lopes, You, Cheng, Marwecki, and
  Baudisch}{Lopes et~al\mbox{.}}{2017}]%
        {lopes_providing_2017}
\bibfield{author}{\bibinfo{person}{Pedro Lopes}, \bibinfo{person}{Sijing You},
  \bibinfo{person}{Lung-Pan Cheng}, \bibinfo{person}{Sebastian Marwecki}, {and}
  \bibinfo{person}{Patrick Baudisch}.} \bibinfo{year}{2017}\natexlab{}.
\newblock \showarticletitle{Providing {Haptics} to {Walls} \& {Heavy} {Objects}
  in {Virtual} {Reality} by {Means} of {Electrical} {Muscle} {Stimulation}}. In
  \bibinfo{booktitle}{\emph{Proceedings of the 2017 {CHI} {Conference} on
  {Human} {Factors} in {Computing} {Systems} - {CHI} '17}}.
  \bibinfo{publisher}{ACM Press}, \bibinfo{address}{Denver, Colorado, USA},
  \bibinfo{pages}{1471--1482}.
\newblock
\showISBNx{978-1-4503-4655-9}
\urldef\tempurl%
\url{https://doi.org/10.1145/3025453.3025600}
\showDOI{\tempurl}


\bibitem[\protect\citeauthoryear{Lécuyer}{Lécuyer}{2009}]%
        {lecuyer_simulating_2009}
\bibfield{author}{\bibinfo{person}{Anatole Lécuyer}.}
  \bibinfo{year}{2009}\natexlab{}.
\newblock \showarticletitle{Simulating {Haptic} {Feedback} {Using} {Vision}:
  {A} {Survey} of {Research} and {Applications} of {Pseudo}-{Haptic}
  {Feedback}}.
\newblock \bibinfo{journal}{\emph{Presence: Teleoperators and Virtual
  Environments}} \bibinfo{volume}{18}, \bibinfo{number}{1}
  (\bibinfo{date}{Feb.} \bibinfo{year}{2009}), \bibinfo{pages}{39--53}.
\newblock
\showISSN{1054-7460, 1531-3263}
\urldef\tempurl%
\url{https://doi.org/10.1162/pres.18.1.39}
\showDOI{\tempurl}


\bibitem[\protect\citeauthoryear{Magnenat-Thalmann, {HyungSeok Kim}, Egges, and
  Garchery}{Magnenat-Thalmann et~al\mbox{.}}{2005}]%
        {magnenat-thalmann_believability_2005}
\bibfield{author}{\bibinfo{person}{N. Magnenat-Thalmann},
  \bibinfo{person}{{HyungSeok Kim}}, \bibinfo{person}{A. Egges}, {and}
  \bibinfo{person}{S. Garchery}.} \bibinfo{year}{2005}\natexlab{}.
\newblock \showarticletitle{Believability and {Interaction} in {Virtual}
  {Worlds}}. In \bibinfo{booktitle}{\emph{11th {International} {Multimedia}
  {Modelling} {Conference}}}. \bibinfo{publisher}{IEEE},
  \bibinfo{address}{Honolulu, HI, USA}, \bibinfo{pages}{2--9}.
\newblock
\showISBNx{978-0-7695-2164-0}
\urldef\tempurl%
\url{https://doi.org/10.1109/MMMC.2005.24}
\showDOI{\tempurl}


\bibitem[\protect\citeauthoryear{Makin, Barnaby, and Roudaut}{Makin
  et~al\mbox{.}}{2019}]%
        {makin_tactile_2019}
\bibfield{author}{\bibinfo{person}{Lawrence Makin}, \bibinfo{person}{Gareth
  Barnaby}, {and} \bibinfo{person}{Anne Roudaut}.}
  \bibinfo{year}{2019}\natexlab{}.
\newblock \showarticletitle{Tactile and kinesthetic feedbacks improve distance
  perception in virtual reality}. In \bibinfo{booktitle}{\emph{Proceedings of
  the 31st {Conference} on l'{Interaction} {Homme}-{Machine} - {IHM} '19}}.
  \bibinfo{publisher}{ACM Press}, \bibinfo{address}{Grenoble, France},
  \bibinfo{pages}{1--9}.
\newblock
\showISBNx{978-1-4503-7026-4}
\urldef\tempurl%
\url{https://doi.org/10.1145/3366550.3372248}
\showDOI{\tempurl}


\bibitem[\protect\citeauthoryear{Marquardt, Nacenta, Young, Carpendale,
  Greenberg, and Sharlin}{Marquardt et~al\mbox{.}}{2009}]%
        {marquardt_haptic_2009}
\bibfield{author}{\bibinfo{person}{Nicolai Marquardt},
  \bibinfo{person}{Miguel~A. Nacenta}, \bibinfo{person}{James~E. Young},
  \bibinfo{person}{Sheelagh Carpendale}, \bibinfo{person}{Saul Greenberg},
  {and} \bibinfo{person}{Ehud Sharlin}.} \bibinfo{year}{2009}\natexlab{}.
\newblock \showarticletitle{The {Haptic} {Tabletop} {Puck}: tactile feedback
  for interactive tabletops}. In \bibinfo{booktitle}{\emph{Proceedings of the
  {ACM} {International} {Conference} on {Interactive} {Tabletops} and
  {Surfaces} - {ITS} '09}}. \bibinfo{publisher}{ACM Press},
  \bibinfo{address}{Banff, Alberta, Canada}, \bibinfo{pages}{85}.
\newblock
\showISBNx{978-1-60558-733-2}
\urldef\tempurl%
\url{https://doi.org/10.1145/1731903.1731922}
\showDOI{\tempurl}


\bibitem[\protect\citeauthoryear{Massie and Salisbury}{Massie and
  Salisbury}{1994}]%
        {massie_phantom_1994}
\bibfield{author}{\bibinfo{person}{Thomas~H Massie} {and} \bibinfo{person}{J~K
  Salisbury}.} \bibinfo{year}{1994}\natexlab{}.
\newblock \showarticletitle{The {PHANTOM} {Haptic} {Interface}: {A} {Device}
  for {Probing} {Virtual} {Objects}}.
\newblock  (\bibinfo{year}{1994}), \bibinfo{pages}{5}.
\newblock


\bibitem[\protect\citeauthoryear{McNeely}{McNeely}{1993}]%
        {mcneely_robotic_1993}
\bibfield{author}{\bibinfo{person}{W.~A. McNeely}.}
  \bibinfo{year}{1993}\natexlab{}.
\newblock \showarticletitle{Robotic graphics: a new approach to force feedback
  for virtual reality}. In \bibinfo{booktitle}{\emph{Proceedings of {IEEE}
  {Virtual} {Reality} {Annual} {International} {Symposium}}}.
  \bibinfo{pages}{336--341}.
\newblock
\urldef\tempurl%
\url{https://doi.org/10.1109/VRAIS.1993.380761}
\showDOI{\tempurl}


\bibitem[\protect\citeauthoryear{Merino, Schwarzl, Kraus, Sedlmair,
  Schmalstieg, and Weiskopf}{Merino et~al\mbox{.}}{2020}]%
        {merino_evaluating_2020}
\bibfield{author}{\bibinfo{person}{Leonel Merino}, \bibinfo{person}{Magdalena
  Schwarzl}, \bibinfo{person}{Matthias Kraus}, \bibinfo{person}{Michael
  Sedlmair}, \bibinfo{person}{Dieter Schmalstieg}, {and}
  \bibinfo{person}{Daniel Weiskopf}.} \bibinfo{year}{2020}\natexlab{}.
\newblock \showarticletitle{Evaluating {Mixed} and {Augmented} {Reality}: {A}
  {Systematic} {Literature} {Review} (2009-2019)}.
\newblock \bibinfo{journal}{\emph{arXiv:2010.05988 [cs]}} (\bibinfo{date}{Oct.}
  \bibinfo{year}{2020}).
\newblock
\urldef\tempurl%
\url{http://arxiv.org/abs/2010.05988}
\showURL{%
\tempurl}
\newblock
\shownote{arXiv: 2010.05988.}


\bibitem[\protect\citeauthoryear{Moline}{Moline}{1997}]%
        {moline_virtual_1997}
\bibfield{author}{\bibinfo{person}{Judi Moline}.}
  \bibinfo{year}{1997}\natexlab{}.
\newblock \bibinfo{booktitle}{\emph{Virtual reality for health care: a
  survey}}.
\newblock \bibinfo{type}{{T}echnical {R}eport}.
\newblock


\bibitem[\protect\citeauthoryear{Nakagaki, Dementyev, Follmer, Paradiso, and
  Ishii}{Nakagaki et~al\mbox{.}}{2016a}]%
        {nakagaki_chainform:_2016}
\bibfield{author}{\bibinfo{person}{Ken Nakagaki}, \bibinfo{person}{Artem
  Dementyev}, \bibinfo{person}{Sean Follmer}, \bibinfo{person}{Joseph~A.
  Paradiso}, {and} \bibinfo{person}{Hiroshi Ishii}.}
  \bibinfo{year}{2016}\natexlab{a}.
\newblock \showarticletitle{{ChainFORM}: {A} {Linear} {Integrated} {Modular}
  {Hardware} {System} for {Shape} {Changing} {Interfaces}}. In
  \bibinfo{booktitle}{\emph{Proceedings of the 29th {Annual} {Symposium} on
  {User} {Interface} {Software} and {Technology} - {UIST} '16}}.
  \bibinfo{publisher}{ACM Press}, \bibinfo{address}{Tokyo, Japan},
  \bibinfo{pages}{87--96}.
\newblock
\showISBNx{978-1-4503-4189-9}
\urldef\tempurl%
\url{https://doi.org/10.1145/2984511.2984587}
\showDOI{\tempurl}


\bibitem[\protect\citeauthoryear{Nakagaki, Vink, Counts, Windham, Leithinger,
  Follmer, and Ishii}{Nakagaki et~al\mbox{.}}{2016b}]%
        {nakagaki_materiable:_2016}
\bibfield{author}{\bibinfo{person}{Ken Nakagaki}, \bibinfo{person}{Luke Vink},
  \bibinfo{person}{Jared Counts}, \bibinfo{person}{Daniel Windham},
  \bibinfo{person}{Daniel Leithinger}, \bibinfo{person}{Sean Follmer}, {and}
  \bibinfo{person}{Hiroshi Ishii}.} \bibinfo{year}{2016}\natexlab{b}.
\newblock \showarticletitle{Materiable: {Rendering} {Dynamic} {Material}
  {Properties} in {Response} to {Direct} {Physical} {Touch} with {Shape}
  {Changing} {Interfaces}}. In \bibinfo{booktitle}{\emph{Proceedings of the
  2016 {CHI} {Conference} on {Human} {Factors} in {Computing} {Systems} - {CHI}
  '16}}. \bibinfo{publisher}{ACM Press}, \bibinfo{address}{Santa Clara,
  California, USA}, \bibinfo{pages}{2764--2772}.
\newblock
\showISBNx{978-1-4503-3362-7}
\urldef\tempurl%
\url{https://doi.org/10.1145/2858036.2858104}
\showDOI{\tempurl}


\bibitem[\protect\citeauthoryear{Nitzsche, Hanebeck, and Schmidt}{Nitzsche
  et~al\mbox{.}}{2003}]%
        {nitzsche_design_2003}
\bibfield{author}{\bibinfo{person}{Norbert Nitzsche}, \bibinfo{person}{Uwe~D.
  Hanebeck}, {and} \bibinfo{person}{G. Schmidt}.}
  \bibinfo{year}{2003}\natexlab{}.
\newblock \showarticletitle{Design issues of mobile haptic interfaces}.
\newblock \bibinfo{journal}{\emph{Journal of Robotic Systems}}
  \bibinfo{volume}{20}, \bibinfo{number}{9} (\bibinfo{date}{Sept.}
  \bibinfo{year}{2003}), \bibinfo{pages}{549--556}.
\newblock
\showISSN{0741-2223, 1097-4563}
\urldef\tempurl%
\url{https://doi.org/10.1002/rob.10105}
\showDOI{\tempurl}


\bibitem[\protect\citeauthoryear{Norman}{Norman}{2013}]%
        {norman_design_2013}
\bibfield{author}{\bibinfo{person}{Donald~A. Norman}.}
  \bibinfo{year}{2013}\natexlab{}.
\newblock \bibinfo{booktitle}{\emph{The design of everyday things}
  (\bibinfo{edition}{revised and expanded edition} ed.)}.
\newblock \bibinfo{publisher}{Basic Books}, \bibinfo{address}{New York, New
  York}.
\newblock
\showISBNx{978-0-465-05065-9}


\bibitem[\protect\citeauthoryear{Optitrack}{Optitrack}{2019}]%
        {optitrack_motion_2019}
\bibfield{author}{\bibinfo{person}{Optitrack}.}
  \bibinfo{year}{2019}\natexlab{}.
\newblock \bibinfo{title}{Motion {Capture} {Systems}}.
\newblock
\newblock
\urldef\tempurl%
\url{http://optitrack.com/index.html}
\showURL{%
\tempurl}


\bibitem[\protect\citeauthoryear{Ortega and Coquillart}{Ortega and
  Coquillart}{2005}]%
        {ortega_prop-based_2005}
\bibfield{author}{\bibinfo{person}{M. Ortega} {and} \bibinfo{person}{S.
  Coquillart}.} \bibinfo{year}{2005}\natexlab{}.
\newblock \showarticletitle{Prop-based haptic interaction with co-location and
  immersion: an automotive application}. In \bibinfo{booktitle}{\emph{{IREE}
  {International} {Worksho} on {Haptic} {Audio} {Visual} {Environments} and
  their {Applications}, 2005.}} \bibinfo{publisher}{IEEE},
  \bibinfo{address}{Ottawa, Canada}, \bibinfo{pages}{23--28}.
\newblock
\showISBNx{978-0-7803-9376-9}
\urldef\tempurl%
\url{https://doi.org/10.1109/HAVE.2005.1545646}
\showDOI{\tempurl}


\bibitem[\protect\citeauthoryear{Pair, Neumann, Piepol, and Swartout}{Pair
  et~al\mbox{.}}{2003}]%
        {pair_flatworld:_2003}
\bibfield{author}{\bibinfo{person}{J. Pair}, \bibinfo{person}{U. Neumann},
  \bibinfo{person}{D. Piepol}, {and} \bibinfo{person}{B. Swartout}.}
  \bibinfo{year}{2003}\natexlab{}.
\newblock \showarticletitle{{FlatWorld}: combining {Hollywood} set-design
  techniques with {VR}}.
\newblock \bibinfo{journal}{\emph{IEEE Computer Graphics and Applications}}
  \bibinfo{volume}{23}, \bibinfo{number}{1} (\bibinfo{date}{Jan.}
  \bibinfo{year}{2003}), \bibinfo{pages}{12--15}.
\newblock
\showISSN{0272-1716}
\urldef\tempurl%
\url{https://doi.org/10.1109/MCG.2003.1159607}
\showDOI{\tempurl}


\bibitem[\protect\citeauthoryear{Pavlik, Vance, and Luecke}{Pavlik
  et~al\mbox{.}}{2013}]%
        {pavlik_interacting_2013}
\bibfield{author}{\bibinfo{person}{Ryan~A. Pavlik}, \bibinfo{person}{Judy~M.
  Vance}, {and} \bibinfo{person}{Greg~R. Luecke}.}
  \bibinfo{year}{2013}\natexlab{}.
\newblock \showarticletitle{Interacting {With} a {Large} {Virtual}
  {Environment} by {Combining} a {Ground}-{Based} {Haptic} {Device} and a
  {Mobile} {Robot} {Base}}. In \bibinfo{booktitle}{\emph{Volume {2B}: 33rd
  {Computers} and {Information} in {Engineering} {Conference}}}.
  \bibinfo{publisher}{ASME}, \bibinfo{address}{Portland, Oregon, USA},
  \bibinfo{pages}{V02BT02A029}.
\newblock
\showISBNx{978-0-7918-5586-7}
\urldef\tempurl%
\url{https://doi.org/10.1115/DETC2013-13441}
\showDOI{\tempurl}


\bibitem[\protect\citeauthoryear{Poyade, Molina-Tanco, Reyes-Lecuona, Langley,
  Frutos, and Flores}{Poyade et~al\mbox{.}}{2012}]%
        {poyade_validation_2012}
\bibfield{author}{\bibinfo{person}{M Poyade}, \bibinfo{person}{L Molina-Tanco},
  \bibinfo{person}{A Reyes-Lecuona}, \bibinfo{person}{A Langley},
  \bibinfo{person}{E Frutos}, {and} \bibinfo{person}{S Flores}.}
  \bibinfo{year}{2012}\natexlab{}.
\newblock \showarticletitle{Validation of a haptic virtual reality simulation
  in the context of industrial maintenance}.
\newblock  (\bibinfo{year}{2012}), \bibinfo{pages}{4}.
\newblock


\bibitem[\protect\citeauthoryear{Praveena, Rakita, Mutlu, and
  Gleicher}{Praveena et~al\mbox{.}}{2020}]%
        {praveena_supporting_2020}
\bibfield{author}{\bibinfo{person}{Pragathi Praveena}, \bibinfo{person}{Daniel
  Rakita}, \bibinfo{person}{Bilge Mutlu}, {and} \bibinfo{person}{Michael
  Gleicher}.} \bibinfo{year}{2020}\natexlab{}.
\newblock \showarticletitle{Supporting {Perception} of {Weight} through
  {Motion}-induced {Sensory} {Conflicts} in {Robot} {Teleoperation}}. In
  \bibinfo{booktitle}{\emph{Proceedings of the 2020 {ACM}/{IEEE}
  {International} {Conference} on {Human}-{Robot} {Interaction}}}.
  \bibinfo{publisher}{ACM}, \bibinfo{address}{Cambridge United Kingdom},
  \bibinfo{pages}{509--517}.
\newblock
\showISBNx{978-1-4503-6746-2}
\urldef\tempurl%
\url{https://doi.org/10.1145/3319502.3374841}
\showDOI{\tempurl}


\bibitem[\protect\citeauthoryear{Provancher, Cutkosky, Kuchenbecker, and
  Niemeyer}{Provancher et~al\mbox{.}}{2005}]%
        {provancher_contact_2005}
\bibfield{author}{\bibinfo{person}{William~R. Provancher},
  \bibinfo{person}{Mark~R. Cutkosky}, \bibinfo{person}{Katherine~J.
  Kuchenbecker}, {and} \bibinfo{person}{Günter Niemeyer}.}
  \bibinfo{year}{2005}\natexlab{}.
\newblock \showarticletitle{Contact {Location} {Display} for {Haptic}
  {Perception} of {Curvature} and {Object} {Motion}}.
\newblock \bibinfo{journal}{\emph{The International Journal of Robotics
  Research}} \bibinfo{volume}{24}, \bibinfo{number}{9} (\bibinfo{date}{Sept.}
  \bibinfo{year}{2005}), \bibinfo{pages}{691--702}.
\newblock
\showISSN{0278-3649, 1741-3176}
\urldef\tempurl%
\url{https://doi.org/10.1177/0278364905057121}
\showDOI{\tempurl}


\bibitem[\protect\citeauthoryear{Pusch and Lécuyer}{Pusch and
  Lécuyer}{2011}]%
        {pusch_pseudo-haptics:_2011}
\bibfield{author}{\bibinfo{person}{Andreas Pusch} {and}
  \bibinfo{person}{Anatole Lécuyer}.} \bibinfo{year}{2011}\natexlab{}.
\newblock \showarticletitle{Pseudo-haptics: from the theoretical foundations to
  practical system design guidelines}. In \bibinfo{booktitle}{\emph{Proceedings
  of the 13th international conference on multimodal interfaces - {ICMI} '11}}.
  \bibinfo{publisher}{ACM Press}, \bibinfo{address}{Alicante, Spain},
  \bibinfo{pages}{57}.
\newblock
\showISBNx{978-1-4503-0641-6}
\urldef\tempurl%
\url{https://doi.org/10.1145/2070481.2070494}
\showDOI{\tempurl}


\bibitem[\protect\citeauthoryear{Putze, Alexandrovsky, Putze, Höffner,
  Smeddinck, and Malaka}{Putze et~al\mbox{.}}{2020}]%
        {putze_breaking_2020}
\bibfield{author}{\bibinfo{person}{Susanne Putze}, \bibinfo{person}{Dmitry
  Alexandrovsky}, \bibinfo{person}{Felix Putze}, \bibinfo{person}{Sebastian
  Höffner}, \bibinfo{person}{Jan~David Smeddinck}, {and}
  \bibinfo{person}{Rainer Malaka}.} \bibinfo{year}{2020}\natexlab{}.
\newblock \showarticletitle{Breaking {The} {Experience}: {Effects} of
  {Questionnaires} in {VR} {User} {Studies}}. In
  \bibinfo{booktitle}{\emph{Proceedings of the 2020 {CHI} {Conference} on
  {Human} {Factors} in {Computing} {Systems}}}. \bibinfo{publisher}{ACM},
  \bibinfo{address}{Honolulu HI USA}, \bibinfo{pages}{1--15}.
\newblock
\showISBNx{978-1-4503-6708-0}
\urldef\tempurl%
\url{https://doi.org/10.1145/3313831.3376144}
\showDOI{\tempurl}


\bibitem[\protect\citeauthoryear{Rakkolainen, Freeman, Sand, Raisamo, and
  Brewster}{Rakkolainen et~al\mbox{.}}{2020}]%
        {rakkolainen_survey_2020}
\bibfield{author}{\bibinfo{person}{Ismo Rakkolainen}, \bibinfo{person}{Euan
  Freeman}, \bibinfo{person}{Antti Sand}, \bibinfo{person}{Roope Raisamo},
  {and} \bibinfo{person}{Stephen Brewster}.} \bibinfo{year}{2020}\natexlab{}.
\newblock \showarticletitle{A {Survey} of {Mid}-{Air} {Ultrasound} {Haptics}
  and {Its} {Applications}}.
\newblock \bibinfo{journal}{\emph{IEEE Transactions on Haptics}}
  (\bibinfo{year}{2020}), \bibinfo{pages}{1--1}.
\newblock
\showISSN{1939-1412, 2329-4051, 2334-0134}
\urldef\tempurl%
\url{https://doi.org/10.1109/TOH.2020.3018754}
\showDOI{\tempurl}


\bibitem[\protect\citeauthoryear{Rangarajan, Davis, and Pucher}{Rangarajan
  et~al\mbox{.}}{2020}]%
        {rangarajan_systematic_2020}
\bibfield{author}{\bibinfo{person}{Karan Rangarajan}, \bibinfo{person}{Heather
  Davis}, {and} \bibinfo{person}{Philip~H. Pucher}.}
  \bibinfo{year}{2020}\natexlab{}.
\newblock \showarticletitle{Systematic {Review} of {Virtual} {Haptics} in
  {Surgical} {Simulation}: {A} {Valid} {Educational} {Tool}?}
\newblock \bibinfo{journal}{\emph{Journal of Surgical Education}}
  \bibinfo{volume}{77}, \bibinfo{number}{2} (\bibinfo{date}{March}
  \bibinfo{year}{2020}), \bibinfo{pages}{337--347}.
\newblock
\showISSN{19317204}
\urldef\tempurl%
\url{https://doi.org/10.1016/j.jsurg.2019.09.006}
\showDOI{\tempurl}


\bibitem[\protect\citeauthoryear{Razzaque, Kohn, and Whitton}{Razzaque
  et~al\mbox{.}}{2001}]%
        {razzaque_eurographics_2001}
\bibfield{author}{\bibinfo{person}{Sharif Razzaque}, \bibinfo{person}{Zachariah
  Kohn}, {and} \bibinfo{person}{Mary~C. Whitton}.}
  \bibinfo{year}{2001}\natexlab{}.
\newblock \bibinfo{booktitle}{\emph{{EUROGRAPHICS} 2001 / {Jonathan} {C}.
  {Roberts} {Short} {Presentation} © {The} {Eurographics} {Association} 2001.
  {Redirected} {Walking}}}.
\newblock


\bibitem[\protect\citeauthoryear{Rietzler, Geiselhart, Gugenheimer, and
  Rukzio}{Rietzler et~al\mbox{.}}{2018}]%
        {rietzler_breaking_2018}
\bibfield{author}{\bibinfo{person}{Michael Rietzler}, \bibinfo{person}{Florian
  Geiselhart}, \bibinfo{person}{Jan Gugenheimer}, {and} \bibinfo{person}{Enrico
  Rukzio}.} \bibinfo{year}{2018}\natexlab{}.
\newblock \showarticletitle{Breaking the {Tracking}: {Enabling} {Weight}
  {Perception} using {Perceivable} {Tracking} {Offsets}}. In
  \bibinfo{booktitle}{\emph{Proceedings of the 2018 {CHI} {Conference} on
  {Human} {Factors} in {Computing} {Systems} - {CHI} '18}}.
  \bibinfo{publisher}{ACM Press}, \bibinfo{address}{Montreal QC, Canada},
  \bibinfo{pages}{1--12}.
\newblock
\showISBNx{978-1-4503-5620-6}
\urldef\tempurl%
\url{https://doi.org/10.1145/3173574.3173702}
\showDOI{\tempurl}


\bibitem[\protect\citeauthoryear{Rietzler, Haas, Dreja, Geiselhart, and
  Rukzio}{Rietzler et~al\mbox{.}}{2019}]%
        {rietzler_virtual_2019}
\bibfield{author}{\bibinfo{person}{Michael Rietzler}, \bibinfo{person}{Gabriel
  Haas}, \bibinfo{person}{Thomas Dreja}, \bibinfo{person}{Florian Geiselhart},
  {and} \bibinfo{person}{Enrico Rukzio}.} \bibinfo{year}{2019}\natexlab{}.
\newblock \showarticletitle{Virtual {Muscle} {Force}: {Communicating}
  {Kinesthetic} {Forces} {Through} {Pseudo}-{Haptic} {Feedback} and {Muscle}
  {Input}}. In \bibinfo{booktitle}{\emph{Proceedings of the 32nd {Annual} {ACM}
  {Symposium} on {User} {Interface} {Software} and {Technology} - {UIST} '19}}.
  \bibinfo{publisher}{ACM Press}, \bibinfo{address}{New Orleans, LA, USA},
  \bibinfo{pages}{913--922}.
\newblock
\showISBNx{978-1-4503-6816-2}
\urldef\tempurl%
\url{https://doi.org/10.1145/3332165.3347871}
\showDOI{\tempurl}


\bibitem[\protect\citeauthoryear{Rubens, Braley, Gomes, Goc, Zhang, Carrascal,
  and Vertegaal}{Rubens et~al\mbox{.}}{2015}]%
        {rubens_bitdrones:_2015}
\bibfield{author}{\bibinfo{person}{Calvin Rubens}, \bibinfo{person}{Sean
  Braley}, \bibinfo{person}{Antonio Gomes}, \bibinfo{person}{Daniel Goc},
  \bibinfo{person}{Xujing Zhang}, \bibinfo{person}{Juan~Pablo Carrascal}, {and}
  \bibinfo{person}{Roel Vertegaal}.} \bibinfo{year}{2015}\natexlab{}.
\newblock \showarticletitle{{BitDrones}: {Towards} {Levitating} {Programmable}
  {Matter} {Using} {Interactive} {3D} {Quadcopter} {Displays}}. In
  \bibinfo{booktitle}{\emph{Proceedings of the 28th {Annual} {ACM} {Symposium}
  on {User} {Interface} {Software} \& {Technology} - {UIST} '15 {Adjunct}}}.
  \bibinfo{publisher}{ACM Press}, \bibinfo{address}{Daegu, Kyungpook, Republic
  of Korea}, \bibinfo{pages}{57--58}.
\newblock
\showISBNx{978-1-4503-3780-9}
\urldef\tempurl%
\url{https://doi.org/10.1145/2815585.2817810}
\showDOI{\tempurl}


\bibitem[\protect\citeauthoryear{Sagayam and Hemanth}{Sagayam and
  Hemanth}{2017}]%
        {sagayam_hand_2017}
\bibfield{author}{\bibinfo{person}{K.~Martin Sagayam} {and}
  \bibinfo{person}{D.~Jude Hemanth}.} \bibinfo{year}{2017}\natexlab{}.
\newblock \showarticletitle{Hand posture and gesture recognition techniques for
  virtual reality applications: a survey}.
\newblock \bibinfo{journal}{\emph{Virtual Reality}} \bibinfo{volume}{21},
  \bibinfo{number}{2} (\bibinfo{date}{June} \bibinfo{year}{2017}),
  \bibinfo{pages}{91--107}.
\newblock
\showISSN{1359-4338, 1434-9957}
\urldef\tempurl%
\url{https://doi.org/10.1007/s10055-016-0301-0}
\showDOI{\tempurl}


\bibitem[\protect\citeauthoryear{Sagheb, Liu, Bahremand, Kidane, and
  LiKamWa}{Sagheb et~al\mbox{.}}{2019}]%
        {sagheb_swish:_2019}
\bibfield{author}{\bibinfo{person}{Shahabedin Sagheb},
  \bibinfo{person}{Frank~Wencheng Liu}, \bibinfo{person}{Alireza Bahremand},
  \bibinfo{person}{Assegid Kidane}, {and} \bibinfo{person}{Robert LiKamWa}.}
  \bibinfo{year}{2019}\natexlab{}.
\newblock \showarticletitle{{SWISH}: {A} {Shifting}-{Weight} {Interface} of
  {Simulated} {Hydrodynamics} for {Haptic} {Perception} of {Virtual} {Fluid}
  {Vessels}}. In \bibinfo{booktitle}{\emph{Proceedings of the 32nd {Annual}
  {ACM} {Symposium} on {User} {Interface} {Software} and {Technology} - {UIST}
  '19}}. \bibinfo{publisher}{ACM Press}, \bibinfo{address}{New Orleans, LA,
  USA}, \bibinfo{pages}{751--761}.
\newblock
\showISBNx{978-1-4503-6816-2}
\urldef\tempurl%
\url{https://doi.org/10.1145/3332165.3347870}
\showDOI{\tempurl}


\bibitem[\protect\citeauthoryear{Samad, Gatti, Hermes, Benko, and Parise}{Samad
  et~al\mbox{.}}{2019}]%
        {samad_pseudo-haptic_2019}
\bibfield{author}{\bibinfo{person}{Majed Samad}, \bibinfo{person}{Elia Gatti},
  \bibinfo{person}{Anne Hermes}, \bibinfo{person}{Hrvoje Benko}, {and}
  \bibinfo{person}{Cesare Parise}.} \bibinfo{year}{2019}\natexlab{}.
\newblock \showarticletitle{Pseudo-{Haptic} {Weight}: {Changing} the
  {Perceived} {Weight} of {Virtual} {Objects} {By} {Manipulating}
  {Control}-{Display} {Ratio}}. In \bibinfo{booktitle}{\emph{Proceedings of the
  2019 {CHI} {Conference} on {Human} {Factors} in {Computing} {Systems} - {CHI}
  '19}}. \bibinfo{publisher}{ACM Press}, \bibinfo{address}{Glasgow, Scotland
  Uk}, \bibinfo{pages}{1--13}.
\newblock
\showISBNx{978-1-4503-5970-2}
\urldef\tempurl%
\url{https://doi.org/10.1145/3290605.3300550}
\showDOI{\tempurl}


\bibitem[\protect\citeauthoryear{Satler, Avizzano, and Ruffaldi}{Satler
  et~al\mbox{.}}{2011}]%
        {satler_control_2011}
\bibfield{author}{\bibinfo{person}{Massimo Satler}, \bibinfo{person}{Carlo~A.
  Avizzano}, {and} \bibinfo{person}{Emanuele Ruffaldi}.}
  \bibinfo{year}{2011}\natexlab{}.
\newblock \showarticletitle{Control of a desktop mobile haptic interface}. In
  \bibinfo{booktitle}{\emph{2011 {IEEE} {World} {Haptics} {Conference}}}.
  \bibinfo{publisher}{IEEE}, \bibinfo{address}{Istanbul},
  \bibinfo{pages}{415--420}.
\newblock
\showISBNx{978-1-4577-0299-0}
\urldef\tempurl%
\url{https://doi.org/10.1109/WHC.2011.5945522}
\showDOI{\tempurl}


\bibitem[\protect\citeauthoryear{Sato}{Sato}{2002}]%
        {sato_spidar_2002}
\bibfield{author}{\bibinfo{person}{M. Sato}.} \bibinfo{year}{2002}\natexlab{}.
\newblock \showarticletitle{{SPIDAR} and virtual reality}. In
  \bibinfo{booktitle}{\emph{Proceedings of the 5th {Biannual} {World}
  {Automation} {Congress}}}, Vol.~\bibinfo{volume}{13}.
  \bibinfo{pages}{17--23}.
\newblock
\urldef\tempurl%
\url{https://doi.org/10.1109/WAC.2002.1049515}
\showDOI{\tempurl}


\bibitem[\protect\citeauthoryear{Savino}{Savino}{2020}]%
        {savino_virtual_2020}
\bibfield{author}{\bibinfo{person}{Gian-Luca Savino}.}
  \bibinfo{year}{2020}\natexlab{}.
\newblock \showarticletitle{Virtual {Smartphone}: {High} {Fidelity}
  {Interaction} with {Proxy} {Objects} in {Virtual} {Reality}}.
\newblock \bibinfo{journal}{\emph{arXiv:2010.00942 [cs]}} (\bibinfo{date}{Oct.}
  \bibinfo{year}{2020}).
\newblock
\urldef\tempurl%
\url{http://arxiv.org/abs/2010.00942}
\showURL{%
\tempurl}
\newblock
\shownote{arXiv: 2010.00942.}


\bibitem[\protect\citeauthoryear{Schmidt, Kovacs, Mehta, Umapathi, Köhler,
  Cheng, and Baudisch}{Schmidt et~al\mbox{.}}{2015}]%
        {schmidt_level-ups_2015}
\bibfield{author}{\bibinfo{person}{Dominik Schmidt}, \bibinfo{person}{Rob
  Kovacs}, \bibinfo{person}{Vikram Mehta}, \bibinfo{person}{Udayan Umapathi},
  \bibinfo{person}{Sven Köhler}, \bibinfo{person}{Lung-Pan Cheng}, {and}
  \bibinfo{person}{Patrick Baudisch}.} \bibinfo{year}{2015}\natexlab{}.
\newblock \showarticletitle{Level-{Ups}: {Motorized} {Stilts} that {Simulate}
  {Stair} {Steps} in {Virtual} {Reality}}. In
  \bibinfo{booktitle}{\emph{Proceedings of the 33rd {Annual} {ACM} {Conference}
  on {Human} {Factors} in {Computing} {Systems} - {CHI} '15}}.
  \bibinfo{publisher}{ACM Press}, \bibinfo{address}{Seoul, Republic of Korea},
  \bibinfo{pages}{2157--2160}.
\newblock
\showISBNx{978-1-4503-3145-6}
\urldef\tempurl%
\url{https://doi.org/10.1145/2702123.2702253}
\showDOI{\tempurl}


\bibitem[\protect\citeauthoryear{Schuemie, van~der Straaten, Krijn, and van~der
  Mast}{Schuemie et~al\mbox{.}}{2001}]%
        {schuemie_research_2001}
\bibfield{author}{\bibinfo{person}{Martijn~J. Schuemie}, \bibinfo{person}{Peter
  van~der Straaten}, \bibinfo{person}{Merel Krijn}, {and}
  \bibinfo{person}{Charles~A.P.G. van~der Mast}.}
  \bibinfo{year}{2001}\natexlab{}.
\newblock \showarticletitle{Research on {Presence} in {Virtual} {Reality}: {A}
  {Survey}}.
\newblock \bibinfo{journal}{\emph{CyberPsychology \& Behavior}}
  \bibinfo{volume}{4}, \bibinfo{number}{2} (\bibinfo{date}{April}
  \bibinfo{year}{2001}), \bibinfo{pages}{183--201}.
\newblock
\showISSN{1094-9313, 1557-8364}
\urldef\tempurl%
\url{https://doi.org/10.1089/109493101300117884}
\showDOI{\tempurl}


\bibitem[\protect\citeauthoryear{Schwind, Knierim, Haas, and Henze}{Schwind
  et~al\mbox{.}}{2019}]%
        {schwind_using_2019}
\bibfield{author}{\bibinfo{person}{Valentin Schwind}, \bibinfo{person}{Pascal
  Knierim}, \bibinfo{person}{Nico Haas}, {and} \bibinfo{person}{Niels Henze}.}
  \bibinfo{year}{2019}\natexlab{}.
\newblock \showarticletitle{Using {Presence} {Questionnaires} in {Virtual}
  {Reality}}. In \bibinfo{booktitle}{\emph{Proceedings of the 2019 {CHI}
  {Conference} on {Human} {Factors} in {Computing} {Systems} - {CHI} '19}}.
  \bibinfo{publisher}{ACM Press}, \bibinfo{address}{Glasgow, Scotland Uk},
  \bibinfo{pages}{1--12}.
\newblock
\showISBNx{978-1-4503-5970-2}
\urldef\tempurl%
\url{https://doi.org/10.1145/3290605.3300590}
\showDOI{\tempurl}


\bibitem[\protect\citeauthoryear{Seifi, Fazlollahi, Oppermann, Sastrillo, Ip,
  Agrawal, Park, Kuchenbecker, and MacLean}{Seifi et~al\mbox{.}}{2019}]%
        {seifi_haptipedia:_2019}
\bibfield{author}{\bibinfo{person}{Hasti Seifi}, \bibinfo{person}{Farimah
  Fazlollahi}, \bibinfo{person}{Michael Oppermann},
  \bibinfo{person}{John~Andrew Sastrillo}, \bibinfo{person}{Jessica Ip},
  \bibinfo{person}{Ashutosh Agrawal}, \bibinfo{person}{Gunhyuk Park},
  \bibinfo{person}{Katherine~J. Kuchenbecker}, {and} \bibinfo{person}{Karon~E.
  MacLean}.} \bibinfo{year}{2019}\natexlab{}.
\newblock \showarticletitle{Haptipedia: {Accelerating} {Haptic} {Device}
  {Discovery} to {Support} {Interaction} \& {Engineering} {Design}}. In
  \bibinfo{booktitle}{\emph{Proceedings of the 2019 {CHI} {Conference} on
  {Human} {Factors} in {Computing} {Systems} - {CHI} '19}}.
  \bibinfo{publisher}{ACM Press}, \bibinfo{address}{Glasgow, Scotland Uk},
  \bibinfo{pages}{1--12}.
\newblock
\showISBNx{978-1-4503-5970-2}
\urldef\tempurl%
\url{https://doi.org/10.1145/3290605.3300788}
\showDOI{\tempurl}


\bibitem[\protect\citeauthoryear{Shaw, Roper, Nilsson, Lawson, Cobb, and
  Miller}{Shaw et~al\mbox{.}}{2019}]%
        {shaw_heat_2019}
\bibfield{author}{\bibinfo{person}{Emily Shaw}, \bibinfo{person}{Tessa Roper},
  \bibinfo{person}{Tommy Nilsson}, \bibinfo{person}{Glyn Lawson},
  \bibinfo{person}{Sue V.~G. Cobb}, {and} \bibinfo{person}{Daniel Miller}.}
  \bibinfo{year}{2019}\natexlab{}.
\newblock \showarticletitle{The {Heat} is {On}: {Exploring} {User} {Behaviour}
  in a {Multisensory} {Virtual} {Environment} for {Fire} {Evacuation}}.
\newblock \bibinfo{journal}{\emph{Proceedings of the 2019 CHI Conference on
  Human Factors in Computing Systems - CHI '19}} (\bibinfo{year}{2019}),
  \bibinfo{pages}{1--13}.
\newblock
\urldef\tempurl%
\url{https://doi.org/10.1145/3290605.3300856}
\showDOI{\tempurl}
\newblock
\shownote{arXiv: 1902.04573.}


\bibitem[\protect\citeauthoryear{Shigeyama, Hashimoto, Yoshida, Narumi,
  Tanikawa, and Hirose}{Shigeyama et~al\mbox{.}}{2019}]%
        {shigeyama_transcalibur:_2019}
\bibfield{author}{\bibinfo{person}{Jotaro Shigeyama}, \bibinfo{person}{Takeru
  Hashimoto}, \bibinfo{person}{Shigeo Yoshida}, \bibinfo{person}{Takuji
  Narumi}, \bibinfo{person}{Tomohiro Tanikawa}, {and}
  \bibinfo{person}{Michitaka Hirose}.} \bibinfo{year}{2019}\natexlab{}.
\newblock \showarticletitle{Transcalibur: {A} {Weight} {Shifting} {Virtual}
  {Reality} {Controller} for {2D} {Shape} {Rendering} based on {Computational}
  {Perception} {Model}}. In \bibinfo{booktitle}{\emph{Proceedings of the 2019
  {CHI} {Conference} on {Human} {Factors} in {Computing} {Systems} - {CHI}
  '19}}. \bibinfo{publisher}{ACM Press}, \bibinfo{address}{Glasgow, Scotland
  Uk}, \bibinfo{pages}{1--11}.
\newblock
\showISBNx{978-1-4503-5970-2}
\urldef\tempurl%
\url{https://doi.org/10.1145/3290605.3300241}
\showDOI{\tempurl}


\bibitem[\protect\citeauthoryear{Simeone, Velloso, and Gellersen}{Simeone
  et~al\mbox{.}}{2015}]%
        {simeone_substitutional_2015}
\bibfield{author}{\bibinfo{person}{Adalberto~L. Simeone},
  \bibinfo{person}{Eduardo Velloso}, {and} \bibinfo{person}{Hans Gellersen}.}
  \bibinfo{year}{2015}\natexlab{}.
\newblock \showarticletitle{Substitutional {Reality}: {Using} the {Physical}
  {Environment} to {Design} {Virtual} {Reality} {Experiences}}. In
  \bibinfo{booktitle}{\emph{Proceedings of the 33rd {Annual} {ACM} {Conference}
  on {Human} {Factors} in {Computing} {Systems} - {CHI} '15}}.
  \bibinfo{publisher}{ACM Press}, \bibinfo{address}{Seoul, Republic of Korea},
  \bibinfo{pages}{3307--3316}.
\newblock
\showISBNx{978-1-4503-3145-6}
\urldef\tempurl%
\url{https://doi.org/10.1145/2702123.2702389}
\showDOI{\tempurl}


\bibitem[\protect\citeauthoryear{Sinclair, Ofek, Gonzalez-Franco, and
  Holz}{Sinclair et~al\mbox{.}}{2019}]%
        {sinclair_capstancrunch:_2019}
\bibfield{author}{\bibinfo{person}{Mike Sinclair}, \bibinfo{person}{Eyal Ofek},
  \bibinfo{person}{Mar Gonzalez-Franco}, {and} \bibinfo{person}{Christian
  Holz}.} \bibinfo{year}{2019}\natexlab{}.
\newblock \showarticletitle{{CapstanCrunch}: {A} {Haptic} {VR} {Controller}
  with {User}-supplied {Force} {Feedback}}. In
  \bibinfo{booktitle}{\emph{Proceedings of the 32nd {Annual} {ACM} {Symposium}
  on {User} {Interface} {Software} and {Technology} - {UIST} '19}}.
  \bibinfo{publisher}{ACM Press}, \bibinfo{address}{New Orleans, LA, USA},
  \bibinfo{pages}{815--829}.
\newblock
\showISBNx{978-1-4503-6816-2}
\urldef\tempurl%
\url{https://doi.org/10.1145/3332165.3347891}
\showDOI{\tempurl}


\bibitem[\protect\citeauthoryear{Sinclair, Pahud, and Benko}{Sinclair
  et~al\mbox{.}}{2014}]%
        {sinclair_touchmover_2014}
\bibfield{author}{\bibinfo{person}{Mike Sinclair}, \bibinfo{person}{Michel
  Pahud}, {and} \bibinfo{person}{Hrvoje Benko}.}
  \bibinfo{year}{2014}\natexlab{}.
\newblock \showarticletitle{{TouchMover} 2.0 - {3D} touchscreen with force
  feedback and haptic texture}. In \bibinfo{booktitle}{\emph{2014 {IEEE}
  {Haptics} {Symposium} ({HAPTICS})}}. \bibinfo{publisher}{IEEE},
  \bibinfo{address}{Houston, TX, USA}, \bibinfo{pages}{1--6}.
\newblock
\showISBNx{978-1-4799-3131-6}
\urldef\tempurl%
\url{https://doi.org/10.1109/HAPTICS.2014.6775425}
\showDOI{\tempurl}


\bibitem[\protect\citeauthoryear{Siu, Gonzalez, Yuan, Ginsberg, and
  Follmer}{Siu et~al\mbox{.}}{2018}]%
        {siu_shapeshift:_2018}
\bibfield{author}{\bibinfo{person}{Alexa~F. Siu}, \bibinfo{person}{Eric~J.
  Gonzalez}, \bibinfo{person}{Shenli Yuan}, \bibinfo{person}{Jason~B.
  Ginsberg}, {and} \bibinfo{person}{Sean Follmer}.}
  \bibinfo{year}{2018}\natexlab{}.
\newblock \showarticletitle{{shapeShift}: {2D} {Spatial} {Manipulation} and
  {Self}-{Actuation} of {Tabletop} {Shape} {Displays} for {Tangible} and
  {Haptic} {Interaction}}. In \bibinfo{booktitle}{\emph{Proceedings of the 2018
  {CHI} {Conference} on {Human} {Factors} in {Computing} {Systems} - {CHI}
  '18}}. \bibinfo{publisher}{ACM Press}, \bibinfo{address}{Montreal QC,
  Canada}, \bibinfo{pages}{1--13}.
\newblock
\showISBNx{978-1-4503-5620-6}
\urldef\tempurl%
\url{https://doi.org/10.1145/3173574.3173865}
\showDOI{\tempurl}


\bibitem[\protect\citeauthoryear{Slater}{Slater}{1999}]%
        {slater_measuring_1999}
\bibfield{author}{\bibinfo{person}{Mel Slater}.}
  \bibinfo{year}{1999}\natexlab{}.
\newblock \showarticletitle{Measuring {Presence}: {A} {Response} to the
  {Witmer} and {Singer} {Presence} {Questionnaire}}.
\newblock \bibinfo{journal}{\emph{Presence: Teleoperators and Virtual
  Environments}} \bibinfo{volume}{8}, \bibinfo{number}{5} (\bibinfo{date}{Oct.}
  \bibinfo{year}{1999}), \bibinfo{pages}{560--565}.
\newblock
\urldef\tempurl%
\url{https://doi.org/10.1162/105474699566477}
\showDOI{\tempurl}
\newblock
\shownote{Publisher: MIT Press.}


\bibitem[\protect\citeauthoryear{Slater, Usoh, and Steed}{Slater
  et~al\mbox{.}}{1994}]%
        {slater_depth_1994}
\bibfield{author}{\bibinfo{person}{Mel Slater}, \bibinfo{person}{Martin Usoh},
  {and} \bibinfo{person}{Anthony Steed}.} \bibinfo{year}{1994}\natexlab{}.
\newblock \showarticletitle{Depth of {Presence} in {Virtual} {Environments}}.
\newblock \bibinfo{journal}{\emph{Presence: Teleoperators and Virtual
  Environments}} \bibinfo{volume}{3}, \bibinfo{number}{2} (\bibinfo{date}{Jan.}
  \bibinfo{year}{1994}), \bibinfo{pages}{130--144}.
\newblock
\showISSN{1054-7460, 1531-3263}
\urldef\tempurl%
\url{https://doi.org/10.1162/pres.1994.3.2.130}
\showDOI{\tempurl}


\bibitem[\protect\citeauthoryear{Steed, Friston, Pawar, and Swapp}{Steed
  et~al\mbox{.}}{2020}]%
        {steed_docking_2020}
\bibfield{author}{\bibinfo{person}{Anthony Steed}, \bibinfo{person}{Sebastian
  Friston}, \bibinfo{person}{Vijay Pawar}, {and} \bibinfo{person}{David
  Swapp}.} \bibinfo{year}{2020}\natexlab{}.
\newblock \showarticletitle{Docking {Haptics}: {Extending} the {Reach} of
  {Haptics} by {Dynamic} {Combinations} of {Grounded} and {Worn} {Devices}}.
\newblock \bibinfo{journal}{\emph{arXiv:2002.06093 [cs]}} (\bibinfo{date}{Feb.}
  \bibinfo{year}{2020}).
\newblock
\urldef\tempurl%
\url{http://arxiv.org/abs/2002.06093}
\showURL{%
\tempurl}
\newblock
\shownote{arXiv: 2002.06093.}


\bibitem[\protect\citeauthoryear{Steinicke, Yon, Campos, and Lecuyer}{Steinicke
  et~al\mbox{.}}{2013}]%
        {steinicke_human_2013}
\bibfield{editor}{\bibinfo{person}{Frank Steinicke}, \bibinfo{person}{Visell
  Yon}, \bibinfo{person}{Jennifer Campos}, {and} \bibinfo{person}{Anatole
  Lecuyer}} (Eds.). \bibinfo{year}{2013}\natexlab{}.
\newblock \bibinfo{booktitle}{\emph{Human walking in virtual environments:
  perception, technology, and applications}}.
\newblock \bibinfo{publisher}{Springer}, \bibinfo{address}{New York,NY}.
\newblock
\showISBNx{978-1-4419-8431-9 978-1-4419-8432-6}
\newblock
\shownote{OCLC: 856865949.}


\bibitem[\protect\citeauthoryear{Strandholt, Dogaru, Nilsson, Nordahl, and
  Serafin}{Strandholt et~al\mbox{.}}{2020}]%
        {strandholt_knock_2020}
\bibfield{author}{\bibinfo{person}{Patrick~L. Strandholt},
  \bibinfo{person}{Oana~A. Dogaru}, \bibinfo{person}{Niels~C. Nilsson},
  \bibinfo{person}{Rolf Nordahl}, {and} \bibinfo{person}{Stefania Serafin}.}
  \bibinfo{year}{2020}\natexlab{}.
\newblock \showarticletitle{Knock on {Wood}: {Combining} {Redirected}
  {Touching} and {Physical} {Props} for {Tool}-{Based} {Interaction} in
  {Virtual} {Reality}}. In \bibinfo{booktitle}{\emph{Proceedings of the 2020
  {CHI} {Conference} on {Human} {Factors} in {Computing} {Systems}}}.
  \bibinfo{publisher}{ACM}, \bibinfo{address}{Honolulu HI USA},
  \bibinfo{pages}{1--13}.
\newblock
\showISBNx{978-1-4503-6708-0}
\urldef\tempurl%
\url{https://doi.org/10.1145/3313831.3376303}
\showDOI{\tempurl}


\bibitem[\protect\citeauthoryear{Strasnick, Holz, Ofek, Sinclair, and
  Benko}{Strasnick et~al\mbox{.}}{2018}]%
        {strasnick_haptic_2018}
\bibfield{author}{\bibinfo{person}{Evan Strasnick}, \bibinfo{person}{Christian
  Holz}, \bibinfo{person}{Eyal Ofek}, \bibinfo{person}{Mike Sinclair}, {and}
  \bibinfo{person}{Hrvoje Benko}.} \bibinfo{year}{2018}\natexlab{}.
\newblock \showarticletitle{Haptic {Links}: {Bimanual} {Haptics} for {Virtual}
  {Reality} {Using} {Variable} {Stiffness} {Actuation}}. In
  \bibinfo{booktitle}{\emph{Proceedings of the 2018 {CHI} {Conference} on
  {Human} {Factors} in {Computing} {Systems} - {CHI} '18}}.
  \bibinfo{publisher}{ACM Press}, \bibinfo{address}{Montreal QC, Canada},
  \bibinfo{pages}{1--12}.
\newblock
\showISBNx{978-1-4503-5620-6}
\urldef\tempurl%
\url{https://doi.org/10.1145/3173574.3174218}
\showDOI{\tempurl}


\bibitem[\protect\citeauthoryear{Strohmeier, Güngör, Herres, Gudea, Fruchard,
  and Steimle}{Strohmeier et~al\mbox{.}}{2020}]%
        {strohmeier_barefoot_2020}
\bibfield{author}{\bibinfo{person}{Paul Strohmeier}, \bibinfo{person}{Seref
  Güngör}, \bibinfo{person}{Luis Herres}, \bibinfo{person}{Dennis Gudea},
  \bibinfo{person}{Bruno Fruchard}, {and} \bibinfo{person}{Jürgen Steimle}.}
  \bibinfo{year}{2020}\natexlab{}.
\newblock \showarticletitle{{bARefoot}: {Generating} {Virtual} {Materials}
  using {Motion} {Coupled} {Vibration} in {Shoes}}. In
  \bibinfo{booktitle}{\emph{Proceedings of the 33rd {Annual} {ACM} {Symposium}
  on {User} {Interface} {Software} and {Technology}}}.
  \bibinfo{publisher}{ACM}, \bibinfo{address}{Virtual Event USA},
  \bibinfo{pages}{579--593}.
\newblock
\showISBNx{978-1-4503-7514-6}
\urldef\tempurl%
\url{https://doi.org/10.1145/3379337.3415828}
\showDOI{\tempurl}


\bibitem[\protect\citeauthoryear{Sun, Yoshida, Narumi, and Hirose}{Sun
  et~al\mbox{.}}{2019}]%
        {sun_pacapa:_2019}
\bibfield{author}{\bibinfo{person}{Yuqian Sun}, \bibinfo{person}{Shigeo
  Yoshida}, \bibinfo{person}{Takuji Narumi}, {and} \bibinfo{person}{Michitaka
  Hirose}.} \bibinfo{year}{2019}\natexlab{}.
\newblock \showarticletitle{{PaCaPa}: {A} {Handheld} {VR} {Device} for
  {Rendering} {Size}, {Shape}, and {Stiffness} of {Virtual} {Objects} in
  {Tool}-based {Interactions}}. In \bibinfo{booktitle}{\emph{Proceedings of the
  2019 {CHI} {Conference} on {Human} {Factors} in {Computing} {Systems} - {CHI}
  '19}}. \bibinfo{publisher}{ACM Press}, \bibinfo{address}{Glasgow, Scotland
  Uk}, \bibinfo{pages}{1--12}.
\newblock
\showISBNx{978-1-4503-5970-2}
\urldef\tempurl%
\url{https://doi.org/10.1145/3290605.3300682}
\showDOI{\tempurl}


\bibitem[\protect\citeauthoryear{Sutherland}{Sutherland}{1965}]%
        {sutherland_ultimate_1965}
\bibfield{author}{\bibinfo{person}{Ivan Sutherland}.}
  \bibinfo{year}{1965}\natexlab{}.
\newblock \showarticletitle{The {Ultimate} {Display}}.
\newblock  (\bibinfo{year}{1965}), \bibinfo{pages}{2}.
\newblock


\bibitem[\protect\citeauthoryear{Suzuki, Hedayati, Zheng, Bohn, Szafir, Do,
  Gross, and Leithinger}{Suzuki et~al\mbox{.}}{2020}]%
        {suzuki_roomshift_2020}
\bibfield{author}{\bibinfo{person}{Ryo Suzuki}, \bibinfo{person}{Hooman
  Hedayati}, \bibinfo{person}{Clement Zheng}, \bibinfo{person}{James Bohn},
  \bibinfo{person}{Daniel Szafir}, \bibinfo{person}{Ellen Yi-Luen Do},
  \bibinfo{person}{Mark~D Gross}, {and} \bibinfo{person}{Daniel Leithinger}.}
  \bibinfo{year}{2020}\natexlab{}.
\newblock \showarticletitle{{RoomShift}: {Room}-scale {Dynamic} {Haptics} for
  {VR} with {Furniture}-moving {Swarm} {Robots}}.
\newblock  (\bibinfo{year}{2020}), \bibinfo{pages}{11}.
\newblock


\bibitem[\protect\citeauthoryear{Suzuki, Yamaoka, Leithinger, Yeh, Gross,
  Kawahara, and Kakehi}{Suzuki et~al\mbox{.}}{2018}]%
        {suzuki_dynablock:_2018}
\bibfield{author}{\bibinfo{person}{Ryo Suzuki}, \bibinfo{person}{Junichi
  Yamaoka}, \bibinfo{person}{Daniel Leithinger}, \bibinfo{person}{Tom Yeh},
  \bibinfo{person}{Mark~D. Gross}, \bibinfo{person}{Yoshihiro Kawahara}, {and}
  \bibinfo{person}{Yasuaki Kakehi}.} \bibinfo{year}{2018}\natexlab{}.
\newblock \showarticletitle{Dynablock: {Dynamic} {3D} {Printing} for {Instant}
  and {Reconstructable} {Shape} {Formation}}. In \bibinfo{booktitle}{\emph{The
  31st {Annual} {ACM} {Symposium} on {User} {Interface} {Software} and
  {Technology} - {UIST} '18}}. \bibinfo{publisher}{ACM Press},
  \bibinfo{address}{Berlin, Germany}, \bibinfo{pages}{99--111}.
\newblock
\showISBNx{978-1-4503-5948-1}
\urldef\tempurl%
\url{https://doi.org/10.1145/3242587.3242659}
\showDOI{\tempurl}


\bibitem[\protect\citeauthoryear{Suzuki, Zheng, Kakehi, Yeh, Do, Gross, and
  Leithinger}{Suzuki et~al\mbox{.}}{2019}]%
        {suzuki_shapebots:_2019}
\bibfield{author}{\bibinfo{person}{Ryo Suzuki}, \bibinfo{person}{Clement
  Zheng}, \bibinfo{person}{Yasuaki Kakehi}, \bibinfo{person}{Tom Yeh},
  \bibinfo{person}{Ellen Yi-Luen Do}, \bibinfo{person}{Mark~D Gross}, {and}
  \bibinfo{person}{Daniel Leithinger}.} \bibinfo{year}{2019}\natexlab{}.
\newblock \showarticletitle{{ShapeBots}: {Shape}-changing {Swarm} {Robots}}.
\newblock  (\bibinfo{year}{2019}), \bibinfo{pages}{13}.
\newblock


\bibitem[\protect\citeauthoryear{Takizawa, Yano, Iwata, Oshiro, and
  Ohkohchi}{Takizawa et~al\mbox{.}}{2017}]%
        {takizawa_encountered-type_2017}
\bibfield{author}{\bibinfo{person}{N. Takizawa}, \bibinfo{person}{H. Yano},
  \bibinfo{person}{H. Iwata}, \bibinfo{person}{Y. Oshiro}, {and}
  \bibinfo{person}{N. Ohkohchi}.} \bibinfo{year}{2017}\natexlab{}.
\newblock \showarticletitle{Encountered-{Type} {Haptic} {Interface} for
  {Representation} of {Shape} and {Rigidity} of {3D} {Virtual} {Objects}}.
\newblock \bibinfo{journal}{\emph{IEEE Transactions on Haptics}}
  \bibinfo{volume}{10}, \bibinfo{number}{4} (\bibinfo{date}{Oct.}
  \bibinfo{year}{2017}), \bibinfo{pages}{500--510}.
\newblock
\showISSN{1939-1412}
\urldef\tempurl%
\url{https://doi.org/10.1109/TOH.2017.2740934}
\showDOI{\tempurl}


\bibitem[\protect\citeauthoryear{Talvas, Marchal, and Lecuyer}{Talvas
  et~al\mbox{.}}{2014}]%
        {talvas_survey_2014}
\bibfield{author}{\bibinfo{person}{Anthony Talvas}, \bibinfo{person}{Maud
  Marchal}, {and} \bibinfo{person}{Anatole Lecuyer}.}
  \bibinfo{year}{2014}\natexlab{}.
\newblock \showarticletitle{A {Survey} on {Bimanual} {Haptic} {Interaction}}.
\newblock \bibinfo{journal}{\emph{IEEE Transactions on Haptics}}
  \bibinfo{volume}{7}, \bibinfo{number}{3} (\bibinfo{date}{July}
  \bibinfo{year}{2014}), \bibinfo{pages}{285--300}.
\newblock
\showISSN{1939-1412}
\urldef\tempurl%
\url{https://doi.org/10.1109/TOH.2014.2314456}
\showDOI{\tempurl}


\bibitem[\protect\citeauthoryear{Teng, Kuo, Wang, Chiang, Huang, Chan, and
  Chen}{Teng et~al\mbox{.}}{2018}]%
        {teng_pupop_2018}
\bibfield{author}{\bibinfo{person}{Shan-Yuan Teng}, \bibinfo{person}{Tzu-Sheng
  Kuo}, \bibinfo{person}{Chi Wang}, \bibinfo{person}{Chi-huan Chiang},
  \bibinfo{person}{Da-Yuan Huang}, \bibinfo{person}{Liwei Chan}, {and}
  \bibinfo{person}{Bing-Yu Chen}.} \bibinfo{year}{2018}\natexlab{}.
\newblock \showarticletitle{{PuPoP}: {Pop}-up {Prop} on {Palm} for {Virtual}
  {Reality}}. In \bibinfo{booktitle}{\emph{The 31st {Annual} {ACM} {Symposium}
  on {User} {Interface} {Software} and {Technology} - {UIST} '18}}.
  \bibinfo{publisher}{ACM Press}, \bibinfo{address}{Berlin, Germany},
  \bibinfo{pages}{5--17}.
\newblock
\showISBNx{978-1-4503-5948-1}
\urldef\tempurl%
\url{https://doi.org/10.1145/3242587.3242628}
\showDOI{\tempurl}


\bibitem[\protect\citeauthoryear{Teng, Lin, Chiang, Kuo, Chan, Huang, and
  Chen}{Teng et~al\mbox{.}}{2019}]%
        {teng_tilepop:_2019}
\bibfield{author}{\bibinfo{person}{Shan-Yuan Teng}, \bibinfo{person}{Cheng-Lung
  Lin}, \bibinfo{person}{Chi-huan Chiang}, \bibinfo{person}{Tzu-Sheng Kuo},
  \bibinfo{person}{Liwei Chan}, \bibinfo{person}{Da-Yuan Huang}, {and}
  \bibinfo{person}{Bing-Yu Chen}.} \bibinfo{year}{2019}\natexlab{}.
\newblock \showarticletitle{{TilePoP}: {Tile}-type {Pop}-up {Prop} for
  {Virtual} {Reality}}.
\newblock  (\bibinfo{year}{2019}), \bibinfo{pages}{11}.
\newblock


\bibitem[\protect\citeauthoryear{Teyssier, Bailly, Pelachaud, and
  Lecolinet}{Teyssier et~al\mbox{.}}{2020}]%
        {teyssier_conveying_2020}
\bibfield{author}{\bibinfo{person}{Marc Teyssier}, \bibinfo{person}{Gilles
  Bailly}, \bibinfo{person}{Catherine Pelachaud}, {and} \bibinfo{person}{Eric
  Lecolinet}.} \bibinfo{year}{2020}\natexlab{}.
\newblock \showarticletitle{Conveying {Emotions} {Through} {Device}-{Initiated}
  {Touch}}.
\newblock \bibinfo{journal}{\emph{IEEE Transactions on Affective Computing}}
  (\bibinfo{year}{2020}), \bibinfo{pages}{1--1}.
\newblock
\showISSN{1949-3045, 2371-9850}
\urldef\tempurl%
\url{https://doi.org/10.1109/TAFFC.2020.3008693}
\showDOI{\tempurl}


\bibitem[\protect\citeauthoryear{Tsagarakis, Horne, and Caldwell}{Tsagarakis
  et~al\mbox{.}}{2005}]%
        {tsagarakis_slip_2005}
\bibfield{author}{\bibinfo{person}{N.~G. Tsagarakis}, \bibinfo{person}{T.
  Horne}, {and} \bibinfo{person}{D.~G. Caldwell}.}
  \bibinfo{year}{2005}\natexlab{}.
\newblock \showarticletitle{{SLIP} {AESTHEASIS}: a portable {2D} slip/skin
  stretch display for the fingertip}. In \bibinfo{booktitle}{\emph{First
  {Joint} {Eurohaptics} {Conference} and {Symposium} on {Haptic} {Interfaces}
  for {Virtual} {Environment} and {Teleoperator} {Systems}. {World} {Haptics}
  {Conference}}}. \bibinfo{pages}{214--219}.
\newblock
\urldef\tempurl%
\url{https://doi.org/10.1109/WHC.2005.117}
\showDOI{\tempurl}


\bibitem[\protect\citeauthoryear{Tsai and Chen}{Tsai and Chen}{2019}]%
        {tsai_elastimpact_2019}
\bibfield{author}{\bibinfo{person}{Hsin-Ruey Tsai} {and}
  \bibinfo{person}{Bing-Yu Chen}.} \bibinfo{year}{2019}\natexlab{}.
\newblock \showarticletitle{{ElastImpact}: 2.{5D} {Multilevel} {Instant}
  {Impact} {Using} {Elasticity} on {Head}-{Mounted} {Displays}}. In
  \bibinfo{booktitle}{\emph{Proceedings of the 32nd {Annual} {ACM} {Symposium}
  on {User} {Interface} {Software} and {Technology}}}.
  \bibinfo{publisher}{ACM}, \bibinfo{address}{New Orleans LA USA},
  \bibinfo{pages}{429--437}.
\newblock
\showISBNx{978-1-4503-6816-2}
\urldef\tempurl%
\url{https://doi.org/10.1145/3332165.3347931}
\showDOI{\tempurl}


\bibitem[\protect\citeauthoryear{Tsai and Rekimoto}{Tsai and Rekimoto}{2018}]%
        {tsai_elasticvr:_2018}
\bibfield{author}{\bibinfo{person}{Hsin-Ruey Tsai} {and} \bibinfo{person}{Jun
  Rekimoto}.} \bibinfo{year}{2018}\natexlab{}.
\newblock \showarticletitle{{ElasticVR}: {Providing} {Multi}-level {Active} and
  {Passive} {Force} {Feedback} in {Virtual} {Reality} {Using} {Elasticity}}. In
  \bibinfo{booktitle}{\emph{Extended {Abstracts} of the 2018 {CHI} {Conference}
  on {Human} {Factors} in {Computing} {Systems} - {CHI} '18}}.
  \bibinfo{publisher}{ACM Press}, \bibinfo{address}{Montreal QC, Canada},
  \bibinfo{pages}{1--4}.
\newblock
\showISBNx{978-1-4503-5621-3}
\urldef\tempurl%
\url{https://doi.org/10.1145/3170427.3186540}
\showDOI{\tempurl}


\bibitem[\protect\citeauthoryear{Tsykunov, Ibrahimov, Vasquez, and
  Tsetserukou}{Tsykunov et~al\mbox{.}}{2019}]%
        {tsykunov_slingdrone:_2019}
\bibfield{author}{\bibinfo{person}{Evgeny Tsykunov}, \bibinfo{person}{Roman
  Ibrahimov}, \bibinfo{person}{Derek Vasquez}, {and} \bibinfo{person}{Dzmitry
  Tsetserukou}.} \bibinfo{year}{2019}\natexlab{}.
\newblock \showarticletitle{{SlingDrone}: {Mixed} {Reality} {System} for
  {Pointing} and {Interaction} {Using} a {Single} {Drone}}. In
  \bibinfo{booktitle}{\emph{25th {ACM} {Symposium} on {Virtual} {Reality}
  {Software} and {Technology} on - {VRST} '19}}. \bibinfo{publisher}{ACM
  Press}, \bibinfo{address}{Parramatta, NSW, Australia}, \bibinfo{pages}{1--5}.
\newblock
\showISBNx{978-1-4503-7001-1}
\urldef\tempurl%
\url{https://doi.org/10.1145/3359996.3364271}
\showDOI{\tempurl}


\bibitem[\protect\citeauthoryear{Tsykunov and Tsetserukou}{Tsykunov and
  Tsetserukou}{2019}]%
        {tsykunov_wiredswarm:_2019}
\bibfield{author}{\bibinfo{person}{Evgeny Tsykunov} {and}
  \bibinfo{person}{Dzmitry Tsetserukou}.} \bibinfo{year}{2019}\natexlab{}.
\newblock \showarticletitle{{WiredSwarm}: {High} {Resolution} {Haptic}
  {Feedback} {Provided} by a {Swarm} of {Drones} to the {User}’s {Fingers}
  for {VR} interaction}. In \bibinfo{booktitle}{\emph{25th {ACM} {Symposium} on
  {Virtual} {Reality} {Software} and {Technology} on - {VRST} '19}}.
  \bibinfo{publisher}{ACM Press}, \bibinfo{address}{Parramatta, NSW,
  Australia}, \bibinfo{pages}{1--2}.
\newblock
\showISBNx{978-1-4503-7001-1}
\urldef\tempurl%
\url{https://doi.org/10.1145/3359996.3364789}
\showDOI{\tempurl}


\bibitem[\protect\citeauthoryear{Ullrich}{Ullrich}{2012}]%
        {ullrich_haptic_2012}
\bibfield{author}{\bibinfo{person}{Sebastian Ullrich}.}
  \bibinfo{year}{2012}\natexlab{}.
\newblock \showarticletitle{Haptic {Palpation} for {Medical} {Simulation} in
  {Virtual} {Environments}}.
\newblock \bibinfo{journal}{\emph{IEEE TRANSACTIONS ON VISUALIZATION AND
  COMPUTER GRAPHICS}} \bibinfo{volume}{18}, \bibinfo{number}{4}
  (\bibinfo{year}{2012}), \bibinfo{pages}{9}.
\newblock


\bibitem[\protect\citeauthoryear{Usoh, Arthur, Whitton, Bastos, Steed, Slater,
  and Brooks}{Usoh et~al\mbox{.}}{1999}]%
        {usoh_walking_1999}
\bibfield{author}{\bibinfo{person}{Martin Usoh}, \bibinfo{person}{Kevin
  Arthur}, \bibinfo{person}{Mary~C. Whitton}, \bibinfo{person}{Rui Bastos},
  \bibinfo{person}{Anthony Steed}, \bibinfo{person}{Mel Slater}, {and}
  \bibinfo{person}{Frederick~P. Brooks}.} \bibinfo{year}{1999}\natexlab{}.
\newblock \showarticletitle{Walking {\textgreater} walking-in-place
  {\textgreater} flying, in virtual environments}. In
  \bibinfo{booktitle}{\emph{Proceedings of the 26th annual conference on
  {Computer} graphics and interactive techniques - {SIGGRAPH} '99}}.
  \bibinfo{publisher}{ACM Press}, \bibinfo{address}{Not Known},
  \bibinfo{pages}{359--364}.
\newblock
\showISBNx{978-0-201-48560-8}
\urldef\tempurl%
\url{https://doi.org/10.1145/311535.311589}
\showDOI{\tempurl}


\bibitem[\protect\citeauthoryear{Usoh, Catena, Arman, and Slater}{Usoh
  et~al\mbox{.}}{2000}]%
        {usoh_using_2000}
\bibfield{author}{\bibinfo{person}{Martin Usoh}, \bibinfo{person}{Ernest
  Catena}, \bibinfo{person}{Sima Arman}, {and} \bibinfo{person}{Mel Slater}.}
  \bibinfo{year}{2000}\natexlab{}.
\newblock \showarticletitle{Using {Presence} {Questionnaires} in {Reality}}.
\newblock \bibinfo{journal}{\emph{Presence: Teleoperators and Virtual
  Environments}} \bibinfo{volume}{9}, \bibinfo{number}{5} (\bibinfo{date}{Oct.}
  \bibinfo{year}{2000}), \bibinfo{pages}{497--503}.
\newblock
\showISSN{1054-7460, 1531-3263}
\urldef\tempurl%
\url{https://doi.org/10.1162/105474600566989}
\showDOI{\tempurl}


\bibitem[\protect\citeauthoryear{Varalakshmi, Thriveni, Venugopal, and
  Patnaik}{Varalakshmi et~al\mbox{.}}{2012}]%
        {varalakshmi_haptics_2012}
\bibfield{author}{\bibinfo{person}{Varalakshmi}, \bibinfo{person}{Thriveni},
  \bibinfo{person}{Venugopal}, {and} \bibinfo{person}{Patnaik}.}
  \bibinfo{year}{2012}\natexlab{}.
\newblock \showarticletitle{Haptics: {State} of the {Art} {Survey}}.
\newblock \bibinfo{journal}{\emph{IJCSI International Journal of Computer
  Science Issues}} (\bibinfo{year}{2012}).
\newblock
\urldef\tempurl%
\url{https://core.ac.uk/download/pdf/25725449.pdf}
\showURL{%
\tempurl}


\bibitem[\protect\citeauthoryear{Villa~Salazar, Pacchierotti, De~Tinguy De
  La~Girouliere, Maciel, and Marchal}{Villa~Salazar et~al\mbox{.}}{2020}]%
        {villa_salazar_altering_2020}
\bibfield{author}{\bibinfo{person}{David~Steeven Villa~Salazar},
  \bibinfo{person}{Claudio Pacchierotti}, \bibinfo{person}{Xavier De~Tinguy De
  La~Girouliere}, \bibinfo{person}{Anderson Maciel}, {and}
  \bibinfo{person}{Maud Marchal}.} \bibinfo{year}{2020}\natexlab{}.
\newblock \showarticletitle{Altering the {Stiffness}, {Friction}, and {Shape}
  {Perception} of {Tangible} {Objects} in {Virtual} {Reality} {Using}
  {Wearable} {Haptics}}.
\newblock \bibinfo{journal}{\emph{IEEE Transactions on Haptics}}
  (\bibinfo{year}{2020}), \bibinfo{pages}{1--1}.
\newblock
\showISSN{1939-1412, 2329-4051, 2334-0134}
\urldef\tempurl%
\url{https://doi.org/10.1109/TOH.2020.2967389}
\showDOI{\tempurl}


\bibitem[\protect\citeauthoryear{Vonach, Gatterer, and Kaufmann}{Vonach
  et~al\mbox{.}}{2017}]%
        {vonach_vrrobot:_2017}
\bibfield{author}{\bibinfo{person}{Emanuel Vonach}, \bibinfo{person}{Clemens
  Gatterer}, {and} \bibinfo{person}{Hannes Kaufmann}.}
  \bibinfo{year}{2017}\natexlab{}.
\newblock \showarticletitle{{VRRobot}: {Robot} actuated props in an infinite
  virtual environment}. In \bibinfo{booktitle}{\emph{2017 {IEEE} {Virtual}
  {Reality} ({VR})}}. \bibinfo{publisher}{IEEE}, \bibinfo{address}{Los Angeles,
  CA, USA}, \bibinfo{pages}{74--83}.
\newblock
\showISBNx{978-1-5090-6647-6}
\urldef\tempurl%
\url{https://doi.org/10.1109/VR.2017.7892233}
\showDOI{\tempurl}


\bibitem[\protect\citeauthoryear{Wang, Huang, Hsu, Lin, Chiu, Hou, and
  Chen}{Wang et~al\mbox{.}}{2020b}]%
        {wang_gaiters_2020}
\bibfield{author}{\bibinfo{person}{Chi Wang}, \bibinfo{person}{Da-Yuan Huang},
  \bibinfo{person}{Shuo-Wen Hsu}, \bibinfo{person}{Cheng-Lung Lin},
  \bibinfo{person}{Yeu-Luen Chiu}, \bibinfo{person}{Chu-En Hou}, {and}
  \bibinfo{person}{Bing-Yu Chen}.} \bibinfo{year}{2020}\natexlab{b}.
\newblock \showarticletitle{Gaiters: {Exploring} {Skin} {Stretch} {Feedback} on
  the {Legs} for {Enhancing} {Virtual} {Reality} {Experiences}}.
\newblock  (\bibinfo{year}{2020}), \bibinfo{pages}{14}.
\newblock


\bibitem[\protect\citeauthoryear{Wang, Guo, Yuru, Weiliang, and Jing}{Wang
  et~al\mbox{.}}{2020a}]%
        {wang_haptic_2020}
\bibfield{author}{\bibinfo{person}{Dangxiao Wang}, \bibinfo{person}{Yuan Guo},
  \bibinfo{person}{Zhang Yuru}, \bibinfo{person}{XY Weiliang}, {and}
  \bibinfo{person}{WWIA Jing}.} \bibinfo{year}{2020}\natexlab{a}.
\newblock \bibinfo{title}{Haptic display for virtual reality: progress and
  challenges {\textbar} {Elsevier} {Enhanced} {Reader}}.
\newblock
\newblock
\urldef\tempurl%
\url{https://doi.org/10.3724/SP.J.2096-5796.2019.0008}
\showDOI{\tempurl}
\newblock
\shownote{ISSN: 2096-5796.}


\bibitem[\protect\citeauthoryear{Wang, Ohnishi, and Xu}{Wang
  et~al\mbox{.}}{2020d}]%
        {wang_multimodal_2020}
\bibfield{author}{\bibinfo{person}{Dangxiao Wang}, \bibinfo{person}{Kouhei
  Ohnishi}, {and} \bibinfo{person}{Weiliang Xu}.}
  \bibinfo{year}{2020}\natexlab{d}.
\newblock \showarticletitle{Multimodal {Haptic} {Display} for {Virtual}
  {Reality}: {A} {Survey}}.
\newblock \bibinfo{journal}{\emph{IEEE Transactions on Industrial Electronics}}
  \bibinfo{volume}{67}, \bibinfo{number}{1} (\bibinfo{date}{Jan.}
  \bibinfo{year}{2020}), \bibinfo{pages}{610--623}.
\newblock
\showISSN{0278-0046, 1557-9948}
\urldef\tempurl%
\url{https://doi.org/10.1109/TIE.2019.2920602}
\showDOI{\tempurl}


\bibitem[\protect\citeauthoryear{Wang, Li, Cao, Luo, Ou, Raiti, Yu, Patel, and
  Shi}{Wang et~al\mbox{.}}{2020c}]%
        {wang_movevr_2020}
\bibfield{author}{\bibinfo{person}{Yuntao Wang}, \bibinfo{person}{Hanchuan Li},
  \bibinfo{person}{Zhengyi Cao}, \bibinfo{person}{Huiyi Luo},
  \bibinfo{person}{Ke Ou}, \bibinfo{person}{John Raiti}, \bibinfo{person}{Chun
  Yu}, \bibinfo{person}{Shwetak Patel}, {and} \bibinfo{person}{Yuanchun Shi}.}
  \bibinfo{year}{2020}\natexlab{c}.
\newblock \showarticletitle{{MoveVR}: {Enabling} {Multiform} {Force} {Feedback}
  in {Virtual} {Reality} using {Household} {Cleaning} {Robot}}.
\newblock  (\bibinfo{year}{2020}), \bibinfo{pages}{12}.
\newblock


\bibitem[\protect\citeauthoryear{Wei, Tsai, Liao, Tsai, Chen, Wang, and
  Chen}{Wei et~al\mbox{.}}{2020}]%
        {wei_elastilinks_2020}
\bibfield{author}{\bibinfo{person}{Tzu-Yun Wei}, \bibinfo{person}{Hsin-Ruey
  Tsai}, \bibinfo{person}{Yu-So Liao}, \bibinfo{person}{Chieh Tsai},
  \bibinfo{person}{Yi-Shan Chen}, \bibinfo{person}{Chi Wang}, {and}
  \bibinfo{person}{Bing-Yu Chen}.} \bibinfo{year}{2020}\natexlab{}.
\newblock \showarticletitle{{ElastiLinks}: {Force} {Feedback} between {VR}
  {Controllers} with {Dynamic} {Points} of {Application} of {Force}}. In
  \bibinfo{booktitle}{\emph{Proceedings of the 33rd {Annual} {ACM} {Symposium}
  on {User} {Interface} {Software} and {Technology}}}.
  \bibinfo{publisher}{ACM}, \bibinfo{address}{Virtual Event USA},
  \bibinfo{pages}{1023--1034}.
\newblock
\showISBNx{978-1-4503-7514-6}
\urldef\tempurl%
\url{https://doi.org/10.1145/3379337.3415836}
\showDOI{\tempurl}


\bibitem[\protect\citeauthoryear{Wexelblat}{Wexelblat}{1993}]%
        {wexelblat_virtual_1993}
\bibfield{author}{\bibinfo{person}{Alan Wexelblat}.}
  \bibinfo{year}{1993}\natexlab{}.
\newblock \bibinfo{title}{Virtual reality: applications and explorations}.
\newblock
\newblock
\urldef\tempurl%
\url{http://libertar.io/lab/wp-content/uploads/2016/02/Virtual.Reality.-.Applications.And_.Explorations.pdf/page=164}
\showURL{%
\tempurl}
\newblock
\shownote{Myron Krueger, Artificial reality 2 An easy entry to Virtual reality
  Chap 7.}


\bibitem[\protect\citeauthoryear{Whitmire, Benko, Holz, Ofek, and
  Sinclair}{Whitmire et~al\mbox{.}}{2018}]%
        {whitmire_haptic_2018}
\bibfield{author}{\bibinfo{person}{Eric Whitmire}, \bibinfo{person}{Hrvoje
  Benko}, \bibinfo{person}{Christian Holz}, \bibinfo{person}{Eyal Ofek}, {and}
  \bibinfo{person}{Mike Sinclair}.} \bibinfo{year}{2018}\natexlab{}.
\newblock \showarticletitle{Haptic {Revolver}: {Touch}, {Shear}, {Texture}, and
  {Shape} {Rendering} on a {Reconfigurable} {Virtual} {Reality} {Controller}}.
  In \bibinfo{booktitle}{\emph{Proceedings of the 2018 {CHI} {Conference} on
  {Human} {Factors} in {Computing} {Systems} - {CHI} '18}}.
  \bibinfo{publisher}{ACM Press}, \bibinfo{address}{Montreal QC, Canada},
  \bibinfo{pages}{1--12}.
\newblock
\showISBNx{978-1-4503-5620-6}
\urldef\tempurl%
\url{https://doi.org/10.1145/3173574.3173660}
\showDOI{\tempurl}


\bibitem[\protect\citeauthoryear{Winther, Ravindran, Svendsen, and
  Feuchtner}{Winther et~al\mbox{.}}{2020}]%
        {winther_design_2020}
\bibfield{author}{\bibinfo{person}{Frederik Winther}, \bibinfo{person}{Linoj
  Ravindran}, \bibinfo{person}{Kasper~Paabol Svendsen}, {and}
  \bibinfo{person}{Tiare Feuchtner}.} \bibinfo{year}{2020}\natexlab{}.
\newblock \showarticletitle{Design and {Evaluation} of a {VR} {Training}
  {Simulation} for {Pump} {Maintenance} {Based} on a {Use} {Case} at
  {Grundfos}}. In \bibinfo{booktitle}{\emph{2020 {IEEE} {Conference} on
  {Virtual} {Reality} and {3D} {User} {Interfaces} ({VR})}}.
  \bibinfo{publisher}{IEEE}, \bibinfo{address}{Atlanta, GA, USA},
  \bibinfo{pages}{738--746}.
\newblock
\showISBNx{978-1-72815-608-8}
\urldef\tempurl%
\url{https://doi.org/10.1109/VR46266.2020.1580939036664}
\showDOI{\tempurl}


\bibitem[\protect\citeauthoryear{Witmer and Singer}{Witmer and Singer}{1998}]%
        {witmer_measuring_1998}
\bibfield{author}{\bibinfo{person}{Bob~G. Witmer} {and}
  \bibinfo{person}{Michael~J. Singer}.} \bibinfo{year}{1998}\natexlab{}.
\newblock \showarticletitle{Measuring {Presence} in {Virtual} {Environments}:
  {A} {Presence} {Questionnaire}}.
\newblock \bibinfo{journal}{\emph{Presence: Teleoperators and Virtual
  Environments}} \bibinfo{volume}{7}, \bibinfo{number}{3} (\bibinfo{date}{June}
  \bibinfo{year}{1998}), \bibinfo{pages}{225--240}.
\newblock
\showISSN{1054-7460, 1531-3263}
\urldef\tempurl%
\url{https://doi.org/10.1162/105474698565686}
\showDOI{\tempurl}


\bibitem[\protect\citeauthoryear{Xia, Herscher, Perlin, and Wigdor}{Xia
  et~al\mbox{.}}{2018}]%
        {xia_spacetime:_2018}
\bibfield{author}{\bibinfo{person}{Haijun Xia}, \bibinfo{person}{Sebastian
  Herscher}, \bibinfo{person}{Ken Perlin}, {and} \bibinfo{person}{Daniel
  Wigdor}.} \bibinfo{year}{2018}\natexlab{}.
\newblock \showarticletitle{Spacetime: {Enabling} {Fluid} {Individual} and
  {Collaborative} {Editing} in {Virtual} {Reality}}. In
  \bibinfo{booktitle}{\emph{The 31st {Annual} {ACM} {Symposium} on {User}
  {Interface} {Software} and {Technology} - {UIST} '18}}.
  \bibinfo{publisher}{ACM Press}, \bibinfo{address}{Berlin, Germany},
  \bibinfo{pages}{853--866}.
\newblock
\showISBNx{978-1-4503-5948-1}
\urldef\tempurl%
\url{https://doi.org/10.1145/3242587.3242597}
\showDOI{\tempurl}


\bibitem[\protect\citeauthoryear{Xia}{Xia}{2016}]%
        {xia_haptics_2016}
\bibfield{author}{\bibinfo{person}{Pingjun Xia}.}
  \bibinfo{year}{2016}\natexlab{}.
\newblock \showarticletitle{Haptics for {Product} {Design} and {Manufacturing}
  {Simulation}}.
\newblock \bibinfo{journal}{\emph{IEEE Transactions on Haptics}}
  \bibinfo{volume}{9}, \bibinfo{number}{3} (\bibinfo{date}{July}
  \bibinfo{year}{2016}), \bibinfo{pages}{358--375}.
\newblock
\showISSN{1939-1412}
\urldef\tempurl%
\url{https://doi.org/10.1109/TOH.2016.2554551}
\showDOI{\tempurl}


\bibitem[\protect\citeauthoryear{Yamaguchi, Kato, Kuroda, Kiyokawa, and
  Takemura}{Yamaguchi et~al\mbox{.}}{2016}]%
        {yamaguchi_non-grounded_2016}
\bibfield{author}{\bibinfo{person}{Kotaro Yamaguchi}, \bibinfo{person}{Ginga
  Kato}, \bibinfo{person}{Yoshihiro Kuroda}, \bibinfo{person}{Kiyoshi
  Kiyokawa}, {and} \bibinfo{person}{Haruo Takemura}.}
  \bibinfo{year}{2016}\natexlab{}.
\newblock \showarticletitle{A {Non}-grounded and {Encountered}-type {Haptic}
  {Display} {Using} a {Drone}}. In \bibinfo{booktitle}{\emph{Proceedings of the
  2016 {Symposium} on {Spatial} {User} {Interaction} - {SUI} '16}}.
  \bibinfo{publisher}{ACM Press}, \bibinfo{address}{Tokyo, Japan},
  \bibinfo{pages}{43--46}.
\newblock
\showISBNx{978-1-4503-4068-7}
\urldef\tempurl%
\url{https://doi.org/10.1145/2983310.2985746}
\showDOI{\tempurl}


\bibitem[\protect\citeauthoryear{Yang, Holz, Ofek, and Wilson}{Yang
  et~al\mbox{.}}{2019}]%
        {yang_dreamwalker:_2019}
\bibfield{author}{\bibinfo{person}{Jackie~(Junrui) Yang},
  \bibinfo{person}{Christian Holz}, \bibinfo{person}{Eyal Ofek}, {and}
  \bibinfo{person}{Andrew~D. Wilson}.} \bibinfo{year}{2019}\natexlab{}.
\newblock \showarticletitle{{DreamWalker}: {Substituting} {Real}-{World}
  {Walking} {Experiences} with a {Virtual} {Reality}}. In
  \bibinfo{booktitle}{\emph{Proceedings of the 32nd {Annual} {ACM} {Symposium}
  on {User} {Interface} {Software} and {Technology} - {UIST} '19}}.
  \bibinfo{publisher}{ACM Press}, \bibinfo{address}{New Orleans, LA, USA},
  \bibinfo{pages}{1093--1107}.
\newblock
\showISBNx{978-1-4503-6816-2}
\urldef\tempurl%
\url{https://doi.org/10.1145/3332165.3347875}
\showDOI{\tempurl}


\bibitem[\protect\citeauthoryear{Ye, Chen, and Chan}{Ye et~al\mbox{.}}{2019}]%
        {ye_pull-ups:_2019}
\bibfield{author}{\bibinfo{person}{Yuan-Syun Ye}, \bibinfo{person}{Hsin-Yu
  Chen}, {and} \bibinfo{person}{Liwei Chan}.} \bibinfo{year}{2019}\natexlab{}.
\newblock \showarticletitle{Pull-{Ups}: {Enhancing} {Suspension} {Activities}
  in {Virtual} {Reality} with {Body}-{Scale} {Kinesthetic} {Force} {Feedback}}.
  In \bibinfo{booktitle}{\emph{Proceedings of the 32nd {Annual} {ACM}
  {Symposium} on {User} {Interface} {Software} and {Technology} - {UIST} '19}}.
  \bibinfo{publisher}{ACM Press}, \bibinfo{address}{New Orleans, LA, USA},
  \bibinfo{pages}{791--801}.
\newblock
\showISBNx{978-1-4503-6816-2}
\urldef\tempurl%
\url{https://doi.org/10.1145/3332165.3347874}
\showDOI{\tempurl}


\bibitem[\protect\citeauthoryear{Yixian, Takashima, Tang, Tanno, Fujita, and
  Kitamura}{Yixian et~al\mbox{.}}{2020}]%
        {yixian_zoomwalls_2020}
\bibfield{author}{\bibinfo{person}{Yan Yixian}, \bibinfo{person}{Kazuki
  Takashima}, \bibinfo{person}{Anthony Tang}, \bibinfo{person}{Takayuki Tanno},
  \bibinfo{person}{Kazuyuki Fujita}, {and} \bibinfo{person}{Yoshifumi
  Kitamura}.} \bibinfo{year}{2020}\natexlab{}.
\newblock \showarticletitle{{ZoomWalls}: {Dynamic} {Walls} that {Simulate}
  {Haptic} {Infrastructure} for {Room}-scale {VR} {World}}. In
  \bibinfo{booktitle}{\emph{Proceedings of the 33rd {Annual} {ACM} {Symposium}
  on {User} {Interface} {Software} and {Technology}}}
  \emph{(\bibinfo{series}{{UIST} '20})}. \bibinfo{publisher}{Association for
  Computing Machinery}, \bibinfo{address}{New York, NY, USA},
  \bibinfo{pages}{223--235}.
\newblock
\showISBNx{978-1-4503-7514-6}
\urldef\tempurl%
\url{https://doi.org/10.1145/3379337.3415859}
\showDOI{\tempurl}


\bibitem[\protect\citeauthoryear{Yokokohji, Hollis, and Kanade}{Yokokohji
  et~al\mbox{.}}{1999}]%
        {yokokohji_wysiwyf_1999}
\bibfield{author}{\bibinfo{person}{Yasuyoshi Yokokohji},
  \bibinfo{person}{Ralph~L. Hollis}, {and} \bibinfo{person}{Takeo Kanade}.}
  \bibinfo{year}{1999}\natexlab{}.
\newblock \showarticletitle{{WYSIWYF} {Display}: {A} {Visual}/{Haptic}
  {Interface} to {Virtual} {Environment}}.
\newblock \bibinfo{journal}{\emph{Presence: Teleoperators and Virtual
  Environments}} \bibinfo{volume}{8}, \bibinfo{number}{4} (\bibinfo{date}{Aug.}
  \bibinfo{year}{1999}), \bibinfo{pages}{412--434}.
\newblock
\showISSN{1054-7460, 1531-3263}
\urldef\tempurl%
\url{https://doi.org/10.1162/105474699566314}
\showDOI{\tempurl}


\bibitem[\protect\citeauthoryear{Yokokohji, Kinoshita, and Yoshikawa}{Yokokohji
  et~al\mbox{.}}{2001}]%
        {yokokohji_path_2001}
\bibfield{author}{\bibinfo{person}{Y. Yokokohji}, \bibinfo{person}{J.
  Kinoshita}, {and} \bibinfo{person}{T. Yoshikawa}.}
  \bibinfo{year}{2001}\natexlab{}.
\newblock \showarticletitle{Path planning for encountered-type haptic devices
  that render multiple objects in {3D} space}. In
  \bibinfo{booktitle}{\emph{Proceedings {IEEE} {Virtual} {Reality} 2001}}.
  \bibinfo{pages}{271--278}.
\newblock
\urldef\tempurl%
\url{https://doi.org/10.1109/VR.2001.913796}
\showDOI{\tempurl}


\bibitem[\protect\citeauthoryear{Yokokohji, Muramori, Sato, and
  Yoshikawa}{Yokokohji et~al\mbox{.}}{2005}]%
        {yokokohji_haptic_2005}
\bibfield{author}{\bibinfo{person}{Yasuyoshi Yokokohji},
  \bibinfo{person}{Nobuhiko Muramori}, \bibinfo{person}{Yuji Sato}, {and}
  \bibinfo{person}{Tsuneo Yoshikawa}.} \bibinfo{year}{2005}\natexlab{}.
\newblock \bibinfo{booktitle}{\emph{Haptic {Display} for {Multiple} {Fingertip}
  {Contacts} {Based} on the {Observation} of {Human} {Grasping} {Behaviors}}}.
\newblock


\bibitem[\protect\citeauthoryear{Yoshida, Sun, and Kuzuoka}{Yoshida
  et~al\mbox{.}}{2020}]%
        {yoshida_pocopo_2020}
\bibfield{author}{\bibinfo{person}{Shigeo Yoshida}, \bibinfo{person}{Yuqian
  Sun}, {and} \bibinfo{person}{Hideaki Kuzuoka}.}
  \bibinfo{year}{2020}\natexlab{}.
\newblock \showarticletitle{{PoCoPo}: {Handheld} {Pin}-based {Shape} {Display}
  for {Haptic} {Rendering} in {Virtual} {Reality}}. In
  \bibinfo{booktitle}{\emph{Proceedings of the 2020 {CHI} {Conference} on
  {Human} {Factors} in {Computing} {Systems}}}. \bibinfo{publisher}{ACM},
  \bibinfo{address}{Honolulu HI USA}, \bibinfo{pages}{1--13}.
\newblock
\showISBNx{978-1-4503-6708-0}
\urldef\tempurl%
\url{https://doi.org/10.1145/3313831.3376358}
\showDOI{\tempurl}


\bibitem[\protect\citeauthoryear{Zenner and Kruger}{Zenner and Kruger}{2017}]%
        {zenner_shifty:_2017}
\bibfield{author}{\bibinfo{person}{Andre Zenner} {and} \bibinfo{person}{Antonio
  Kruger}.} \bibinfo{year}{2017}\natexlab{}.
\newblock \showarticletitle{Shifty: {A} {Weight}-{Shifting} {Dynamic} {Passive}
  {Haptic} {Proxy} to {Enhance} {Object} {Perception} in {Virtual} {Reality}}.
\newblock \bibinfo{journal}{\emph{IEEE Transactions on Visualization and
  Computer Graphics}} \bibinfo{volume}{23}, \bibinfo{number}{4}
  (\bibinfo{date}{April} \bibinfo{year}{2017}), \bibinfo{pages}{1285--1294}.
\newblock
\showISSN{1077-2626, 1941-0506, 2160-9306}
\urldef\tempurl%
\url{https://doi.org/10.1109/TVCG.2017.2656978}
\showDOI{\tempurl}


\bibitem[\protect\citeauthoryear{Zenner and Krüger}{Zenner and
  Krüger}{2019}]%
        {zenner_drag::_2019}
\bibfield{author}{\bibinfo{person}{André Zenner} {and}
  \bibinfo{person}{Antonio Krüger}.} \bibinfo{year}{2019}\natexlab{}.
\newblock \showarticletitle{Drag:on: {A} {Virtual} {Reality} {Controller}
  {Providing} {Haptic} {Feedback} {Based} on {Drag} and {Weight} {Shift}}. In
  \bibinfo{booktitle}{\emph{Proceedings of the 2019 {CHI} {Conference} on
  {Human} {Factors} in {Computing} {Systems} - {CHI} '19}}.
  \bibinfo{publisher}{ACM Press}, \bibinfo{address}{Glasgow, Scotland Uk},
  \bibinfo{pages}{1--12}.
\newblock
\showISBNx{978-1-4503-5970-2}
\urldef\tempurl%
\url{https://doi.org/10.1145/3290605.3300441}
\showDOI{\tempurl}


\bibitem[\protect\citeauthoryear{Zhao}{Zhao}{2009}]%
        {zhao_survey_2009}
\bibfield{author}{\bibinfo{person}{QinPing Zhao}.}
  \bibinfo{year}{2009}\natexlab{}.
\newblock \showarticletitle{A survey on virtual reality}.
\newblock \bibinfo{journal}{\emph{Science in China Series F: Information
  Sciences}} \bibinfo{volume}{52}, \bibinfo{number}{3} (\bibinfo{date}{March}
  \bibinfo{year}{2009}), \bibinfo{pages}{348--400}.
\newblock
\showISSN{1009-2757, 1862-2836}
\urldef\tempurl%
\url{https://doi.org/10.1007/s11432-009-0066-0}
\showDOI{\tempurl}


\bibitem[\protect\citeauthoryear{Zhao and Follmer}{Zhao and Follmer}{2018}]%
        {zhao_functional_2018}
\bibfield{author}{\bibinfo{person}{Yiwei Zhao} {and} \bibinfo{person}{Sean
  Follmer}.} \bibinfo{year}{2018}\natexlab{}.
\newblock \showarticletitle{A {Functional} {Optimization} {Based} {Approach}
  for {Continuous} {3D} {Retargeted} {Touch} of {Arbitrary}, {Complex}
  {Boundaries} in {Haptic} {Virtual} {Reality}}. In
  \bibinfo{booktitle}{\emph{Proceedings of the 2018 {CHI} {Conference} on
  {Human} {Factors} in {Computing} {Systems} - {CHI} '18}}.
  \bibinfo{publisher}{ACM Press}, \bibinfo{address}{Montreal QC, Canada},
  \bibinfo{pages}{1--12}.
\newblock
\showISBNx{978-1-4503-5620-6}
\urldef\tempurl%
\url{https://doi.org/10.1145/3173574.3174118}
\showDOI{\tempurl}


\bibitem[\protect\citeauthoryear{Zhao, Kim, Wang, Le~Goc, and Follmer}{Zhao
  et~al\mbox{.}}{2017}]%
        {zhao_robotic_2017}
\bibfield{author}{\bibinfo{person}{Yiwei Zhao}, \bibinfo{person}{Lawrence~H.
  Kim}, \bibinfo{person}{Ye Wang}, \bibinfo{person}{Mathieu Le~Goc}, {and}
  \bibinfo{person}{Sean Follmer}.} \bibinfo{year}{2017}\natexlab{}.
\newblock \showarticletitle{Robotic {Assembly} of {Haptic} {Proxy} {Objects}
  for {TangibleInteraction} and {Virtual} {Reality}}. In
  \bibinfo{booktitle}{\emph{Proceedings of the {Interactive} {Surfaces} and
  {Spaces} on {ZZZ} - {ISS} '17}}. \bibinfo{publisher}{ACM Press},
  \bibinfo{address}{Brighton, United Kingdom}, \bibinfo{pages}{82--91}.
\newblock
\showISBNx{978-1-4503-4691-7}
\urldef\tempurl%
\url{https://doi.org/10.1145/3132272.3134143}
\showDOI{\tempurl}


\bibitem[\protect\citeauthoryear{Zhou and Deng}{Zhou and Deng}{2009}]%
        {zhou_virtual_2009}
\bibfield{author}{\bibinfo{person}{Ning-Ning Zhou} {and}
  \bibinfo{person}{Yu-Long Deng}.} \bibinfo{year}{2009}\natexlab{}.
\newblock \showarticletitle{Virtual reality: {A} state-of-the-art survey}.
\newblock \bibinfo{journal}{\emph{International Journal of Automation and
  Computing}} \bibinfo{volume}{6}, \bibinfo{number}{4} (\bibinfo{date}{Nov.}
  \bibinfo{year}{2009}), \bibinfo{pages}{319--325}.
\newblock
\showISSN{1476-8186, 1751-8520}
\urldef\tempurl%
\url{https://doi.org/10.1007/s11633-009-0319-9}
\showDOI{\tempurl}


\bibitem[\protect\citeauthoryear{Ziat, Rolison, Shirtz, Wilbern, and
  Balcer}{Ziat et~al\mbox{.}}{2014}]%
        {ziat_enhancing_2014}
\bibfield{author}{\bibinfo{person}{Mounia Ziat}, \bibinfo{person}{Taylor
  Rolison}, \bibinfo{person}{Andrew Shirtz}, \bibinfo{person}{Daniel Wilbern},
  {and} \bibinfo{person}{Carrie~Anne Balcer}.} \bibinfo{year}{2014}\natexlab{}.
\newblock \showarticletitle{Enhancing virtual immersion through tactile
  feedback}. In \bibinfo{booktitle}{\emph{Proceedings of the adjunct
  publication of the 27th annual {ACM} symposium on {User} interface software
  and technology - {UIST}'14 {Adjunct}}}. \bibinfo{publisher}{ACM Press},
  \bibinfo{address}{Honolulu, Hawaii, USA}, \bibinfo{pages}{65--66}.
\newblock
\showISBNx{978-1-4503-3068-8}
\urldef\tempurl%
\url{https://doi.org/10.1145/2658779.2659116}
\showDOI{\tempurl}


\bibitem[\protect\citeauthoryear{Zielasko and Riecke}{Zielasko and
  Riecke}{2020}]%
        {zielasko_either_2020}
\bibfield{author}{\bibinfo{person}{Daniel Zielasko} {and}
  \bibinfo{person}{Bernhard~E Riecke}.} \bibinfo{year}{2020}\natexlab{}.
\newblock \showarticletitle{Either {Give} {Me} a {Reason} to {Stand} or an
  {Opportunity} to {Sit} in {VR}}.
\newblock  (\bibinfo{year}{2020}), \bibinfo{pages}{3}.
\newblock


\bibitem[\protect\citeauthoryear{Zilles and Salisbury}{Zilles and
  Salisbury}{1995}]%
        {zilles_constraint-based_1995}
\bibfield{author}{\bibinfo{person}{C.~B. Zilles} {and} \bibinfo{person}{J.~K.
  Salisbury}.} \bibinfo{year}{1995}\natexlab{}.
\newblock \showarticletitle{A constraint-based god-object method for haptic
  display}. In \bibinfo{booktitle}{\emph{In {International} {Conference} on
  {Intelligent} {Robots} and {Systems}}}. \bibinfo{pages}{146--151}.
\newblock


\bibitem[\protect\citeauthoryear{Zimmermann}{Zimmermann}{2008}]%
        {zimmermann_virtual_2008}
\bibfield{author}{\bibinfo{person}{Peter Zimmermann}.}
  \bibinfo{year}{2008}\natexlab{}.
\newblock \showarticletitle{Virtual {Reality} {Aided} {Design}. {A} survey of
  the use of {VR} in automotive industry}.
\newblock  (\bibinfo{date}{Jan.} \bibinfo{year}{2008}).
\newblock
\urldef\tempurl%
\url{https://doi.org/10.1007/978-1-4020-8200-9_13}
\showDOI{\tempurl}


\end{thebibliography}
